\documentclass[twocolumn,prd,superscriptaddress,altaffilletter,nofootinbib]{revtex4-1}
\usepackage{amsmath}
\usepackage{amssymb}
\usepackage{amsfonts}
\usepackage{comment}
\usepackage{graphicx}
\usepackage{hyperref}
\usepackage[nolist]{acronym}
\usepackage[usenames,dvipsnames]{xcolor}
\usepackage{xspace}
\usepackage{tikz}
\usepackage{siunitx}
\usepackage{multirow}
\usepackage{bm}
\usepackage{romannum}

\usetikzlibrary{arrows,shapes,trees,decorations.pathreplacing}
\DeclareSIUnit \parsec {pc}

\allowdisplaybreaks

\newcommand{\e}{\mathrm{e}}
\newcommand{\likelihood}{p(\{\bm{d}_i\}_{i=1}^N | \bm{\theta})}
\newcommand{\posterior}{p( \bm{\theta} | \{\bm{d}_i\}_{i=1}^N)}
\newcommand{\prior}{p(\bm{\theta})}
\newcommand{\vtheta}{\bm{\theta}}
\newcommand{\di}{\bm{d}_i}
\newcommand{\diN}{\{\bm{d}_i\}_{i=1}^N}
\newcommand{\phic}{\phi_{\mathrm{c}}}
\newcommand{\tc}{t_{\mathrm{c}}}
\newcommand{\flow}{f_{\mathrm{low}}}
\newcommand{\fstop}{f_{\mathrm{stop}}}

\newcommand{\PN}{\mathrm{PN}}
\newcommand{\fref}{f_{\mathrm{ref}}}
\newcommand{\ffref}{\left(\frac{f}{f_{\mathrm{ref}}}\right)}
\newcommand{\fhigh}{f_{\mathrm{high}}}

\newcommand{\Mc}{\mathcal{M}}
\newcommand{\thetaex}{\bm{\theta}_{\mathrm{ex}}}
\newcommand{\thetain}{\bm{\theta}_{\mathrm{in}}}

\newcommand{\bigzero}{{\bf 0}}
\newcommand{\Msun}{M_{\odot}}

\newcommand{\etamin}{\eta_{\mathrm{min}}}
\newcommand{\chimax}{\chi_{\mathrm{max}}}
\newcommand{\vpsi}{\bm{\psi}}
\newcommand{\Mct}{\mathcal{M}_{\mathrm{t}}}

\newcommand{\chieff}{\chi_{\mathrm{eff}}}

\newcommand{\tabref}[1]{Tab. \ref{#1}}
\newcommand{\figref}[1]{Fig.~\ref{#1}}
\newcommand{\secref}[1]{Sec.~\ref{#1}}
\newcommand{\appref}[1]{Appendix \ref{#1}}

\begin{document}

\title{Rapid Parameter Estimation of Gravitational Waves from Binary Neutron Star Coalescence using Focused Reduced Order Quadrature}

\author{Soichiro Morisaki}
\affiliation{Institute for Cosmic Ray Research, The University of Tokyo, 5-1-5 Kashiwanoha, Kashiwa, Chiba 277-8582, Japan}
\author{Vivien Raymond}
\affiliation{Gravity Exploration Institute, School of Physics and Astronomy, Cardiff University, The Parade, Cardiff CF24 3AA, UK}

\begin{abstract}

Rapid parameter estimation of gravitational waves from binary neutron star coalescence, in particular accurate sky localisation in minutes after the initial detection stage, is crucial for the success of multi-messenger observations.
One of the techniques to speed up the parameter estimation, which has been applied for the production analysis of the LIGO-Virgo collaboration, is reduced order quadrature (ROQ).
While it speeds up parameter estimation significantly, the time required is still on the order of hours.
Focusing on the fact that the parameter-estimation follow-up can be tuned with the information available at the detection stage, we improve the ROQ technique and develop a new technique, which we designate focused reduced order quadrature (FROQ).
We find that FROQ speeds up the parameter estimation by a factor of $\mathcal{O}(10^3)$ to $\mathcal{O}(10^4)$ and enables providing accurate source properties such as the location of a source in several tens of minutes after detection.

\end{abstract}

\maketitle

\section{Introduction} \label{Introduction}

On August 17, 2017, the LIGO-Virgo collaboration \cite{Harry:2010zz, TheVirgo:2014hva} succeeded in the first direct detection of gravitational-waves emitted by binary neutron star (BNS) coalescence in its second observation run (O2), and designated this event GW170817 \cite{TheLIGOScientific:2017qsa}.
The associated gamma-ray burst, which was later designated GRB170817A, was also detected by Fermi-GBM and INTEGRAL \cite{Goldstein:2017mmi, Savchenko:2017ffs, Monitor:2017mdv}.
These coincident detections triggered broadband electromagnetic follow-up observations ranging from radio to gamma-ray band \cite{GBM:2017lvd, Coulter:2017wya, Drout:2017ijr, Kasliwal:2017ngb, Cowperthwaite:2017dyu, Tanvir:2017pws, Evans:2017mmy, Arcavi:2017xiz, Utsumi:2017cti, Troja:2017nqp, Hallinan:2017woc}, which provided us a lot of fruitful astrophysical information \cite{Langlois:2017dyl, Creminelli:2017sry, Ezquiaga:2017ekz, Baker:2017hug, Abbott:2017xzu, Drout:2017ijr, Kasliwal:2017ngb, Cowperthwaite:2017dyu, Tanvir:2017pws, Utsumi:2017cti, Tanaka:2017qxj, Troja:2017nqp, Hallinan:2017woc, Alexander:2017aly, Margutti:2017cjl, Mooley:2017enz}.
The optical counterpart was found by the observation with the Swope telescope \cite{Coulter:2017wya} and the host galaxy was identified, which enabled a measurement of the Hubble constant in a way independent from the cosmic ladder \cite{Abbott:2017xzu}.
The near-infrared, optical and ultraviolet observations allowed us to learn the production of heavy elements at the event site \cite{Drout:2017ijr, Kasliwal:2017ngb, Cowperthwaite:2017dyu, Tanvir:2017pws, Utsumi:2017cti, Tanaka:2017qxj}.
The radio, X-ray and gamma-ray observations allowed us to learn the jet structure of ultra-relativistic jet possibly originating from the merger \cite{Troja:2017nqp, Hallinan:2017woc, Alexander:2017aly, Margutti:2017cjl, Mooley:2017enz}.
GW170817 became the first successful example of multi-messenger observations. 

One of the key ingredients for the success of multi-messenger observations is rapid parameter estimation of gravitational-wave sources.
The most important information for the follow-up observations is the 3-dimensional location of a gravitational-wave source, which is estimated with gravitational-wave data from multiple detectors \cite{Fairhurst:2009tc}.
Gravitational-wave sources detected by the LIGO-Virgo collaboration are localised in seconds after their detections by the Bayestar software \cite{Singer:2015ema, Singer:2016eax}.
The masses and spins of two colliding bodies, which can be estimated from the waveform of gravitational waves, are also helpful to determine how much the follow-up observations should be prioritized. 
The LIGO-Virgo collaboration calculates the probabilities of the system being BNS, neutron star black hole binary, binary black hole, massgap or non-astrophysical noise based on the classification that the object whose mass is less than $3\Msun$ is a neutron star, between $3\Msun$ and $5\Msun$ massgap, and larger than $5\Msun$ a black hole \cite{Kapadia:2019uut}. 
The probabilities of the system having more than one neutron stars and having the electromagnetic counterparts are also calculated based on the masses and spins \cite{Chatterjee:2019avs}.
The information is sent out to the follow-up observation community in minutes after the detections.
While these initial analyses are quite rapid, they are based on approximations and sacrifice the accuracy.
Therefore, they are finally updated by more detailed parameter estimation analyses performed by the LALInference \cite{Veitch:2014wba} or Bilby \cite{Ashton:2018jfp} software. 

The detailed parameter estimation is performed with stochastic sampling algorithms such as Markov-Chain Monte Carlo (MCMC) \cite{Metropolis:1953am, 10.1093/biomet/57.1.97} and nested sampling \cite{skilling2006}.
While they are efficient methods to explore a high-dimensional parameter space, they require millions of sequential likelihood evaluations, which are computationally costly.
At each likelihood evaluation, a gravitational-wave template waveform is calculated in frequency domain, and its correlation with gravitational-wave data is calculated.
Since a BNS signal is longer and goes up to higher frequency than that for heavier binaries, the number of frequency bins needs to be much larger to represent the waveforms accurately.
It makes the parameter estimation for the BNS events much more computationally costly, and the analysis time can be a few weeks, or even years, depending on the analysis setup without approximate methods \cite{Canizares:2014fya, Smith:2016qas}.
This long analysis time is not acceptable for the purpose of multi-messenger observations.
For example, the ultraviolet and blue optical emissions from GW170817 faded away in the time scale of a day \cite{Cowperthwaite:2017dyu, Arcavi:2017xiz}.
We thus need a technique to speed up the parameter estimation of BNS signals.

There have been various techniques proposed to speed up the gravitational-wave parameter estimation \cite{Canizares:2014fya, Smith:2016qas, Pankow:2015cra, Lange:2018pyp, Wysocki:2019grj, Vinciguerra:2017ngf, Zackay:2018qdy, Talbot:2019okv, Smith:2019ucc}.
One of these techniques, which has been applied for the production analysis of the LIGO-Virgo collaboration \cite{LIGOScientific:2018mvr, Abbott:2020uma}, is reduced order quadrature (ROQ) \cite{Canizares:2014fya, Smith:2016qas}.
Its basic idea is to approximate gravitational-wave templates as linear combinations of reduced basis vectors, which are much fewer than the frequency bins.
With this approximation, the likelihood evaluation, and hence the parameter estimation, is sped up by the ratio between the number of the original frequency bins and the number of the reduced basis vectors.

In \cite{Canizares:2014fya}, the authors constructed ROQ basis vectors of the TaylorF2 \cite{Buonanno:2009zt} waveform model for non-spinning binaries with masses between $1\Msun$ and $4\Msun$. 
This study shows that the ROQ basis vectors speed up parameter estimation with the lower frequency cutoff of $20\si{\hertz}$ by a factor of $\sim 70$, which reduces the analysis time from a couple of weeks to hours.
In \cite{Smith:2016qas}, the authors extended the technique to the IMRPhenomPv2 \cite{Hannam:2013oca} waveform model, a phenomenological waveform including orbital precession effects \cite{Thorne:1980ru}, and constructed their basis vectors with the lower frequency cutoff of $20\si{\hertz}$.
With the basis, they succeeded in speeding up parameter estimation of relatively heavy BNS signals (e.g. $1.7\Msun$-$1.7\Msun$) by a factor of $300$ (See Table \Romannum{1} in \cite{Smith:2016qas}).
It reduces the analysis time from half a year to half a day.
The basis can also be applied to typical BNS mergers (i.e. $1.4\Msun$-$1.4\Msun$) with the mass-frequency scaling described in their paper, while the lower frequency cutoff needs to be increased in this case.
The ROQ basis vectors of IMRPhenomDNRT and IMRPhenomPv2NRT waveform models \cite{Dietrich:2018uni}, which are phenomenological waveforms \cite{Khan:2015jqa, Hannam:2013oca} including tidal effects of neutron stars \cite{Dietrich:2017aum}, were also constructed for parameter estimation of GW190425 \cite{pDNRT_roqdata, pv2NRT_roqdata, Abbott:2020uma}, which speed up parameter estimation by similar factors.
While they are significant speedups, the analysis time is on the order of hours and still not sufficiently short for the follow-up observations of the rapidly dimming ultraviolet and blue optical counterparts.
In addition, there is an interesting possibility that some ejected neutrons remain free and their decays power the electromagnetic emission in the timescale of $\sim30$ minutes \cite{Ishii:2018yjg}, for which the analysis time of hours is far too long.

In this work, we improve the ROQ technique to speed up the parameter estimation of BNS signals further, and designated our improved technique as focused reduced order quadrature (FROQ). 
The basic idea of FROQ is to tune the parameter-estimation follow-up with the information available at the detection.
For the detections of gravitational waves from compact binary coalescences, the data are matched filtered with theoretically expected gravitational-wave waveforms \cite{Thorne:300yrs, Allen:2005fk}.
The set of template waveforms used for matched filtering is called template bank.
In order not to miss the signals from the population of the binaries, the template bank contains a lot of template waveforms for various masses and spins \cite{Owen:1998dk, Brown:2012qf, Harry:2009ea, Roy:2017oul}.
Matched filter signal-to-noise ratio is calculated for each mass and spin in the template bank.
To mitigate the effect of the non-Gaussian nature of the instrumental noise, some additional quantities such as $\chi^2$ \cite{Allen:2004gu} are also calculated.
Finally, those quantities are combined to form detection statistic, which quantifies the significance of the signal, and the detection is claimed if the detection statistic maximized over the template bank exceeds a threshold \cite{Messick:2016aqy, Usman:2015kfa}.

The masses and spins maximizing the detection statistic, which we call ``trigger values", can guide the exploration in the parameter space.
Actually, some combinations of the trigger values, such as chirpmass $\Mc$ (See \eqref{eq:chirpmass}), are very reliable \cite{Biscoveanu:2019ugx}, and we can focus on a very narrow range of them in the follow-up parameter estimation.
For this focused exploration, we can use reduced basis vectors constructed over the narrow parameter space while the previous studies \cite{Canizares:2014fya, Smith:2016qas} used those constructed over a broad parameter space.
Since the waveforms over the narrow parameter space are quite similar, the number of required basis vectors can be significantly reduced, and hence the parameter estimation is significantly sped up.
In this paper, we formulate a method to restrict parameter space based on the trigger values and investigate how much the parameter estimation can be sped up.

The paper is organized as follows.
In \secref{sec:basics}, we introduce the standard Bayesian inference used for gravitational-wave parameter estimation studies and review the previous studies on the ROQ technique.
In \secref{sec:froq}, we formulate a method to restrict parameter space based on the trigger values and show how much the parameter space restriction reduces the number of basis vectors.
In \secref{sec:injection}, we study the performance of FROQ with the LIGO-Virgo's O2 public data.
\secref{sec:conclusion} is devoted to the conclusion.

Throughout this paper, we apply the geometric unit system, $c=G=1$.
The Fourier transform of a continuous time series, $x(t)$, is given by
\begin{equation}
\tilde{x}(f) \equiv \int^\infty_{- \infty} x(t) \e^{- 2 \pi i f t} dt.
\end{equation}
In real experiments, we measure a time series at discrete and finite time samples, $x[m]\equiv x(m \Delta t + t_0)~(m=0, 1, \dots, M-1)$, where $1/\Delta t$ is the sampling rate, $t_0$ is the start time of the discrete series and $M$ is the number of the samples.
The Fourier transform of the discrete time series is given by
\begin{equation}
\tilde{x}[l] = \Delta t \sum^{M-1}_{m=0} x[m] \e^{- 2 \pi i l m / M}.
\end{equation}

\section{Bayesian inference and reduced order quadrature} \label{sec:basics}

In this section, we review the standard Bayesian inference applied for gravitational-wave parameter estimation and the ROQ technique.

\subsection{Bayesian inference}

In the Bayesian inference, the probability of source parameters, which is referred to as posterior probability density function, is calculated via the Bayes' theorem, 
\begin{equation}
\posterior \propto \prior \likelihood, \label{eq:bayestheorem}
\end{equation}
where $\vtheta$ is a vector of source parameters and $\diN$ is a set of data from $N$ detectors.
$\prior$ is referred to as the prior probability density function of $\vtheta$, which encodes our prior knowledge or belief on $\vtheta$, and $\likelihood$ is referred to as likelihood. 

The data are given by the sum of the signal, $\bm{h}_i (\bm{\theta})$, and noise, $\bm{n}_i$,
\begin{equation}
\di = \bm{h}_i (\bm{\theta}) + \bm{n}_i.
\end{equation}
Under the assumption that the noise is stationary and Gaussian, and instrumental noises of different detectors are uncorrelated, the likelihood is given by \cite{Creighton:2011zz}
\begin{align}
&\likelihood \nonumber \\
&= A \mathrm{exp}\left[-\frac{1}{2} \sum_{i=1}^N (\di - \bm{h}_i (\bm{\theta}), \di - \bm{h}_i (\bm{\theta}))_i\right], \label{eq:likelihood}
\end{align}
where $A$ is a normalization constant.
The inner product, $(\bm{x}, \bm{y})_i$, of arbitrary data, $\bm{x}$ and $\bm{y}$, is defined by
\begin{equation}
(\bm{x},\bm{y})_i \equiv 4 \Re \left[ \frac{1}{T} \sum^{L-1}_{l=0} \frac{\tilde{x}^*[l + l_0] \tilde{y}[l + l_0]}{S_{n,i}[l + l_0]} \right], \label{eq:inner_product}
\end{equation}
where $T$ represents the duration of data and $l_0$ corresponds to the lower frequency cutoff in the analysis, $\flow=l_0 / T$.
$S_{n,i}[l]$ is the one-sided power spectral density (PSD) of the $i$-th detector and defined by
\begin{equation}
S_{n,i}[l] \equiv \frac{2}{T} \mathrm{E}\left[ |\tilde{n}_i[l]|^2 \right],
\end{equation}
where $E[\ast]$ represents the ensemble average of $\ast$.

The source parameters we consider in this work are $\vtheta = (\alpha, \delta, \iota, \psi, r, \tc, \phic, m_1, m_2, \chi_1, \chi_2)$.
$\alpha$ and $\delta$ are the right ascension and declination of the binary respectively.
$\iota$ and $\psi$ are the inclination angle of the binary's orbital plane and the polarization angle of the gravitational waves, which determine the direction of the binary's orbital angular momentum.
$r$ is the luminosity distance to the binary.
$\tc$ is the time at which the coalescence part of gravitational waves arrives at the geocenter and $\phic$ is the coalescence phase.
$m_1$ and $m_2$ are the component masses of the binary.
$\chi_1$ and $\chi_2$ are the dimensionless spins defined by $\chi_k \equiv |\bm{S}_{k}| / m^2_k~(k=1,2)$, where $\bm{S}_{k}$ is the spin angular momentum of the $k$-th colliding body.
In this work, we assume that the spins are aligned with the orbital angular momentum and neglect the effects of the orbital precession \cite{Apostolatos:1994mx}.
We also neglect tidal deformabilities of the colliding bodies \cite{Flanagan:2007ix}.
We leave testing our technique with these additional parameters as the future work. 

What we wish to calculate from the posterior is the marginalized posterior density function,
\begin{equation}
p( \bm{\theta}_1 | \{\bm{d}_i\}_{i=1}^N) = \int \posterior d \bar{\bm{\theta}}_1, \label{eq:marginalized}
\end{equation}
where $\bm{\theta}_1$ is a vector of parameters we are interested in and $\bar{\bm{\theta}}_1$ is a vector of the remaining parameters.
We are also often interested in the expectation value of a function of the source parameters,
\begin{equation}
\left< f(\bm{\theta}) \right> = \int f(\bm{\theta}) \posterior d \bm{\theta}. \label{eq:expectation}
\end{equation}
Stochastic sampling techniques such as MCMC and nested sampling are efficient ways to perform such high-dimensional integrations.
In the stochastic sampling, thousands of random samples following the posterior, which we refer to ``posterior samples", are generated, and \eqref{eq:marginalized} and \eqref{eq:expectation} are calculated as the histogram of $\bm{\theta}_1$ and the mean of $f(\bm{\theta})$ respectively.

While the stochastic sampling techniques are quite efficient, they require millions of sequential likelihood computations.
As seen in \eqref{eq:likelihood}, each likelihood calculation requires waveform evaluations at $L$ frequency bins, which is computationally costly for large $L$.
If a BNS signal is analyzed with the lower frequency cutoff of $\flow$, $L$ needs to be larger than $\sim \fstop \tau(\flow)$, where $\fstop$ is the maximum frequency of the signal and $\tau(\flow)$ is the duration of the signal from $\flow$.
For a typical BNS signal, $\fstop = \mathcal{O}(10^3)\si{\hertz}$ and $\tau = \mathcal{O}(10^2)\si{\second}$ with $\flow=20\si{\hertz}$, and $L$ is at least in the order of $\mathcal{O}(10^5)$.
Therefore, the parameter estimation on a BNS signal is computationally very costly, and the analysis time can be a few weeks, or even years, depending on the analysis setup \cite{Canizares:2014fya, Smith:2016qas}.

\subsection{Reduced order quadrature} \label{sec:roq}

The costly likelihood evaluations can be sped up by the ROQ technique.
ROQ starts from constructing the reduced basis vectors of the waveforms and the squares of their norms with the greedy algorithm \cite{Field:2011mf}.
Then, the reduced basis vectors are related with the subset of the frequency bins, $\{l^{\mathrm{lin}}_j\}^{J-1}_{j=0}$ and $\{l^{\mathrm{quad}}_k\}^{K-1}_{k=0}$, with the empirical interpolation algorithm (See the algorithm 2 of \cite{Field:2013cfa}).
Finally, the waveforms and the squares of their norms are represented in the following form,
\begin{align}
&\tilde{h}_i[l + l_0] \simeq \e^{-2 \pi i (l + l_0) t_{\mathrm{c},i} / T} \sum^{J-1}_{j=0} B_{lj} \tilde{h}^{t_{\mathrm{c},i}=0}_i [l^{\mathrm{lin}}_j], \label{eq:linearbasis} \\
&\left| \tilde{h}_i[l + l_0] \right|^2 \simeq  \sum^{K-1}_{k=0} C_{lk} \left| \tilde{h}_i[l^{\mathrm{quad}}_k] \right|^2, \label{eq:quadraticbasis}
\end{align}
for $l=0,1,\dots,L-1$.
$B_{lj}$ and $C_{lk}$ are components of $L\times J$ and $L\times K$ matrices respectively.
Here, we drop the input source parameters to $\tilde{h}_i$ for ease of notation.
$t_{\mathrm{c},i}$ represents the time at which the coalescence part of gravitational waves arrives at the $i$-th detector, and $\tilde{h}^{t_{\mathrm{c},i}=0}_i$ is the time-shifted signal whose $t_{\mathrm{c},i}$ is zero.

The calculation of \eqref{eq:likelihood} requires the calculations of the inner products, $(\bm{d}_i, \bm{h}_i)_i$ and $(\bm{h}_i, \bm{h}_i)_i$.
With \eqref{eq:linearbasis} and \eqref{eq:quadraticbasis}, they are reduced to
\begin{align}
&(\bm{d}_i, \bm{h}_i)_i \simeq \Re\left[ \sum^{J-1}_{j=0} \tilde{h}^{t_{\mathrm{c},i}=0}_i[l^{\mathrm{lin}}_j] \omega_{j,i}(t_{\mathrm{c},i}) \right],  \\
&(\bm{h}_i, \bm{h}_i)_i=\sum^{K-1}_{k=0} \left| \tilde{h}_i [l^{\mathrm{quad}}_k] \right|^2 \Omega_k, 
\end{align}
where
\begin{align}
&\omega_{j,i}(t_{\mathrm{c},i}) = \frac{4}{T} \sum^{L-1}_{l=0} \frac{\tilde{d}^*_i[l + l_0] B_{lj}}{S_{\mathrm{n},i}[l + l_0]} \e^{-2 \pi i (l + l_0) t_{\mathrm{c},i} / T}, \label{eq:linearweight} \\
&\Omega_k = \frac{4}{T} \sum^{L-1}_{l=0} \frac{C_{lk}}{S_{\mathrm{n},i}[l + l_0]}. \label{eq:quadraticweight}
\end{align}
Since $\omega_{j,i}(t)$ and $\Omega_k$ can be computed before the stochastic sampling, each likelihood evaluation now requires only $J + K$ waveform evaluations.
Therefore, the parameter estimation can be sped up by a factor of $\sim L / (J + K)$.

In \cite{Canizares:2014fya}, the authors applied the ROQ technique to the TaylorF2 \cite{Buonanno:2009zt} waveform model, which is calculated based on the Post-Newtonian (PN) expansion \cite{Blanchet:2013haa}.
They considered the mass region of $1\Msun \leq m_1,~m_2 \leq 4\Msun$ and neglected spin effects.
For TaylorF2, $K=1$ since the dependence of $\left| \tilde{h}_i[l] \right|^2$ on $l$ is trivial, $\left| \tilde{h}_i[l] \right|^2 \propto l^{-7/3}$.
It was shown that $J\sim3000$ and the parameter estimation can be sped up by a factor of $\sim70$ for $\flow=20\si{\hertz}$, which reduces the analysis time from a couple of weeks to hours.

In \cite{Smith:2016qas}, the authors extended the ROQ technique to the IMRPhenomPv2 \cite{Hannam:2013oca} waveform model, which includes orbital precession effects due to the misalignment of spins and orbital angular momentum \cite{Thorne:1980ru}.
They divided mass region based on chirpmass,
\begin{equation}
\Mc = \frac{(m_1 m_2)^{\frac{3}{5}}}{(m_1 + m_2)^{\frac{1}{5}}}, \label{eq:chirpmass} 
\end{equation}
and found that the numbers of the basis vectors constructed over $1.4<\Mc<2.6$ with $\flow=20\si{\hertz}$ are $J=1253$ and $K=487$.
It speeds up parameter estimation of relatively heavy BNS signals (e.g. $1.7\Msun$-$1.7\Msun$) by a factor of $300$ and reduces the analysis time from around half a year to half a day.
The same basis can also be used to typical BNS signals (i.e. $1.4\Msun$-$1.4\Msun$) with the range of $\Mc$ and the frequency cutoffs being simultaneously scaled, while the lower frequency cutoff needs to be increased in this case.
While mass regions, waveform models and physics they consider or take into account are different, the basis sizes are $\mathcal{O}(10^2)$ and the analysis times are on the order of hours for the BNS case.

\section{Focused reduced order quadrature} \label{sec:froq}

In this section, we formulate our improved technique, focused reduced order quadrature (FROQ), which speeds up the parameter estimation of BNS signals further.
The key ingredient of FROQ is the reduction of the parameter space based on the trigger values.
Our strategy for the parameter space reduction is to rely on some combinations of masses and spins, and explore only within their narrow range around their trigger values.
We discuss which combinations of masses and spins should be applied and how to determine their range in this section.

First, we introduce the waveform model we apply for this discussion.
Since the merger part of a typical BNS signal is at high frequency, $f > 1000\si{\hertz}$, and outside the sensitive band of the LIGO and Virgo detectors, we ignore that part and only consider the inspiral part, which is well described by the PN expansion.
Following \cite{Cutler:1994ys, Poisson:1995ef}, we apply the so-called restricted $n$-PN waveform,
\begin{equation}
\tilde{h}^{(n\PN)}(f) \equiv \ffref^{-\frac{7}{6}} \e^{-i \Phi^{(n\PN)}(f)},
\end{equation}
where
\begin{align}
&\Phi^{(n\PN)}(f) = \nonumber \\
& \sum^{2n}_{\substack{k=0, \\ k\neq5,8}} \psi^k \ffref^{\frac{k-5}{3}}  + \sum^{2n}_{k=0} \psi^k_{\mathrm{log}} \ffref^{\frac{k-5}{3}} \log\ffref \nonumber \\
&+ \psi^5 + \psi^8 \ffref.
\end{align}
$\fref$ is a reference frequency, and we apply $\fref=200\si{\hertz}$, which is the same as that used in \cite{Ohme:2013nsa}.
The coefficients, $\psi^k$ and $\psi^k_{\mathrm{log}}$, depend on masses and spins.
The explicit expressions of the non-zero coefficients up to the $2\PN$ order are given by
\begin{align}
&\psi^0(\Mc) = \frac{3}{4}(8 \pi \Mc \fref)^{-\frac{5}{3}}, \\
&\psi^2(\Mc, \eta) = \frac{20}{9} \left( \frac{743}{336} + \frac{11}{4} \eta \right) \eta^{-\frac{2}{5}} (\pi \Mc \fref)^{\frac{2}{3}} \psi^0, \\
&\psi^3(\Mc, \eta, \chi_1, \chi_2) = (4 \beta - 16 \pi) \eta^{-\frac{3}{5}} (\pi \Mc \fref) \psi^0, \\
&\psi^4 (\Mc, \eta, \chi_1, \chi_2) = 10\left(\frac{3058673}{1016064} + \frac{5429}{1008} \eta + \frac{617}{144} \eta^2 - \sigma\right) \nonumber \\
&~~~~~~~~~~~~~~~~~~~~~~~~~~~~~~\times \eta^{-\frac{4}{5}} ( \pi \Mc f)^{4/3} \psi^0,
\end{align}
where
\begin{align}
&\eta = \frac{m_1 m_2}{(m_1 + m_2)^2}, \\
&\beta = \frac{1}{12} \sum^2_{k=1} \left[113 \left(\frac{m_k}{M}\right)^2 + 75 \eta \right] \chi_k, \\
&\sigma = \frac{79}{8} \eta \chi_1 \chi_2.
\end{align}
$\psi^5$ and $\psi^8$ are constant phase and time respectively and correspond to $\phi_c$ and $\tc$.
The observed signal at each detector, $\tilde{h}_i(f)$, is calculated by multiplying $\tilde{h}^{(n\PN)}(f)$ by the scaling factor depending on $\iota$ and $r$, the beam pattern functions of detectors, and the factor accounting for the time delay from the geocenter,

Our analysis in this section is based on the Fisher analysis, which is applied in the previous studies \cite{Cutler:1994ys, Poisson:1995ef, Ohme:2013nsa}.
The Fisher matrix is given by
\begin{equation}
\Gamma^{(n\PN)}_{pq} \equiv \frac{1}{(\bm{h}^{(n\PN)}, \bm{h}^{(n\PN)})} \left(\frac{\partial \bm{h}^{(n\PN)}}{\partial \psi^{p}},\frac{\partial \bm{h}^{(n\PN)}}{\partial \psi^{q}}\right),
\end{equation}
where $p$ or $q$ runs over integers for which $\psi^p$ or $\psi^q$ is non-zero.
For example, $p$ and $q$ run over $0,2,3,5,8$ for $n=1.5$ and $0,2,3,4,5,8$ for $n=2$.
The inner product in this section is defined with continuous Fourier components,
\begin{equation}
(\bm{x}, \bm{y}) = 4 \Re \left[ \int^{\fhigh}_{\flow} \frac{\tilde{x}^*(f) \tilde{y}(f)}{S_{\mathrm{n}}(f)} df \right]. \label{eq:ip_continuous}
\end{equation}

$\flow$ and $\fhigh$ are the lower and higher frequency cutoffs.
$\fhigh$ is the minimum among the highest frequency of the signal and the higher frequency limit of the detector's sensitive band.
Since the highest frequency of a BNS signal is so high, $>1000\si{\hertz}$, and outside the sensitive band of the LIGO and Virgo detectors, $\fhigh$ should be the latter, which does not depend on masses and spins.
Therefore, the Fisher matrix is a constant matrix.
$S_{\mathrm{n}}(f)$ is a representative PSD of a ground-based detector.
In reality, different ground-based detectors have different PSDs.
However, their dependence on the frequency is similar because all of the ground-based detectors suffer from similar kinds of noise, such as seismic noise, thermal noise and quantum noise of laser.
Therefore, we assume that the PSDs from different detectors are same up to multiplicative factors, $c_i S_{i,n}(f) \equiv S_{n}(f)~(i=1,2,\dots, N)$, throughout this section.
We also note that the Fisher matrix defined here is normalized by $(\bm{h}^{(n\PN)}, \bm{h}^{(n\PN)})$ and should be regarded as the Fisher matrix for the signal-to-noise ratio (SNR) of 1.

Since we are interested in mass and spin parameters, we project the constant time and phase out of the Fisher matrix.
The Fisher matrix after the projection is given by \cite{Owen:1995tm}
\begin{equation}
\tilde{\Gamma}^{(n\PN)}_{\alpha \beta} \equiv \Gamma^{(n\PN)}_{\alpha \beta} - \Gamma^{(n\PN)}_{\alpha a} \gamma^{a b} \Gamma^{(n\PN)}_{b \beta},
\end{equation}
where $\gamma^{ab}$ denotes the inverse of the submatrix of $\Gamma^{(n\PN)}$ corresponding to $\psi^5$ and $\psi^8$.
Then, the covariance matrix of $\psi^\alpha$ for the SNR of 1 is given by the inverse of the Fisher matrix, $\left( \tilde{\Gamma}^{(n\PN)} \right)^{-1}$.

\subsection{The best measurable combinations}

First, we discuss which combinations of masses and spins should be applied.
To speed up the parameter estimation significantly, they should be the combinations whose marginalized posterior distributions are the narrowest.
That is, they should be the best measurable combinations of masses and spins.
Such combinations correspond to the eigenvectors of the covariance matrix with the smallest eigenvalues, or the eigenvectors of the Fisher matrix with the largest eigenvalues \cite{Ohme:2013nsa}.

Diagonalizing the Fisher matrix, we get
\begin{equation}
U \tilde{\Gamma}^{(1.5\PN)} U^{T} = 
 \left( 
  \begin{array}{ccc}
   \lambda_1 & 0 & 0 \\
   0 & \lambda_2 & 0 \\
   0 & 0 & \lambda_3
  \end{array}  
  \right),
\end{equation}
where $\lambda_1 > \lambda_2 > \lambda_3 > 0$.
Here, to incorporate the spin effects, we take into account the terms up to the $1.5\PN$ term.
The combinations to parametrize the directions of the eigenvectors are
\begin{equation}
\left( 
  \begin{array}{c}
 \mu^{1} \\
  \mu^{2} \\
  \mu^{3}
  \end{array}  
  \right)
 = U \left( 
  \begin{array}{c}
 \psi^{0} \\
  \psi^{2} \\
  \psi^{3}
  \end{array}  
  \right).
\end{equation}
We focus on the two best measurable combinations, $\mu^1$ and $\mu^2$.

\begin{figure} 
        \begin{center}
                \includegraphics[width = \columnwidth]{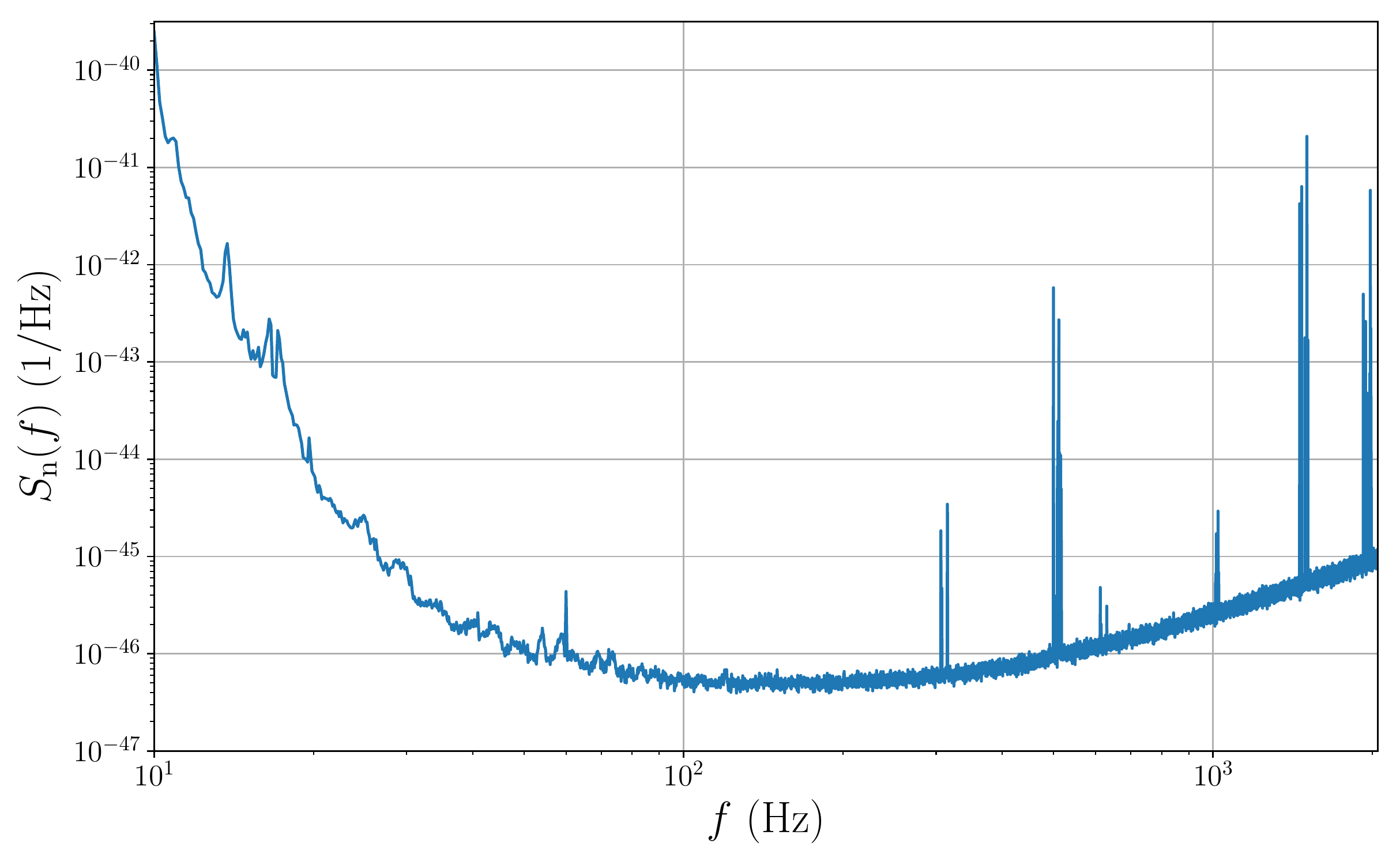}
                \caption{A representative PSD of the LIGO-Livingston detector in the O2. This is estimated with the $1000\si{\second}$ data from 07:16:40 UTC on August 19, 2017.} \label{fig:psd}
        \end{center}
\end{figure}

\begin{figure}
\includegraphics[width=0.8\columnwidth]{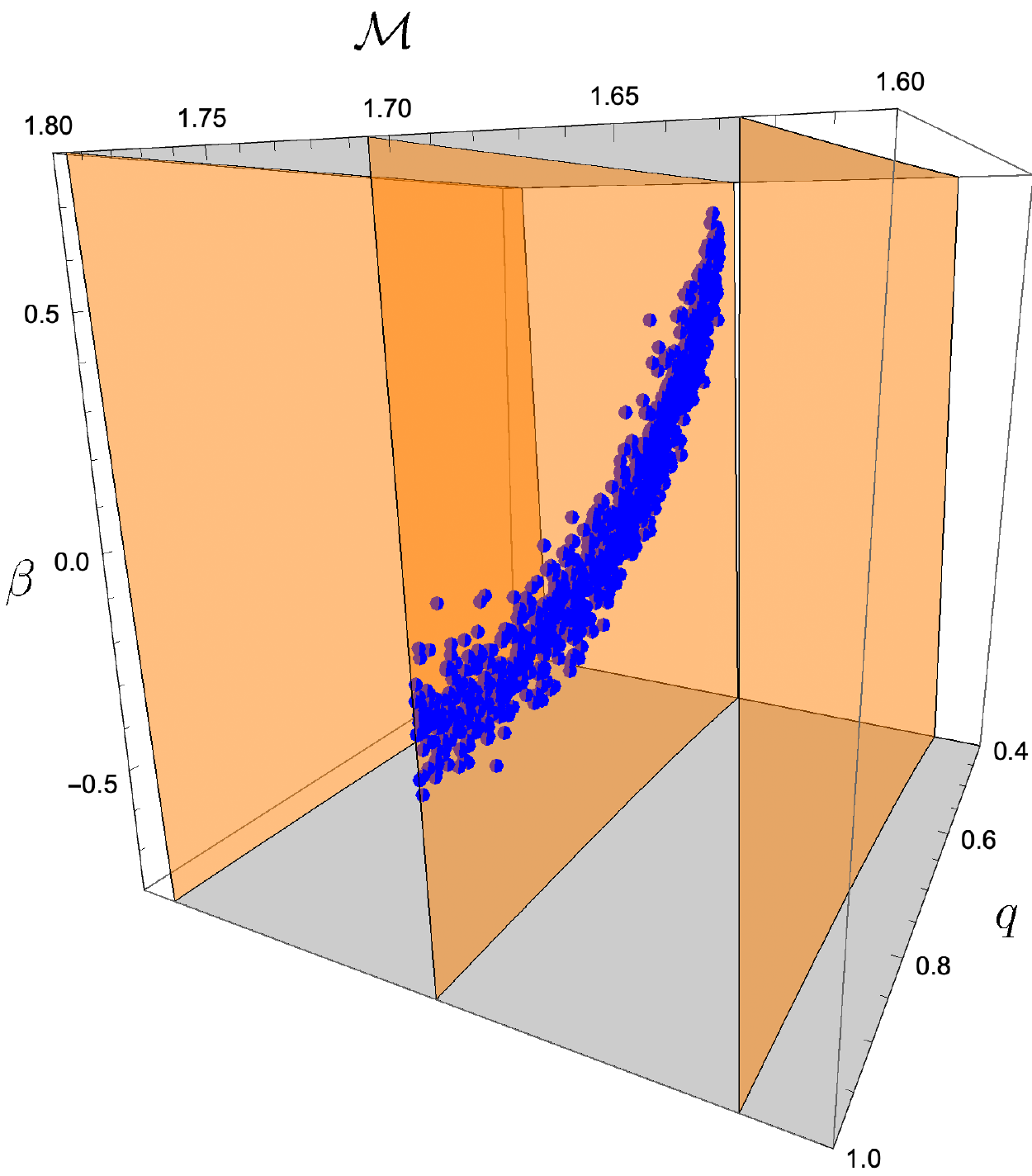}
\includegraphics[width=0.8\columnwidth]{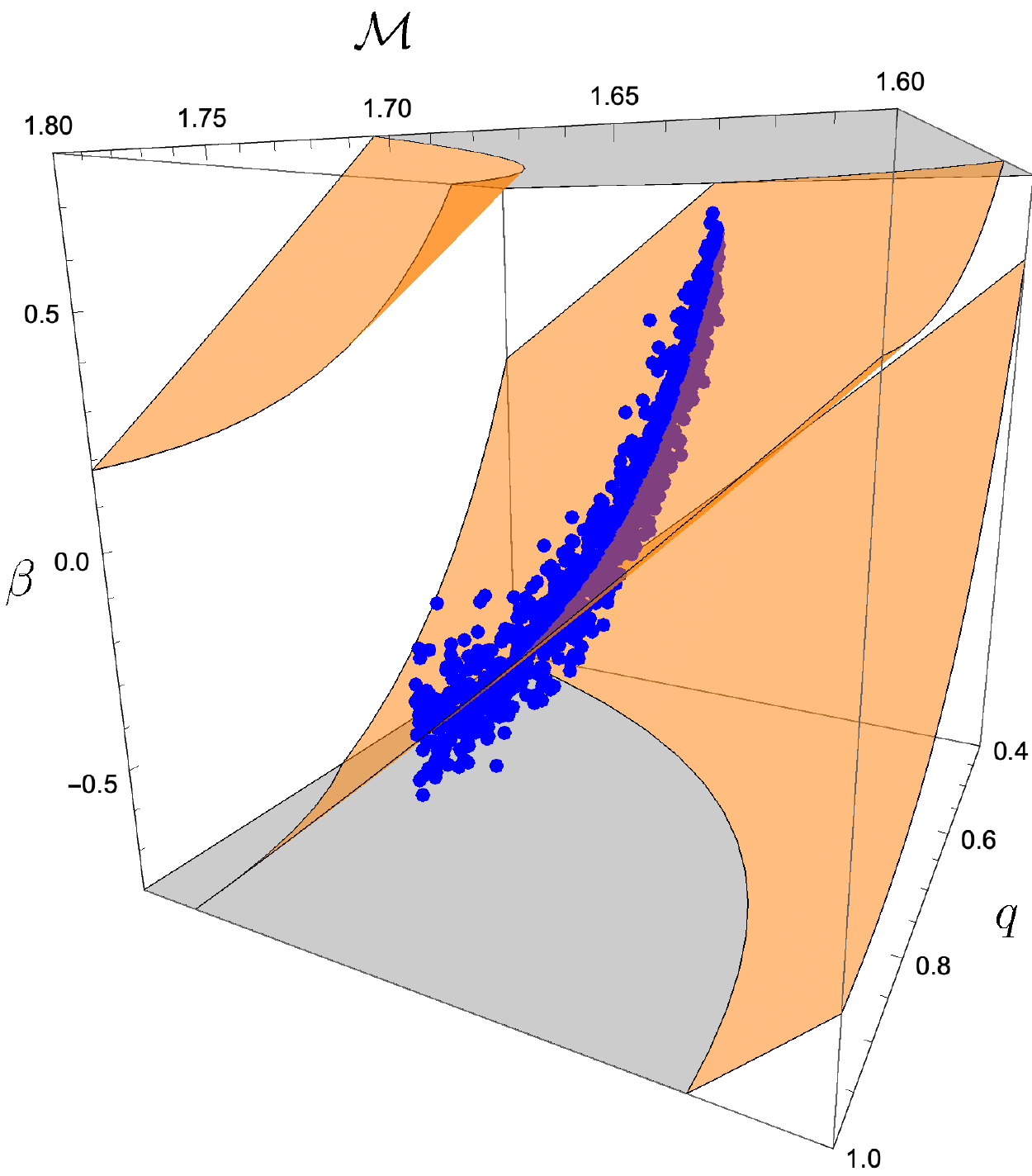}
\caption{Posterior samples from the parameter estimation on a BNS signal artificially injected into the O2 data (blue dots) and the planes defined by $\mu^1=\mathrm{const.}$ (upper figure) or $\mu^2=\mathrm{const.}$ (lower figure).} 
  \label{fig:mu1mu2planes}
\end{figure}

With a representative PSD of the LIGO-Livingston detector in the O2, which is shown in \figref{fig:psd}, and the frequency cutoffs of $\flow=20\si{\hertz}$ and $\fhigh=1024\si{\hertz}$, $\mu^1$ and $\mu^2$ are given by
\begin{align}
\mu^1 &= 0.974\psi^0 + 0.209\psi^2 + 0.0840\psi^3, \\
\mu^2 &= -0.221\psi^0 + 0.823\psi^2 + 0.524\psi^3. 
\end{align}
To visually show what these parameters are, we show $\mu^1$-constant and $\mu^2$-constant planes with posterior samples from a BNS signal in \figref{fig:mu1mu2planes}.
The posterior samples are on the characteristic curve in $\Mc$-$q$-$\beta$ space due to the parameter degeneracy, and the figure shows that $\mu^1$ and $\mu^2$ are parameters parametrizing the two directions orthogonal to the curve.
 
\subsection{Requirements on $\mu^1-\mu^2$ ranges} \label{sec:requirements}

Given the trigger values of masses and spins, we determine the range of $\mu^1$ and $\mu^2$.
While they should be as narrow as possible, they need to be broad enough to incorporate the parameter space over which the posterior is not negligibly small.
Especially, we need to take into account the following errors: the systematic errors of trigger values and the statistical errors due to instrumental noise.
In the calculation of these errors, we take into account the terms up to the $2\PN$ order since it affects the error estimate by a non-negligible factor \cite{Poisson:1995ef}.

\subsubsection{The systematic errors of trigger values}

Let $\vpsi$ denote $(\psi^0,\psi^2,\psi^3,\psi^4)^{\mathrm{T}}$ and $\vpsi_{\mathrm{t}}$ denote $\vpsi$ calculated with trigger values.
Since the template bank is a discrete set of template waveforms, $\vpsi_{\mathrm{t}}$ is in general different from that maximizing the SNR, which is denoted by $\hat{\vpsi}$.
The fraction of the SNR lost due to this offset is given by the so-called mismatch,
\begin{align}
&MM =\nonumber \\
&~~ 1 - \frac{\underset{\psi^5,\psi^8}{\mathrm{max}}(\bm{h}^{(\mathrm{2\PN})}(\vpsi_{\mathrm{t}}, \psi^5, \psi^8), \bm{h}^{\mathrm{(2\PN)}}(\hat{\vpsi},\psi'^5, \psi'^8)) }{(\bm{h}^{\mathrm{(2\PN)}}, \bm{h}^{\mathrm{(2\PN)}})}.
\end{align}
If $\hat{\psi}^\alpha - \psi^\alpha_{\mathrm{t}}$ is small enough, this can be approximated by
\begin{equation}
MM \simeq \frac{1}{2} \tilde{\Gamma}^{(\mathrm{2\PN})}_{\alpha \beta} (\hat{\psi}^\alpha - \psi^\alpha_{\mathrm{t}}) (\hat{\psi}^\beta - \psi^\beta_{\mathrm{t}}). \label{eq:mismatch}
\end{equation}
Typically, the template bank is constructed so that the mismatch does not exceed $0.03$ \cite{Brown:2012qf, DalCanton:2017ala, Mukherjee:2018yra}.
Therefore, we have the following constraint, 
\begin{equation}
\tilde{\Gamma}^{(\mathrm{2\PN})}_{\alpha \beta} (\hat{\psi}^\alpha - \psi^\alpha_{\mathrm{t}}) (\hat{\psi}^\beta - \psi^\beta_{\mathrm{t}}) < 0.06 \label{eq:systematic1}
\end{equation}
if the template bank covers the whole parameter space we consider in the parameter estimation.

The template bank in the BNS mass region typically only covers the low-spin parameter region, $-0.05\leq\chi_1,\chi_2\leq0.05$ \cite{DalCanton:2017ala, Mukherjee:2018yra}, while the parameter estimation can be performed over high-spin parameter region, e.g. $-0.7\leq\chi_1,\chi_2<0.7$.
In this case, the mismatch can exceed $0.03$.
Nevertheless, we do not expect the mismatch is so large, in which case the signal is not detected in the first place.
If the mismatch is $1- (0.1)^{\frac{1}{3}} \simeq 0.536$, the observable volume, and hence the detection rate, drops to $10\%$ compared to that in the optimal case.
Therefore, we assume that the mismatch is smaller than $0.536$.

\begin{figure} 
        \begin{center}
                \includegraphics[width = \columnwidth]{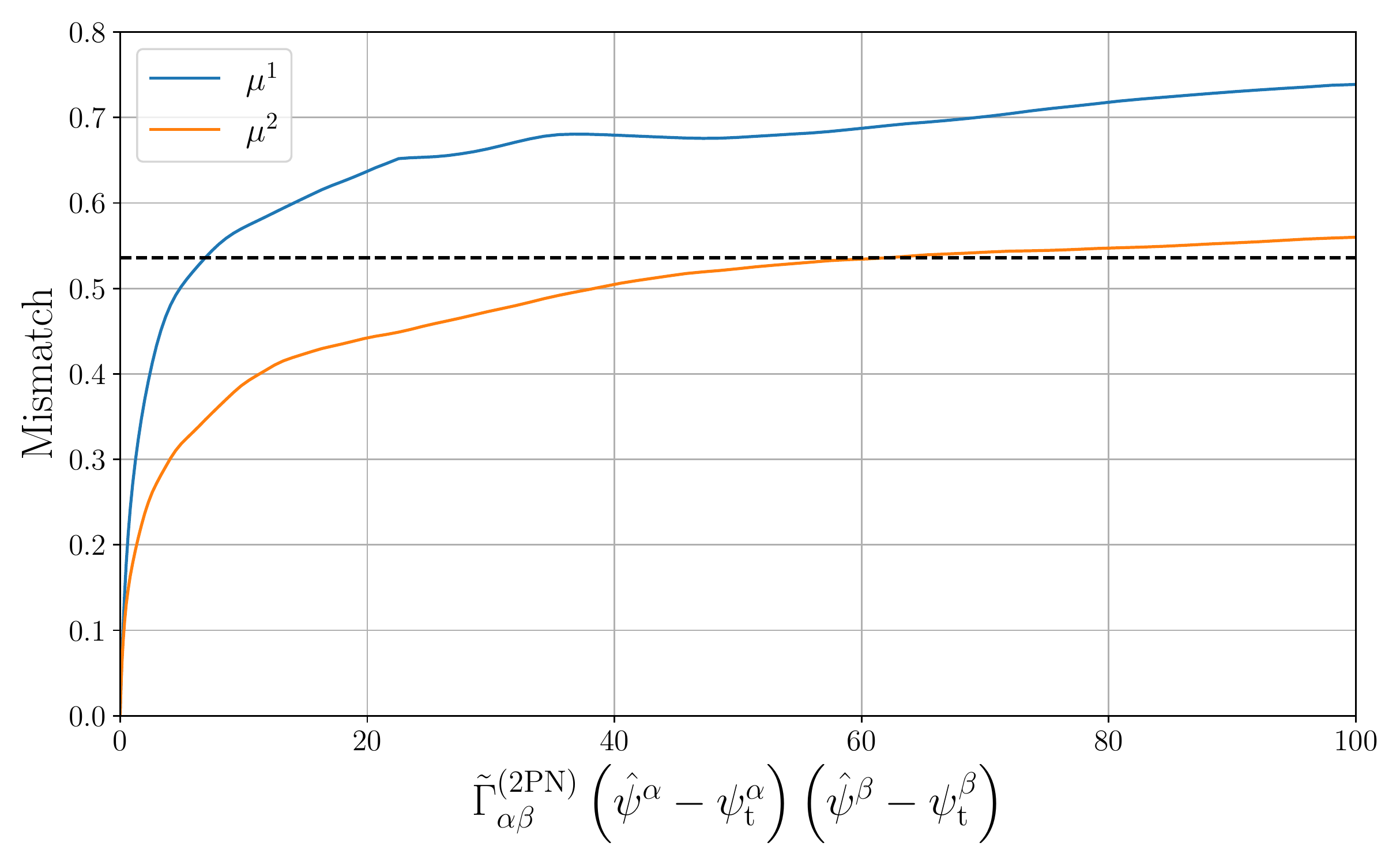}
                \caption{Mismatches calculated along the directions of $\mu^1$ (blue) and $\mu^2$ (orange). They are calculated with $\flow=20\si{\hertz}$, $\fhigh=1024\si{\hertz}$ and the O2 LIGO-Livingston PSD shown in \figref{fig:psd}. The dashed line represents the target mismatch of $1- (0.1)^{\frac{1}{3}} \simeq 0.536$.}
                \label{fig:mismatch}
        \end{center}
\end{figure}

Since \eqref{eq:mismatch} is not valid at such high mismatch, we calculate the mismatch without the approximations.
Fortunately, the mismatch depends only on $\hat{\psi}^\alpha - \psi^\alpha_{\mathrm{t}}$, and we do not need to calculate it at each $\vpsi_{\mathrm{t}}$.
While the mismatch is a function of the four coefficients, $\psi^0, \psi^2, \psi^3$ and $\psi^4$, we calculate it only along the $\mu^1$ and $\mu^2$ directions in the four-dimensional parameter space.
Then, we investigate how the mismatch increases as $\tilde{\Gamma}^{(\mathrm{2\PN})}_{\alpha \beta} (\hat{\psi}^\alpha - \psi^\alpha_{\mathrm{t}}) (\hat{\psi}^\beta - \psi^\beta_{\mathrm{t}})$ increases, and estimate its upper limit corresponding to the mismatch of $0.536$.
\figref{fig:mismatch} shows the mismatches calculated with $\flow=20\si{\hertz}$, $\fhigh=1024\si{\hertz}$ and the O2 LIGO-Livingston PSD shown in \figref{fig:psd}.
This shows that the mismatch upper limit of $0.536$ is translated into the following constraint,
\begin{equation}
\tilde{\Gamma}^{(\mathrm{2\PN})}_{\alpha \beta} (\hat{\psi}^\alpha - \psi^\alpha_{\mathrm{t}}) (\hat{\psi}^\beta - \psi^\beta_{\mathrm{t}}) < 63.7. \label{eq:systematic2}
\end{equation}

\subsubsection{The statistical errors due to instrumental noise} \label{sec:statistical}

The posterior distribution has a finite width due to the statistical errors, and we need to take it into account in determining the range of $\mu^1$ and $\mu^2$.
In the limit that the SNR is high, the posterior marginalized over the parameters except for $\thetain\equiv(m_1,m_2,\chi_1,\chi_2)$ can be approximated by
\begin{align}
&\int \posterior d \thetaex \nonumber \\
&\propto p(\thetain) \mathrm{exp}\left[-\frac{1}{2} \rho^2_{\mathrm{net}} \tilde{\Gamma}^{(2\PN)}_{\alpha \beta} (\psi^\alpha - \hat{\psi}^\alpha) (\psi^\beta - \hat{\psi}^\beta) \right], \label{eq:marginalized_posterior}
\end{align}
where $\thetaex \equiv (\alpha, \delta, \iota, \psi, r, \tc, \phic)$, $p(\thetain)$ is the prior on $\thetain$ and $\rho_{\mathrm{net}}$ is the network signal-to-noise ratio, which is defined by $\rho^2_{\mathrm{net}}=\sum_i c_i (\bm{h}_i,\bm{h}_i)$.
Here, we assume that the prior can be decomposed into the prior on $\thetain$ and $\thetaex$, $p(\bm{\theta})=p(\bm{\thetain})p(\bm{\thetaex})$.
Also, we assume that the prior on $\tc$ and $\phic$ is uniform and does not have any correlations with the other parameters.
They are valid in most of the parameter estimation studies.
The derivation of \eqref{eq:marginalized_posterior} is given in \appref{sec:derivation}.

If the prior is neglected, the posterior can be regarded as the four-dimensional Gaussian distribution in the $\psi^0$-$\psi^2$-$\psi^3$-$\psi^4$ space.
The region encompassing the probability of $p$ is given by
\begin{equation}
\tilde{\Gamma}^{(2\PN)}_{\alpha \beta} (\psi^\alpha - \hat{\psi}^\alpha) (\psi^\beta - \hat{\psi}^\beta) < \left(\frac{N_{\chi^2_4}(p)}{\rho_{\mathrm{net}}}\right)^2.   \label{eq:statistical}
\end{equation}
$N_{\chi^2_4}(p)$ is the percent point function of $\chi^2$ with $4$ degrees of freedom.
$\rho_{\mathrm{net}}$ should be small enough for a conservative estimate.
Since a detection requires the network SNR of $\gtrsim12$ \cite{Fairhurst:2010is}, we apply $\rho_{\mathrm{net}}=12$.
On the value of $p$, we apply $p=0.999$, which is conservative enough since we are typically interested in $90\%$ credible intervals of parameters.
Then, \eqref{eq:statistical} is reduced to
\begin{equation}
\tilde{\Gamma}^{(2\PN)}_{\alpha \beta} (\psi^\alpha - \hat{\psi}^\alpha) (\psi^\beta - \hat{\psi}^\beta) < 0.128. \label{eq:statistical1}
\end{equation}
The corrections from the matter effects of colliding bodies or the merger and ringdown part of the waveform are negligible because the systematic errors from them on the inference of masses and spins are smaller than the statistical errors at the SNR of $12$ and in the mass range we consider \cite{Dudi:2018jzn}.
It means that \eqref{eq:statistical1} can be safely used for waveforms incorporating those effects.

\subsubsection{The prior constraint} \label{sec:prior_constraint}

As seen in \eqref{eq:marginalized_posterior}, the posterior distribution is also affected by the prior, $p(\thetain)$.
Especially, the range in which $p(\thetain)$ is non-zero is crucial in determining the $\mu^1$-$\mu^2$ ranges.
Since we focus on BNS signals and the causality upper limit of the mass of a neutron star is $\sim 2.9~\Msun$ \cite{Rhoades:1974fn, Kalogera:1996ci}, we apply the following mass prior range,
\begin{equation}
0~\Msun < m_1,m_2 < 3~\Msun. \label{eq:mass_prior}
\end{equation}
We also impose the following limit on the spins, 
\begin{equation}
-\chimax < \chi_1,~\chi_2 < \chimax. \label{eq:spin_prior}
\end{equation}
In this work, we consider two different upper limits.
One is $\chimax=0.05$, which is applied in the parameter estimation on GW170817 \cite{TheLIGOScientific:2017qsa, Abbott:2018wiz} and GW190425 \cite{Abbott:2020uma}.
This is motivated by the fact that even PSR J0737-3039A \cite{Burgay:2003jj}, which is one of the observed binary neutron star system that will merge within a Hubble time and contains the most extremely spinning pulsar among them, will have $|\chi| \lesssim 0.04$ at the merger.
The other is $\chimax=0.7$, which is motivated by the fact that the maximum spin parameter of a uniformly rotating star is $\sim 0.7$ for various realistic nuclear equations of state \cite{Lo:2010bj}.
We call these priors {\it low-spin prior} and {\it high-spin prior} respectively.

\subsection{Calculation of a $\mu^1-\mu^2$ range} \label{sec:mu1mu2calculation}

The range of $\mu^1$ and $\mu^2$ is determined by calculating their minimums and maximums under the constraints introduced in \secref{sec:requirements}.
Here, we explain how to calculate them.
Combining the constraints, \eqref{eq:systematic1}, \eqref{eq:systematic2} and \eqref{eq:statistical1}, we have
\begin{align}
&\tilde{\Gamma}^{(\mathrm{2\PN})}_{\alpha \beta} (\psi^\alpha - \psi^\alpha_{\mathrm{t}}) (\psi^\beta - \psi^\beta_{\mathrm{t}}) < R^2, \label{eq:error_circle} \\
&R=R_{\mathrm{sys}} + R_{\mathrm{stat}}.
\end{align}
$R^2_{\mathrm{sys}}=0.06$ in the case where the parameter space we consider in the parameter estimation is narrower than that covered by the template bank and $R^2_{\mathrm{sys}}=63.7$ otherwise, and $R^2_{\mathrm{stat}}= 0.128$.
Since chirp mass is very close to its trigger value under the constraint, the mass constraint \eqref{eq:mass_prior} can be approximately reduced to the constraints on $\eta$,
\begin{equation}
\eta > \etamin(\Mc_{\mathrm{t}}), \label{eq:massratio_prior}
\end{equation}
where $\Mct$ is the trigger value of chirp mass and $\etamin(\Mc_{\mathrm{t}})$ is the smallest value of $\eta$ under \eqref{eq:mass_prior} and $\Mc=\Mc_{\mathrm{t}}$.

For the calculations of the minimums and maximums of $\mu^1$ and $\mu^2$, we densely sample $\eta$ within the constraint \eqref{eq:massratio_prior}, calculate the minimums and maximums of $\mu^1$ and $\mu^2$ at each $\eta$, and finally calculate the minimums and maximums among them.
At each $\eta$, we make a few approximations to simplify the problem.
First, since $\Mc \simeq \Mc_{\mathrm{t}}$, we apply the approximations that
\begin{align}
\psi^2(\Mc, \eta) &\simeq \psi^2(\Mc_{\mathrm{t}}, \eta), \nonumber \\
\psi^3(\Mc, \eta, \chi_1, \chi_2) &\simeq \psi^3(\Mc_{\mathrm{t}}, \eta, \chi_1, \chi_2), \\
\psi^4(\Mc, \eta, \chi_1, \chi_2) &\simeq \psi^4(\Mc_{\mathrm{t}}, \eta, \chi_1, \chi_2). \nonumber
\end{align}
Since $\eta$ is fixed, $\psi^2$ becomes constant.
\eqref{eq:error_circle} is then reduced to
\begin{equation}
\tilde{\Gamma}^{(2\PN)}_{\alpha' \beta'} (\psi^{\alpha'} - \psi^{\alpha'}_{\mathrm{t}} - \Delta \psi^{\alpha'}) (\psi^{\beta'} - \psi^{\beta'}_{\mathrm{t}}  - \Delta \psi^{\beta'}) < R'^2, \label{threedim_ellipsoid}
\end{equation}
where $\alpha', \beta'=0, 3, 4$.
$\Delta \psi^{\alpha'}$ and $R'^2$ are given by
\begin{align}
&\Delta \psi^{\alpha'} = - \gamma'^{\alpha' \beta'} \Gamma^{(2\PN)}_{\beta' 2} (\psi^2 - \psi^2_{\mathrm{t}} ), \\
&R'^2 = R^2 - (\Gamma^{(2\PN)}_{22} - \Gamma^{(2\PN)}_{2 \alpha'} \gamma'^{\alpha' \beta'} \Gamma^{(2\PN)}_{\beta' 2}) (\psi^2 - \psi^2_{\mathrm{t}} )^2,
\end{align}
where $\gamma'$ is the inverse matrix of $\Gamma^{(2\PN)}_{\alpha' \beta'}$.

Second, since the spin constraint \eqref{eq:spin_prior} are complicated in the $\psi$ coordinates, we instead impose the following constraints,
\begin{align}
\psi^3(\Mc_{\mathrm{t}}, \eta, -\chimax, -\chimax) < \psi^3 < \psi^3(\Mc_{\mathrm{t}}, \eta, \chimax, \chimax), \\
\psi^4(\Mc_{\mathrm{t}}, \eta, \chimax, \chimax) < \psi^4 < \psi^4(\Mc_{\mathrm{t}}, \eta, -\chimax, \chimax).
\end{align}
They represent a broader parameter space than the original prior constraint and broaden the range of $\mu^1$ and $\mu^2$.
Since our primary goal is to provide a broad enough range, this does not cause any problems.

With these approximations, the problem is reduced to that of calculating extremums of the linear function inside a $3$-dimensional ellipsoid with the simple boundaries, and it can be analytically solved with the standard Lagrange multiplier method.

\subsection{Construction of $\mu^1-\mu^2$ ranges} \label{sec:mu1mu2range}

Given the trigger values, $(m_{1,\mathrm{t}}, m_{2,\mathrm{t}}, \chi_{1,\mathrm{t}}, \chi_{2,\mathrm{t}})$, the $\mu^1$-$\mu^2$ range can be calculated with the algorithm introduced in \secref{sec:mu1mu2calculation},
\begin{widetext}
\begin{align}
\mu^1_{\mathrm{t}} - \delta^{-} \mu^1(m_{1,\mathrm{t}}, m_{2,\mathrm{t}}, \chi_{1,\mathrm{t}}, \chi_{2,\mathrm{t}}) \leq &\mu^1 \leq \mu^1_{\mathrm{t}} + \delta^{+} \mu^1(m_{1,\mathrm{t}}, m_{2,\mathrm{t}}, \chi_{1,\mathrm{t}}, \chi_{2,\mathrm{t}}), \\
\mu^2_{\mathrm{t}} - \delta^{-} \mu^2(m_{1,\mathrm{t}}, m_{2,\mathrm{t}}, \chi_{1,\mathrm{t}}, \chi_{2,\mathrm{t}}) \leq &\mu^2 \leq \mu^2_{\mathrm{t}} + \delta^{+} \mu^2(m_{1,\mathrm{t}}, m_{2,\mathrm{t}}, \chi_{1,\mathrm{t}}, \chi_{2,\mathrm{t}}),
\end{align}
\end{widetext}
where $\mu^1_{\mathrm{t}}$ and $\mu^2_{\mathrm{t}}$ are the trigger values of $\mu^1$ and $\mu^2$.
On the other hand, the ROQ basis construction is computationally costly and needs to be done offline.
Therefore, we need to construct a lot of prior ranges accommodating all the possible trigger values and construct corresponding ROQ basis sets beforehand.
In this section, we introduce a method for the offline $\mu^1$-$\mu^2$ range construction.

\begin{figure} 
        \begin{center}
                \includegraphics[width = \columnwidth]{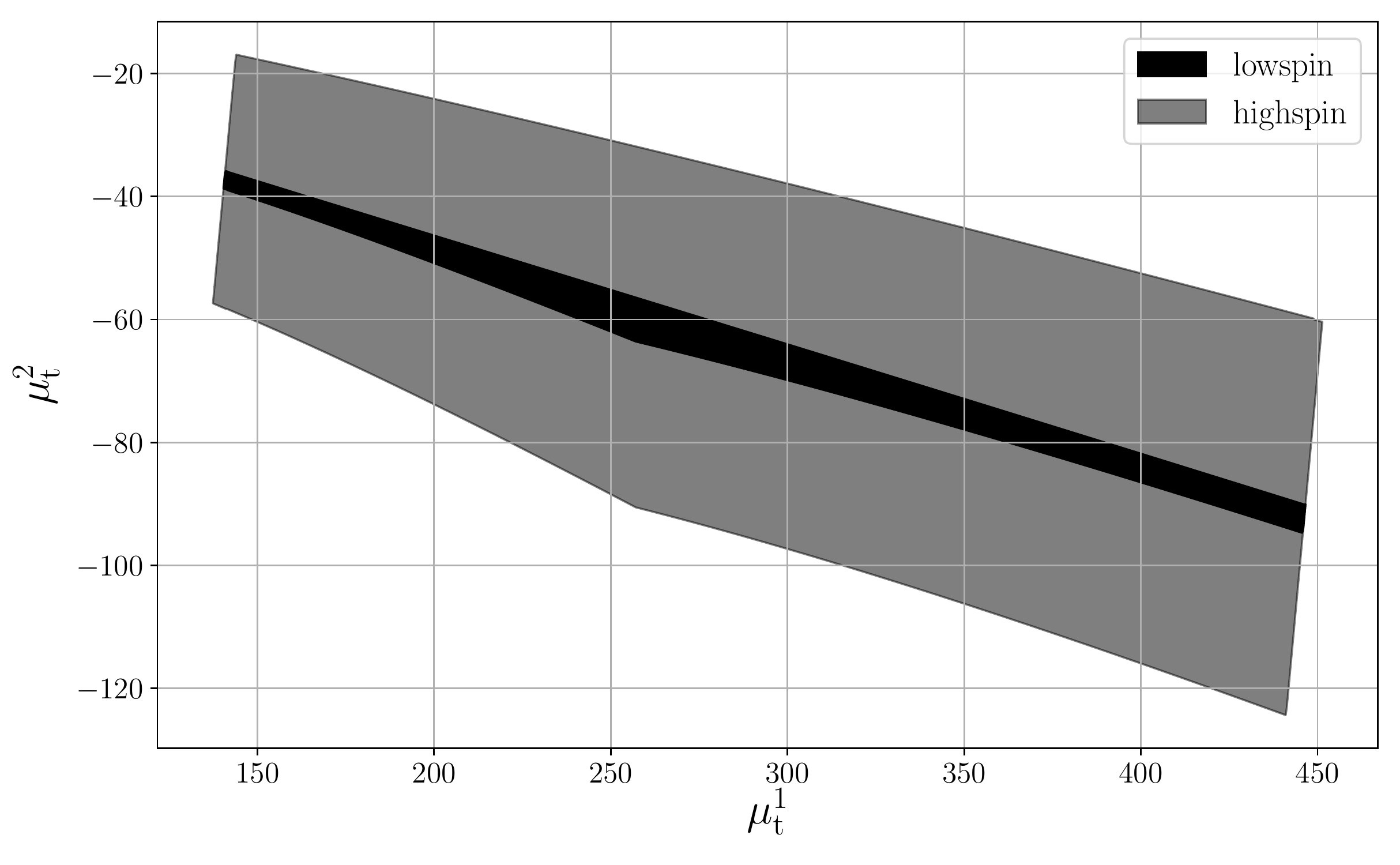}
                \caption{$\mu^1$-$\mu^2$ space covered by the low-spin (black) and high-spin (gray) template bank. The mass range of $1\Msun \leq m_{1,\mathrm{t}}, m_{2,\mathrm{t}} \leq 2 \Msun$ is applied, and the spin ranges of $-0.05 \leq \chi_{1, \mathrm{t}},\chi_{2,\mathrm{t}} \leq 0.05$ and $-0.7 \leq \chi_{1,\mathrm{t}},\chi_{2,\mathrm{t}} \leq 0.7$ are assumed for the low-spin and high-spin template banks respectively.}
                \label{mu1mu2space}
        \end{center}
\end{figure}

First, we discuss the range of the trigger values.
We consider the case where BNS signals are searched for with the template bank covering the following mass region,
\begin{equation}
1\Msun \leq m_{1,\mathrm{t}}, m_{2,\mathrm{t}} \leq 2 \Msun.
\end{equation}
Following the previous studies \cite{DalCanton:2017ala, Mukherjee:2018yra}, where template banks for BNS signals are constructed only over low-spin parameter space, we consider a {\it low-spin template bank} with the following spin range,
\begin{equation}
-0.05 \leq \chi_{1, \mathrm{t}},\chi_{2,\mathrm{t}} \leq 0.05.
\end{equation}
This narrow spin range is motivated by the observations of binary pulsars as explained in \secref{sec:prior_constraint}.
Since the template bank can be constructed over high-spin region \cite{Roulet:2019hzy}, we also consider a {\it high-spin template bank} with the following spin range,
\begin{equation}
-0.7 \leq \chi_{1,\mathrm{t}},\chi_{2,\mathrm{t}} \leq 0.7.
\end{equation}
\figref{mu1mu2space} shows $\mu^1_{\mathrm{t}}$-$\mu^2_{\mathrm{t}}$ space covered by the low-spin and high-spin template bank.
We assign a $\mu^1$-$\mu^2$ range to each $(\mu^1_{\mathrm{t}}, \mu^2_{\mathrm{t}})$ in the space.
When a signal is detected, we calculate the trigger values of $\mu^1$ and $\mu^2$, and use the $\mu^1$-$\mu^2$ range assigned to it and the ROQ basis vectors constructed over the range.

The construction of the $\mu^1$-$\mu^2$ ranges starts from $\mu^1_{\mathrm{t}}=\mu^1_{\mathrm{t,min}}$, where $\mu^1_{\mathrm{t,min}}$ is the minimum value of $\mu^1_{\mathrm{t}}$ in the $\mu^1$-$\mu^2$ space covered by the template bank.
First, the following values are calculated,
\begin{align}
\delta^- \mu^1(\mu^1_{\mathrm{t}}) = \max_{\mu^1_{\mathrm{t}}\text{ fixed}} \delta^-  \mu^1(m_{1,\mathrm{t}}, m_{2,\mathrm{t}}, \chi_{1,\mathrm{t}}, \chi_{2,\mathrm{t}}), \\
\delta^+ \mu^1(\mu^1_{\mathrm{t}}) = \max_{\mu^1_{\mathrm{t}}\text{ fixed}} \delta^+  \mu^1(m_{1,\mathrm{t}}, m_{2,\mathrm{t}}, \chi_{1,\mathrm{t}}, \chi_{2,\mathrm{t}}),
\end{align}
where the maximization is performed over $m_{1,\mathrm{t}}, m_{2,\mathrm{t}}, \chi_{1,\mathrm{t}}, \chi_{2,\mathrm{t}}$ with $\mu^1_{\mathrm{t}}$ being fixed.
They are calculated by generating samples with $\mu^1_{\mathrm{t}}$ fixed, calculating $\delta^{(+,-)}\mu^1$ for them and taking the maximums.
Then, $\mu^1$-$\mu^2$ ranges are constructed within $\mu^1_{\mathrm{t}} \leq \mu^1 \leq \mu^1_{\mathrm{t}} + \Delta \mu^1$, where $\Delta \mu^1= \epsilon_1 \left(\delta^- \mu^1(\mu^1_{\mathrm{t}})  + \delta^+ \mu^1(\mu^1_{\mathrm{t}})\right)$ and $\epsilon_1$ is a user-specified constant.

The construction starts from the minimum of $\mu^2_{\mathrm{t}}$ at $\mu^1_{\mathrm{t}}$.
At each step, the following values are numerically calculated
\begin{align}
\delta^- \mu^2(\mu^1_{\mathrm{t}}, \mu^2_{\mathrm{t}}) = \max_{\mu^1_{\mathrm{t}},\mu^2_{\mathrm{t}}\text{ fixed}} \delta^-  \mu^2(m_{1,\mathrm{t}}, m_{2,\mathrm{t}}, \chi_{1,\mathrm{t}}, \chi_{2,\mathrm{t}}), \\
\delta^+ \mu^2(\mu^1_{\mathrm{t}}, \mu^2_{\mathrm{t}}) = \max_{\mu^1_{\mathrm{t}},\mu^2_{\mathrm{t}}\text{ fixed}} \delta^+  \mu^2(m_{1,\mathrm{t}}, m_{2,\mathrm{t}}, \chi_{1,\mathrm{t}}, \chi_{2,\mathrm{t}}).
\end{align}
Then, the $\mu^1$-$\mu^2$ range defined by
\begin{align}
\mu^1_{\mathrm{t}} - \delta^- \mu^1(\mu^1_{\mathrm{t}}) \leq &\mu^1 \leq \mu^1_{\mathrm{t}} + \Delta \mu^1 + \delta^+ \mu^1(\mu^1_{\mathrm{t}}), \\
\mu^2_{\mathrm{t}} - \delta^- \mu^2(\mu^1_{\mathrm{t}}, \mu^2_{\mathrm{t}}) \leq &\mu^2 \leq \mu^2_{\mathrm{t}} + \Delta \mu^2 + \delta^+ \mu^2(\mu^1_{\mathrm{t}}, \mu^2_{\mathrm{t}}),
\end{align}
is assigned to the trigger values within
\begin{equation}
\mu^1_{\mathrm{t}} \leq \mu^1 \leq \mu^1_{\mathrm{t}} + \Delta \mu^1,~\mu^2_{\mathrm{t}} \leq \mu^2 \leq \mu^2_{\mathrm{t}} + \Delta \mu^2.
\end{equation}
Here, $\Delta \mu^2 \equiv \epsilon_2 \left(\delta^- \mu^2(\mu^1_{\mathrm{t}}, \mu^2_{\mathrm{t}}) + \delta^+ \mu^2(\mu^1_{\mathrm{t}}, \mu^2_{\mathrm{t}})  \right)$ and $\epsilon_2$ is a user-specified constant.
$\mu^2_{\mathrm{t}}$ is incremented by $\Delta \mu^2$ at the end of each step, and the process continues until $\mu^2_{\mathrm{t}}$ reaches its maximum.
After the construction is done, $\mu^1_{\mathrm{t}}$ is incremented by $\Delta \mu^1$ and the process is repeated until $\mu^1_{\mathrm{t}}$ reaches it maximum.

We constructed $\mu^1$-$\mu^2$ ranges with $\epsilon_1=\epsilon_2=0.1$, $\flow=20\si{\hertz}$, $\fhigh=1024\si{\hertz}$ and the O2 LIGO-Livingston PSD shown in \figref{fig:psd}.
Here we consider the following 3 cases.
The first is the {\it low-spin case}, where signals are searched for with the low-spin template bank and the parameter estimation is performed with the low-spin prior.
The second is the {\it high-spin case}, where signals are searched for with the high-spin template bank and the parameter estimation is performed with the high-spin prior.
The third is the {\it broader-spin case}, where signals are searched for with the the low-spin template bank and the parameter estimation is performed with the high-spin prior.
\eqref{eq:systematic1} is applied for the first two cases and \eqref{eq:systematic2} is applied for the last case.
We found that we need $\sim 1.5 \times 10^5$, $\sim 2.1 \times 10^5$ and $\sim 1.2 \times 10^3$ ranges for the low-spin, high-spin and broader-spin cases respectively.

\subsection{ROQ basis construction} \label{sec:roqbasis}

Finally, the ROQ bases are constructed over the $\mu^1$-$\mu^2$ ranges.
In this work, we constructed ROQ bases of the TaylorF2 waveform model implemented in LALSuite \cite{lalsuite}.
We applied the lower frequency cutoff of $20\si{\hertz}$.
The higher frequency cutoff for each $\mu^1$-$\mu^2$ range is determined so that it is higher than the highest frequency of the waveforms in the range.
The frequency resolution for each $\mu^1$-$\mu^2$ range is determined so that it is finer than the inverse of the longest duration of the waveforms in the range.
For the basis construction, we applied the same method applied in \cite{Smith:2016qas}.
The bases were constructed so that the interpolation errors, which is defined by (21) of \cite{Field:2013cfa}, are less than $10^{-6}$.

As explained in \secref{sec:roq}, the quadratic basis size, $K$, is equal to $1$ for TaylorF2.
The linear basis sizes $J$ for representative trigger values are shown in \tabref{tab:basis_sizes}.
The basis sizes are $\mathcal{O}(10)$ for the low-spin and high-spin case, and $\mathcal{O}(10^2)$ for the broader-spin case.
Compared to the basis size of $3000$ from the previous study \cite{Canizares:2014fya} considering the same waveform model, our basis sizes are smaller by a factor of $\mathcal{O}(10)$ to $\mathcal{O}(100)$, and our ROQ base vectors speed up parameter estimation further by same factors.

In total, our technique can speed up the parameter estimation of BNS signals by a factor of $\mathcal{O}(10^3)$ to $\mathcal{O}(10^4)$ for the low-spin and high-spin case, and $\mathcal{O}(10^3)$ for the broader-spin case.
We note that the TaylorF2 waveform model is chosen just as a demonstration, and the same procedure can be applied to waveform models including the merger and ringdown part of the signal or the matter effects of colliding bodies.
For example, we also constructed the ROQ basis of the IMRPhenomD waveform model \cite{Khan:2015jqa} for the case of the first row in the table.
The basis sizes $(J, K)$ are $(21, 2)$, $(48, 3)$ and $(113, 3)$, which means the speed-up factors are $27000$, $12000$ and $5400$ for the low-spin, high-spin and broader-spin cases respectively.

\begin{table*}
\centering
\begin{tabular}{c | c c c | c c c | c c c }

\multirow{2}{*}{$m_{1,\mathrm{t}}=m_{2,\mathrm{t}}$}
& \multicolumn{3}{c |}{Low-spin}
& \multicolumn{3}{c |}{High-spin}
& \multicolumn{3}{c}{Broader-spin} \\
  
        & Prior range \!& Size \!& Speedup 
        & Prior range \!& Size \!& Speedup
        & Prior range \!& Size \!& Speedup \\
\hline
\multirow{2}{*}{$1\Msun$} 
& $446.10 \leq \mu^1 \leq 446.38$ & \multirow{2}{*}{21} & \multirow{2}{*}{28000}
& $445.41 \leq \mu^1 \leq 446.75$ & \multirow{2}{*}{63} & \multirow{2}{*}{9800}
& $444.34 \leq \mu^1 \leq 447.95$ & \multirow{2}{*}{153} & \multirow{2}{*}{4100}  \\
 
& $-95.0 \leq \mu^2 \leq -89.9$ & &
& $-98.5 \leq \mu^2 \leq -86.5$ & & 
& $-126.9 \leq \mu^2 \leq -58.8$ & & \\ \hline
\multirow{2}{*}{$1.4\Msun$}
& $254.79 \leq \mu^1 \leq 255.09$ & \multirow{2}{*}{21} & \multirow{2}{*}{12000}
& $254.31 \leq \mu^1 \leq 255.36$ & \multirow{2}{*}{43} & \multirow{2}{*}{5800}
& $253.15 \leq \mu^1 \leq 256.53$ & \multirow{2}{*}{129} & \multirow{2}{*}{2000}  \\

& $-60.8 \leq \mu^2 \leq -55.5$ & &
& $-62.9 \leq \mu^2 \leq -52.9$ & & 
& $-95.0 \leq \mu^2 \leq -25.8$ & &  \\ \hline
\multirow{2}{*}{$2\Msun$}
& $140.49 \leq \mu^1 \leq 140.76$ & \multirow{2}{*}{21} & \multirow{2}{*}{4600}
& $140.19 \leq \mu^1 \leq 141.05$ & \multirow{2}{*}{37} & \multirow{2}{*}{2600}
& $138.94 \leq \mu^1 \leq 142.14$ & \multirow{2}{*}{107} & \multirow{2}{*}{930} \\
 
& $-39.9 \leq \mu^2 \leq -35.5$ & &
& $-41.4 \leq \mu^2 \leq  -32.6$ & & 
& $-64.1 \leq \mu^2 \leq -14.2$ & & 
\end{tabular}
\caption{Prior ranges, basis sizes $J$, and speed-up factors $L / (J + K)$ for the TaylorF2 waveform model and representative trigger masses. Here we assume $m_{1,\mathrm{t}}=m_{2,\mathrm{t}}$ and $\chi_{1,\mathrm{t}} = \chi_{2, \mathrm{t}} = 0$. The lower frequency cutoff of the waveform is $20\si{\hertz}$. We only show the linear basis size since the quadratic basis size is alway $1$ for TaylorF2. The definitions of the low-spin, high-spin and broader-spin cases are explained in \secref{sec:mu1mu2range}. For the calculation of the speed-up factors, we estimate $L$ by $L\simeq \fstop \tau$, where $\fstop$ is the highest frequency and $\tau$ is the duration from $20\si{\hertz}$ of the BNS waveform with the trigger source parameters.}
\label{tab:basis_sizes}
\end{table*}

\section{Performance} \label{sec:injection}

For the application to rapid updates of source locations and properties, the main concerns are the run time and the accuracy of the parameter estimation. 
While the use of narrow prior range of $\mu^1$ and $\mu^2$ significantly speeds up the parameter estimation, it can bias the inference if the range is not broad enough to include the true values.
To investigate them, we artificially injected BNS signals into the publicly available LIGO-Virgo's O2 data \cite{Abbott:2019ebz} and performed parameter estimation of them with FROQ.
About $2000$ BNS signals were injected by intervals of $30$ seconds into data taken between 07:00:00 UTC and 23:39:00 UTC on August 19, 2017. 
This period was selected because the LIGO-Hanford, LIGO-Livingston and Virgo detectors were all running in observing mode during the period.
In addition, the analysis presented here is not biased by real gravitational-wave events since no significant gravitational-wave events have been detected during the period \cite{LIGOScientific:2018mvr}.

To simulate the detections of the signals, we searched for the injected signals with the matched filter technique. 
For filtering the data, we utilized a compact binary coalescence detection software, GstLAL \cite{Messick:2016aqy, Sachdev:2019vvd}, and the template bank used in the O2 search \cite{Mukherjee:2018yra}.
Since we focus on BNS signals, we clipped the BNS region ($1\Msun\leq m_{1,\mathrm{t}},m_{2,\mathrm{t}}\leq 2\Msun$) from the template bank.
The template bank is a low-spin template bank covering only $-0.05 \leq \chi_{1,\mathrm{t}}, \chi_{2,\mathrm{t}} \leq 0.05$.
This narrow spin range is motivated by the observations of binary pulsars as explained in \secref{sec:prior_constraint}.
For each injected signal, matched filter SNRs within a $0.2\si{\second}$ time window centered on its arrival time were calculated.
The width of the window is determined so that it is broader than the measurement error of the time.
We then calculate the network SNR, which is the square root of square sum of matched filter SNRs from the 3 detectors, and the second largest SNR among the SNRs from the 3 detectors.
The signal was assumed to be detected and deserved further parameter estimation investigations if the network SNR and the second largest SNR exceeded $12$ and $5.5$ respectively for a template waveform in the bank, which is similar to the criterion applied in \cite{Fairhurst:2010is} and same as that applied in \cite{Chan:2018csa}.
The source parameters maximizing the network SNR were used as input trigger values to the FROQ-accelerated parameter estimation.

The waveforms of the injected signals are TaylorF2 without the tidal effects of colliding bodies implemented in LALSuite, and they are recovered with the same waveform model in the parameter estimation.
In real observations the matter effects or the effects from the merger and ringdown part of the signal should be taken into account for events with high signal-to-noise ratios.
However, our primary goal is to check whether the use of the narrow prior range biases the parameter estimation, and this simplified setup is enough for our purpose.
We also note that we can apply the same procedure to waveform models including the matter effects or the merger and ringdown part of the signal for real events. 
The gain in the SNR from including these effects is negligible for the mass range we consider, and the localization errors presented in \secref{sec:skymap} should be quantitatively correct.
The waveform model we use abruptly terminates at twice the orbital frequency of the innermost stable circular orbit.
The systematic errors from the abrupt termination are expected to be smaller than the statistical errors for the mass range we consider and moderate signal-to-noise ratios of $\sim 10$ \cite{Mandel:2014tca}.

To test our technique in a broad parameter range, we distributed masses and spins of the injected signals uniformly.
To test the low-spin and broader-spin ROQ bases, we prepare two injection sets with different range of masses and spins: {\it narrow injection set} and {\it broad injection set}.
Injected signals in the narrow injection set have masses and spins within $1\Msun \leq m_1,m_2 \leq 2\Msun,~-0.05 \leq \chi_1,\chi_2 \leq 0.05$, and the parameter estimation on them was performed with the low-spin ROQ bases.
Injected signals in the broad injection set have masses and spins within $0 \leq m_1,m_2 \leq 3\Msun,~0.87\Msun \leq \Mc \leq 1.74 \Msun,~-0.7 \leq \chi_1,\chi_2 \leq 0.7$, and the parameter estimation on them was performed with the broader-spin ROQ bases.
The lower and upper bounds of chirp mass are chirp masses for $m_1=m_2=1\Msun$ and $m_1=m_2=2\Msun$.

For each set of masses and spins, $(\alpha, \delta)$ and $(\iota, \psi)$ were distributed isotropically, and $\phic$ was distributed uniformly.
$r$ was sampled from a probability distribution proportional to $r^2$ so that the sources are distributed uniformly over the volume\footnote{Here the cosmological corrections are neglected since the distances to the detectable sources are much shorter than the cosmological distance.}.
The expected SNRs were calculated with the representative PSDs of the 3 detectors and the sampling of these parameters was repeated if the injected signal is not expected to pass the detection criterion.
After matched filtering the data, we detected $1065$ and $491$ injected signals for the narrow and broad injection sets respectively.

The parameter estimation on these detected signals was performed with the MCMC samplers implemented in LALInference \cite{Veitch:2014wba}.
$4$ independent MCMC chains ran for each signal and each chain terminated when it generated $500$ statistically independent posterior samples.
The priors on $(\alpha,\delta)$ and $(\iota, \psi)$ are isotropic priors.
The prior on $r$ is proportional to $r^2$ so that the prior is uniform over the volume.
The priors on the remaining parameters, $\tc$, $\phic$, $m_1$, $m_2$, $\chi_1$ and $\chi_2$, are uniform.
The prior range of $\tc$ is limited to a $0.2\si{\second}$ window centered on the time of an injected signal to accelerate the parameter estimation.
We have chosen $\mu^1$, $\mu^2$, $q \equiv m_2 / m_1$ and $\chi_2$ as sampling parameters.
The prior is multiplied by the Jacobian determinant of the transformation from $(\mu^1,\mu^2,q,\chi_2)$ to $(m_1, m_2, \chi_1,\chi_2)$.

\subsection{Run time} \label{sec:runtime}

To investigate the time required for the FROQ-accelerated parameter estimation, we measured end-to-end wall clock times of the runs.
The most computationally costly parts are the computations of the ROQ weights, \eqref{eq:linearweight} and \eqref{eq:quadraticweight}, and the MCMC sampling.
Therefore, we measured the sum of their wall clock times as approximated end-to-end wall clock times.
All of the runs were performed on 18-core Intel Xenon E5-2695 CPUs with the clock rate of $2.1\si{\giga}\si{\hertz}$.
We have found that $50\%$ and $90\%$ runs for the narrow injection set finished within $16$ minutes and $29$ minutes respectively.
On the other hand, $50\%$ and $90\%$ runs for the broad injection set finished within $27$ minutes and $64$ minutes respectively.
Therefore, we conclude that FROQ can reduce the analysis time of parameter estimation of BNS signals to several tens of minutes.
Especially, the analysis for the narrow injection set, where the low-spin prior is applied, is fast enough even for the follow-up observations of the electromagnetic radiation in $\sim30$ minutes predicted in \cite{Ishii:2018yjg}.

We note that the waveform generation is no longer the dominant cost of the FROQ-accelerated parameter estimation.
With the low-spin FROQ basis, the waveform needs to be calculated only at $\simeq 20$ frequency samples as shown in \tabref{tab:basis_sizes}, and each waveform generation takes $\simeq 3\times10^{-5}~\si{s}$.
For the low-spin run with the median run time, the average number of waveform generations  across the MCMC chains is $\simeq 2.8 \times 10^6$.
With these numbers, the total time of the waveform generations is estimated to be $\simeq 100~\si{s}$, which is much shorter than the run time of $16$ minutes.
This indicates that the optimization in other respects is required for further speedup.

\subsection{Accuracy} \label{sec:accuracy}

To assess the accuracy of the parameter estimation, we check whether true parameter values are encompassed within credible intervals with the level of $p$ with probability of $p$ \cite{doi:10.1198/106186006X136976, 2018arXiv180406788T}.
In our case, the test is to check whether $50\%$ credible regions encompass true parameter values for $50\%$ of the injected signals, $90\%$ credible regions encompass true parameter values for $90\%$ of the injected signals, and so on.
The tool to visually check whether it is the case is a so-called {\it p-p plot}, where a fraction of the injected signals whose source parameter values are encompassed in credible intervals is plotted as a function of the interval's level.
In the ideal case, it becomes a diagonal line.
This method has been used for validating various gravitational-wave parameter estimation infrastructures \cite{Sidery:2013zua, Veitch:2014wba, Berry:2014jja, Pankow:2015cra, Singer:2015ema, Biwer:2018osg, DelPozzo:2018dpu, Smith:2019ucc, Romero-Shaw:2020owr}.

\begin{figure}
\includegraphics[width=0.85\columnwidth]{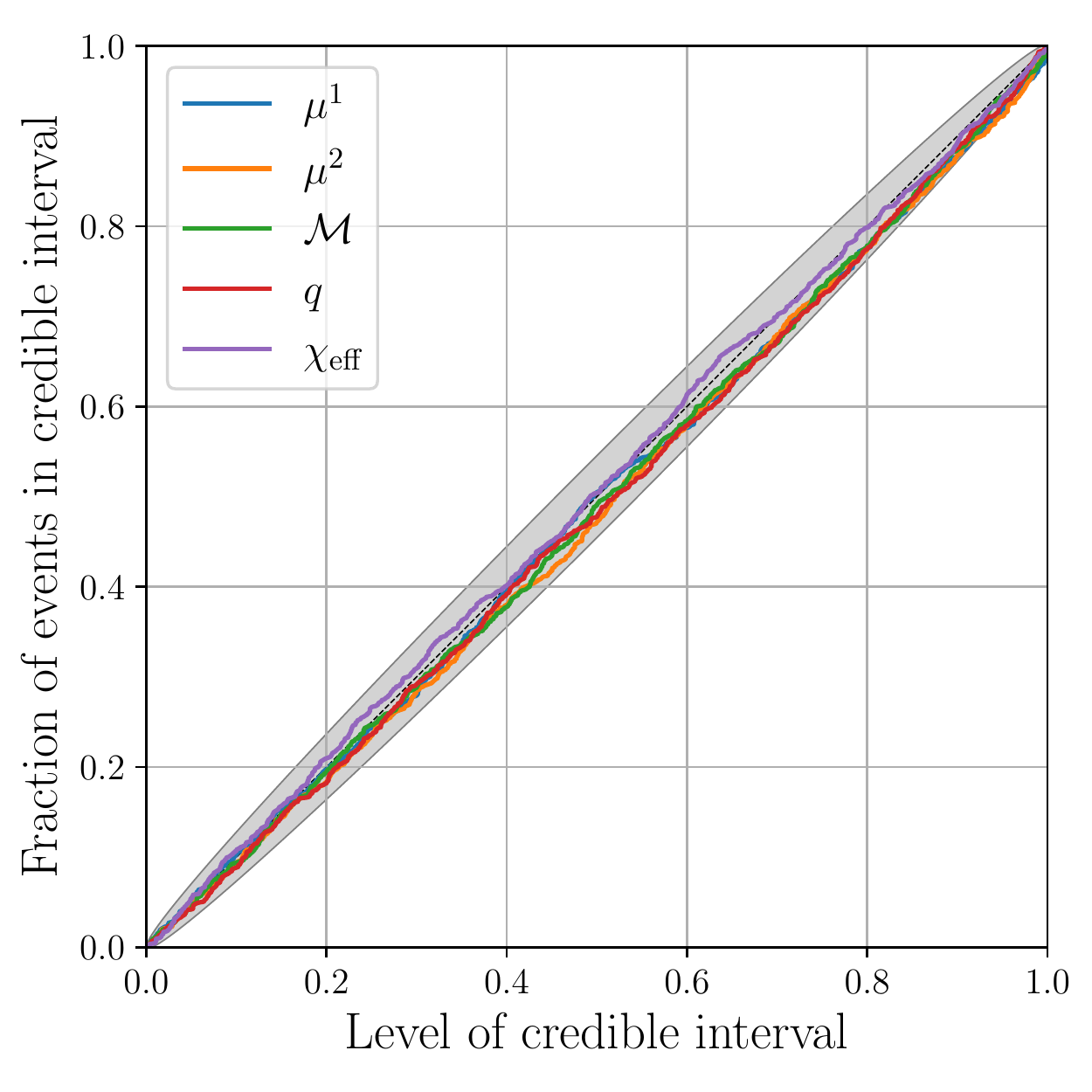}
\includegraphics[width=0.85\columnwidth]{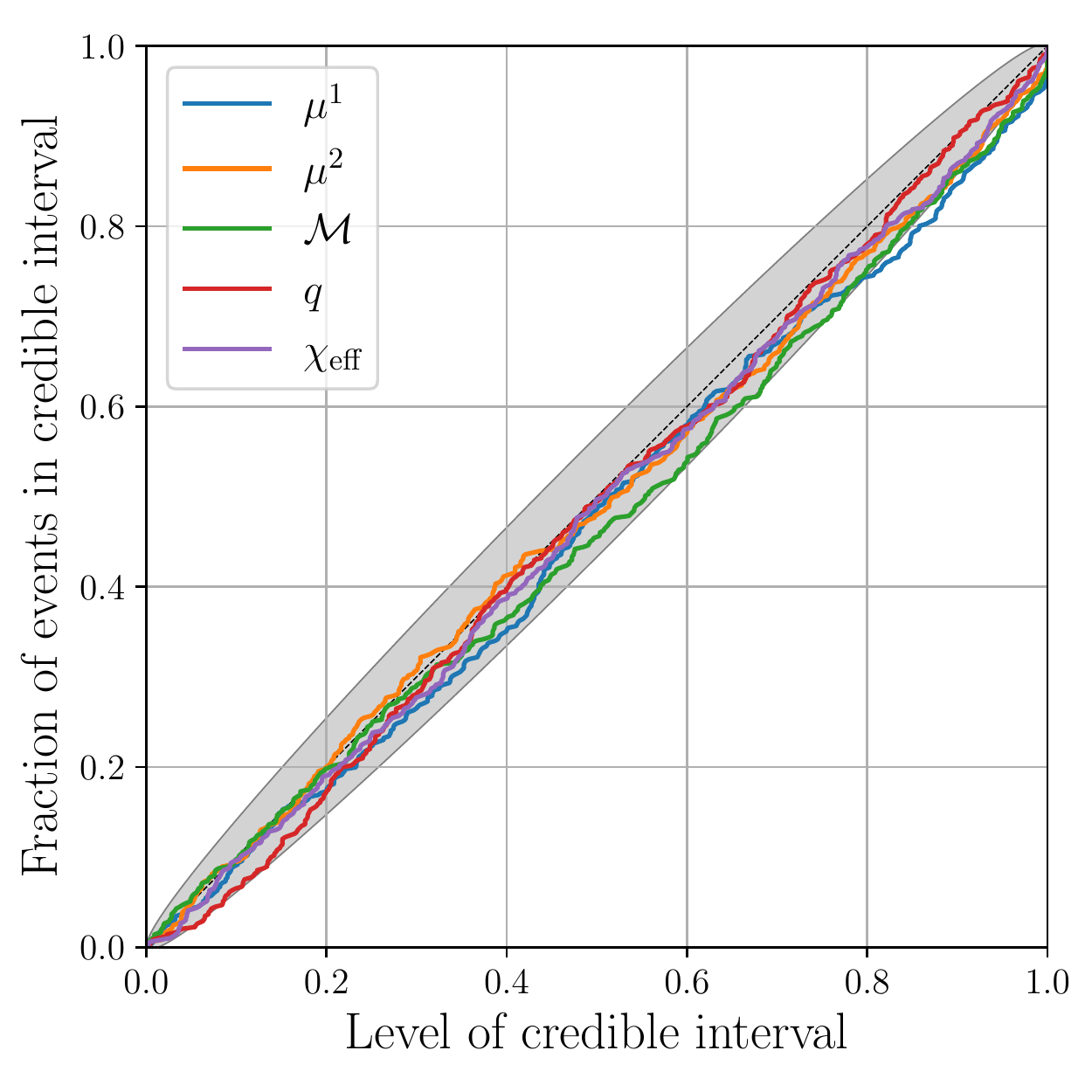}
\caption{Fraction of events encompassed in credible interval as a function of its level for $\mu^1$ (blue), $\mu^2$ (orange), $\Mc$ (green), $q$ (red) and $\chieff$ (purple). The diagonal dotted line indicates the ideal case where the credible interval encompasses events with a probability of the credible level. The gray region indicates the $3$-$\sigma$ ($99.7\%$) confidence band of the statistical error due to the finite number of the event samples. The upper one is for the narrow injection set and the lower one for the broad injection set.}
  \label{fig:p_p_plot}
\end{figure}

The p-p plots for representative parameters are shown in \figref{fig:p_p_plot}.
We select $\mu^1$, $\mu^2$, chirp mass $\Mc$, mass ratio $q$, and effective aligned spin $\chieff\equiv (m_1 \chi_1 + m_2 \chi_2) / (m_1 + m_2)$ \cite{Ajith:2009bn, Santamaria:2010yb} as representative parameters.
The statistical error due to the finite number of the event samples needs to be taken into account, and the gray region shows $3$-$\sigma$ ($99.7\%$) confidence interval of the error.
The p-p plots for the narrow injection set are inside the error region in most of the range, which indicates that the FROQ technique is stable and accurate enough in the case where the template bank covers the whole parameter space we consider in the parameter estimation.
On the other hand, the p-p plots of $\mu^1$ and $\Mc$ for the broad injection set tend to be outside the error region at $p \gtrsim 0.8$.
This is caused because the error estimate \eqref{eq:systematic2} does not provide broad enough range of $\mu^1$ and the prior range does not encompass its true value in some cases.
This may be solved by using a template bank covering the high-spin region or performing additional filtering with high-spin template waveforms after the detections.
On the other hand, the deviation from the diagonal line is not so significant as seen in the figure.
Especially, the p-p plots of $\mu^2$, $q$ and $\chieff$ are inside the error region in most of the range.
In addition, it is shown that the estimates of sources' locations are more accurate than those from Bayestar in \secref{sec:skymap}.
Therefore, we conclude that the FROQ-accelerated parameter estimation can at least provide more accurate information on the parameters than the trigger values.

\subsection{Rapid skymap update} \label{sec:skymap}

Our primary goal is to update the estimate of a source location very quickly.
Here, we investigate how much the FROQ-accelerated parameter estimation improves the initial estimate of a source location produced by the Bayestar software.

\begin{figure}
\includegraphics[width=\columnwidth]{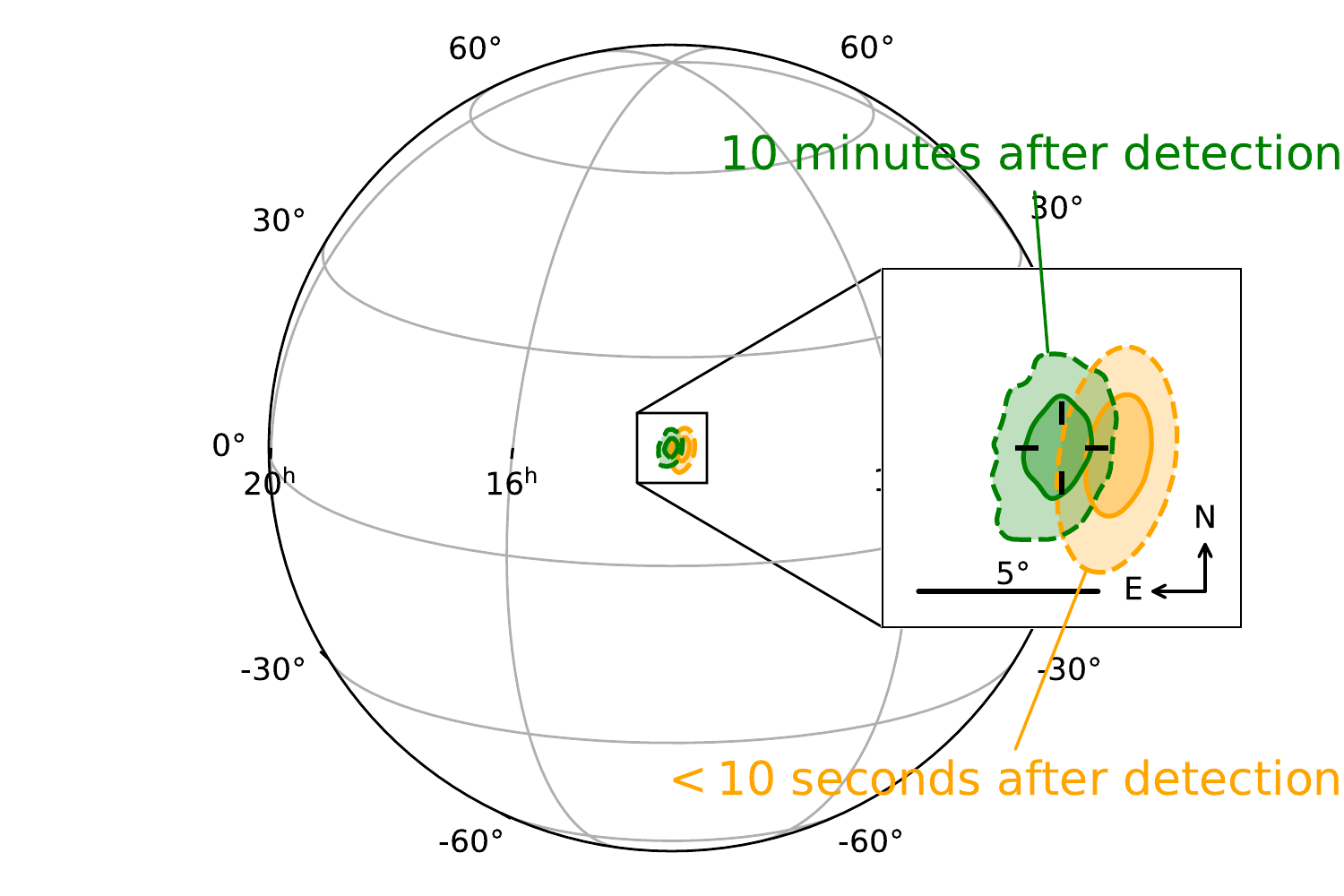}
\caption{Two-dimensional location estimates of an injected signal in the narrow injection set calculated by the Bayestar software (orange) and the FROQ-accelerated parameter estimation (green). The solid line represents the $50\%$ credible region and the dashed line represents the $90\%$ credible region. The luminosity distance of the injected signal is $\simeq 45\si{\mega}\si{\parsec}$.}
  \label{fig:skymap_comparison}
\end{figure}

\figref{fig:skymap_comparison} shows the $50\%$ and $90\%$ credible regions of source's direction given by the Bayestar software and the FROQ-accelerated parameter estimation for an injected signal in the narrow injection set, whose luminosity distance is $\simeq 45\si{\mega}\si{\parsec}$.
As seen in the figure, the latter skymap captures the true location closer to its center and can be generated in about $10$ minutes after detection.
Following \cite{Singer:2014qca}, we calculate the searched area, the area of the smallest credible region encompassing the signal's true location, for each skymap.
The searched area is $\simeq$17\,deg$^2$ for Bayestar and $\simeq$4.2$\times$10$^{-2}$\,deg$^2$ for FROQ-accelerated parameter estimation, which means the FROQ-accelerated parameter estimation significantly reduces the burdens required to find the electromagnetic counterpart.

\begin{figure}
\includegraphics[width=\columnwidth]{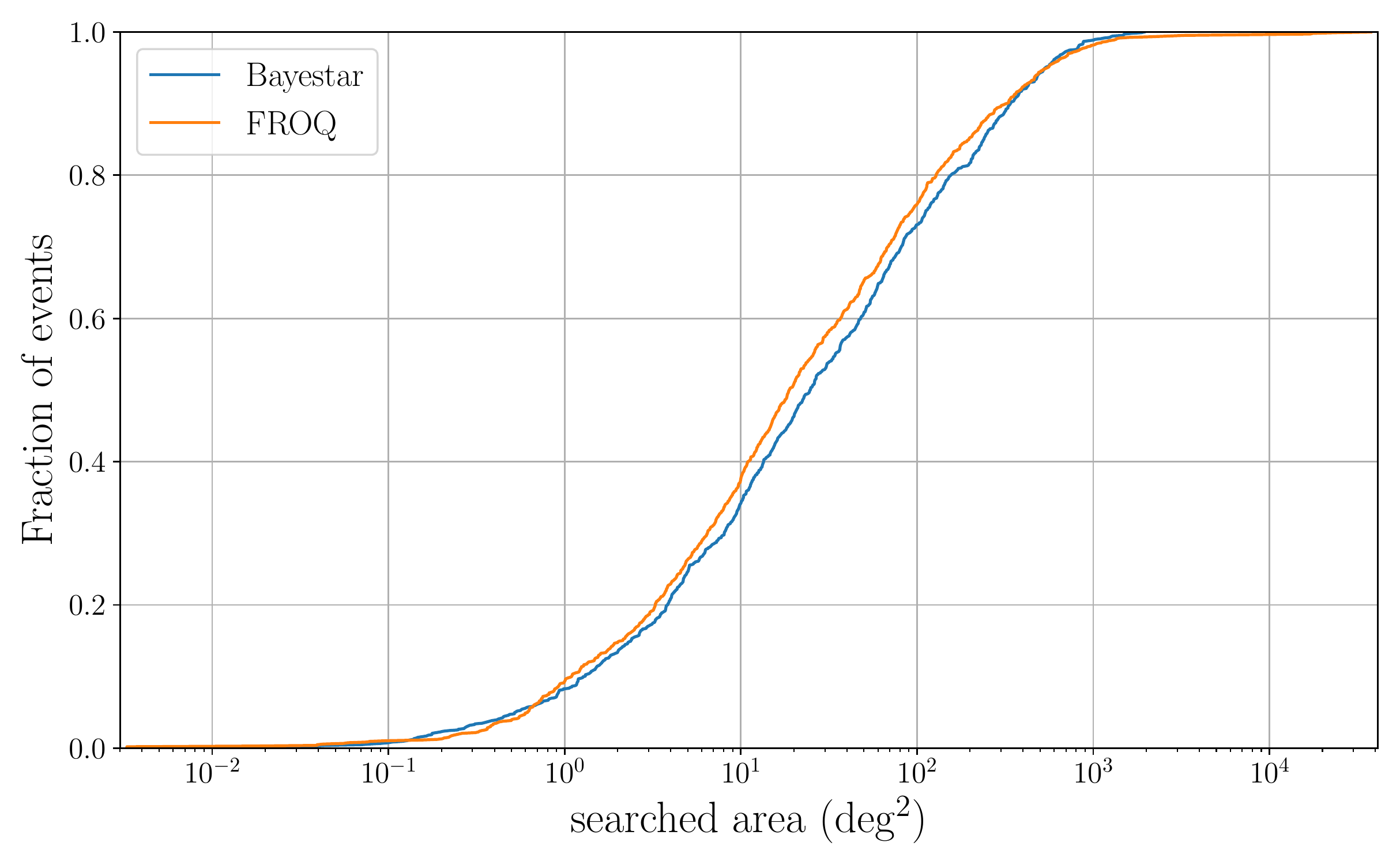}
\includegraphics[width=\columnwidth]{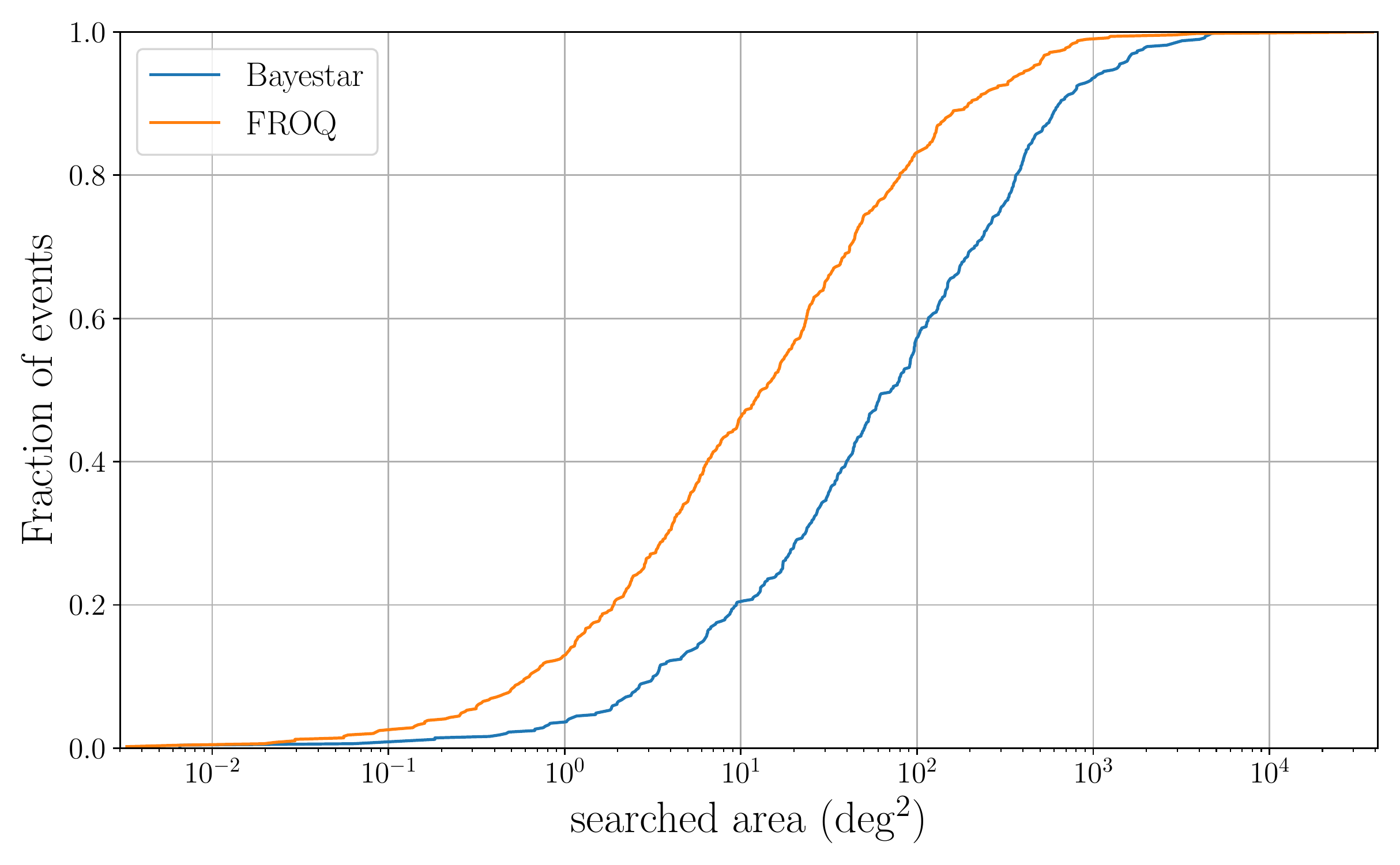}
\caption{Cumulative distributions of the searched area, the smallest credible regions encompassing the signals' true locations, from the initial Bayestar analysis (blue) and the follow-up FROQ-accelerated parameter estimation (orange). The upper one is for the narrow injection study and the lower one for the broad injection study.}
  \label{fig:cumulative_area}
\end{figure}

To show that the searched area is systematically reduced by the FROQ-accelerated parameter estimation, we plot their cumulative distributions in \figref{fig:cumulative_area}.
The upper figure for the narrow injection study shows the slight systematic reduction of the searched area.
For example, the median searched area is 25\,deg$^2$ for Bayestar and 19\,deg$^2$ for FROQ.
This can be explained by the SNR loss at the initial analysis due to the mismatch between a trigger waveform and a true waveform.
The improvement is more drastic for the broad injection study as seen in the lower figure.
The median searched area is 72\,deg$^2$ for Bayestar and 13\,deg$^2$ for FROQ.
This is because the parameter space covered by the template bank does dot encompass the true parameters and the initial SNR loss is more significant.
In both cases, the FROQ-accelerated parameter estimation can reduce the searched area in several tens of minutes and helps the follow-up observations.

\section{Conclusion} \label{sec:conclusion}

In this paper we have presented a technique to speed up the parameter estimation of gravitational waves from binary neutron star coalescence and designated it focused reduced order quadrature (FROQ).
Our technique is based on the reduced order quadrature technique and further improves it by utilizing the trigger values of masses and spins provided by a detection software for compact binary coalescence signals.
We have found that our technique can speed up the parameter estimation by a factor of $\mathcal{O}(10^3)$ to $\mathcal{O}(10^4)$, which is $\mathcal{O}(10)$ to $\mathcal{O}(10^2)$ times faster than the previous reduced order quadrature technique as shown in \secref{sec:roqbasis}.
In addition to being quite fast, it is very accurate and can be used to update the initial estimates of the source parameters as shown in \secref{sec:accuracy}.
Our technique allows for significant improvements of the initial estimates of source locations within several tens of minutes and will be able to help the follow-up observations of the gravitational-wave signals as shown in \secref{sec:runtime} and \secref{sec:skymap}.

Finally we discuss possible extensions of our work.
One direction is the extension to neutron star black hole binary and binary black hole coalescence.
For such massive events, the effects from the merger and black hole ringdown part of the signal, which are ignored in \secref{sec:froq}, need to be taken into account.
The dependence of the higher frequency cutoff on the source parameters also needs to be taken into account since it is inside the sensitive band of the detectors.
They can affect the best measurable combinations of the source parameters and their error estimates.
This extension is necessary for this technique to be applied to rapid source classifications.

The extension to more general waveforms including orbital precession and modes from higher-order multipoles \cite{Thorne:1980ru} are also important.
Since these effects can bias the initial trigger values, it needs to be quantified and taken into account.
Since these effects can break the degeneracy and improve the estimate of the distance \cite{Vitale:2018wlg, Tagoshi:2014xsa}, this extension can be helpful for the follow-up observations.

We also note that our scheme of the $\mu^1$-$\mu^2$ range construction discussed in \secref{sec:mu1mu2range} is not efficient in a sense that the resultant ranges are highly overlapping.
This problem arises because we use only one ROQ basis in the parameter estimation.
Instead, we can use multiple ROQ bases for different parameter ranges and use one of them depending on the current location in the parameter space at each sampling step.
This can further reduce the basis sizes and speed up the parameter estimation since each ROQ basis can be constructed in the parameter range narrower than required from the error estimates.
We leave these extensions and improvements as the future work.

\acknowledgements{
We thank Christopher Berry, Kipp Cannon, Kazumi Kashiyama, Chunglee Kim, Kyohei Kawaguchi, Yasuyuki Okumura, Toshikazu Shigeyama, Rory Smith, Tadayuki Takahashi and Naoki Yoshida for useful discussions and comments.
We thank Leo Tsukada for careful reading of the manuscript and helpful comments.
We also thank Heather Fong for sharing the template bank used for the O2 GstLAL search and Cody Messick for instructions on GstLAL and providing a patch for code changes in GstLAL required for our study.
This work is supported by JSPS KAKENHI Grant Number 19J13840 (S.M.).
The authors are grateful for computational resources provided by the LIGO Laboratory and supported by National Science Foundation Grants PHY-0757058 and PHY-0823459.
This research has made use of data obtained from the Gravitational Wave Open Science Center (https://www.gw-openscience.org), a service of LIGO Laboratory, the LIGO Scientific Collaboration and the Virgo Collaboration.
LIGO is funded by the U.S. National Science Foundation. Virgo is funded by the French Centre National de Recherche Scientifique (CNRS), the Italian Istituto Nazionale della Fisica Nucleare (INFN) and the Dutch Nikhef, with contributions by Polish and Hungarian institutes.
}

\appendix

\section{The derivation of \eqref{eq:marginalized_posterior}} \label{sec:derivation}

Our derivation is largely based on Appendix A of \cite{Singer:2015ema}.
The signal observed at the $i$-th detector is given by
\begin{equation}
\tilde{h}_i(f) = \frac{\rho_i}{\sqrt{c_i (\bm{h}', \bm{h}')}} \e^{i(\phi_i - 2 \pi f t_i)} \tilde{h}' (f;\vpsi').
\end{equation}
$\phi_i$ and $t_i$ are constant phase and time at the $i$-th detector.
$\bm{h}'$ is given by
\begin{align}
&\tilde{h}'(f;\bm{\psi}') = \ffref^{-\frac{7}{6}} \e^{-i \Phi'(f)}, \\
&\Phi' = \psi^0 \ffref^{-\frac{5}{3}} + \psi^2 \ffref^{-1} \nonumber \\
&~~~~~~~~~~+ \psi^3 \ffref^{-\frac{2}{3}} + \psi^4 \ffref^{-\frac{1}{3}},
\end{align}
and $\vpsi'=(\psi^0, \psi^2, \psi^3, \psi^4)$.
The inner product is defined by \eqref{eq:ip_continuous}, and here we assume that the PSD of the $i$-th detector is given by $S_{\mathrm{n},i}(f)=S_{\mathrm{n}}(f) / c_i$ as explained in \secref{sec:froq}.
The (optimal) SNR at the $i$-th detector is given by $\sqrt{c_i (\bm{h}_i, \bm{h}_i)} = \rho_i$.
The parameters parametrizing the signal are
\begin{equation}
\bm{s}\equiv (\ln \rho_1, \phi_1, t_1, \dots, \ln \rho_N, \phi_N, t_N, \psi^0, \psi^2, \psi^3, \psi^4)
\end{equation}

In the high-SNR limit, the likelihood can be approximated by the Gaussian distribution \cite{Cutler:1994ys, Poisson:1995ef},
\begin{equation}
\likelihood \simeq B \mathrm{exp}\left[ - \frac{1}{2} G_{AB} (s^A - \hat{s}^A) (s^B - \hat{s}^B) \right],
\end{equation}
where $s^A$ is a component of $\bm{s}$ and $B$ is a normalization constant.
$\hat{s}^A$ represents the value of $s^A$ maximizing the likelihood and
\begin{equation}
G_{AB} \equiv \sum^N_{i=1} c_i \left(\frac{\partial \bm{h}_i}{\partial s^A}, \frac{\partial \bm{h}_i}{\partial s^B}\right),
\end{equation}
where the derivative is evaluated at $s^A = \hat{s}^A$.

Following \cite{Singer:2015ema}, we apply the following transformation,
\begin{align}
\ln\rho_N &\rightarrow \overline{\ln\rho} = \frac{\sum_i \hat{\rho}^2_i \ln\rho_i}{\rho^2_{\mathrm{net}}}, \\
\phi_N &\rightarrow \bar{\phi} = \frac{\sum_i \hat{\rho}^2_i \phi_i}{\rho^2_{\mathrm{net}}}, \\
t_N &\rightarrow \bar{t} = \frac{\sum_i \hat{\rho}^2_i t_i}{\rho^2_{\mathrm{net}}}
\end{align}
and
\begin{align}
\ln \rho_i &\rightarrow \delta \ln\rho_i = \ln\rho_i - \overline{\ln\rho}, \\
\phi_i &\rightarrow \delta \phi_i = \phi_i - \bar{\phi}, \\
t_i &\rightarrow \delta t_i = t_i - \bar{t}
\end{align}
for $i = 1, 2, \dots, N-1$.
$\rho_{\mathrm{net}}$ is the network SNR defined by $\rho^2_{\mathrm{net}}=\sum^N_{i=1} \hat{\rho}^2_i$.
Transforming $G$ with the corresponding Jacobian, $J$, we get
\begin{equation}
J^T G J =
\left(
\begin{array}{cc}
X & \bigzero \\
\bigzero & \rho^2_{\mathrm{net}} Y
\end{array}
\right).
\end{equation}
$X$ is a $3(N-1)$-dimensional square matrix for the subset, $(\delta \ln \rho_1, \delta \phi_1, \delta t_1, \dots, \delta \ln \rho_{N-1}, \delta \phi_{N-1}, \delta t_{N-1})$, and $\rho^2_{\mathrm{net}} Y$ is a square matrix for the remaining parameters, $(\overline{\ln \rho}, \bar{\phi}, \bar{t}, \psi^0, \psi^2, \psi^3, \psi^4)$.

$Y$ has the following structure,
\begin{equation}
    Y = \bordermatrix{
        ~ & \overline{\ln\rho} & \bar{\phi} & \bar{t} & \bm{\psi}' \cr
        \overline{\ln\rho} & 1 & 0 & 0 & \bm{0}^{\mathrm{T}} \cr
        \bar{\phi} & 0 & * & * & * \cr
        \bar{t} & 0 & * & * & * \cr
        \bm{\psi}' & \bm{0} & * & * & *
    }.
\end{equation}
$-\bar{\phi}$ and $2 \pi \fref \bar{t}$ correspond to $\psi^5$ and $\psi^8$ respectively.
Multiplying $\bar{\phi}$ and $\bar{t}$ by $-1$ and $2 \pi \fref$ respectively and reordering the parameters, we get
\begin{equation}
\tilde{J}^T G \tilde{J} = \bordermatrix{
~ & \bm{\Psi} & \tilde{\bm{\psi}} \cr
\bm{\Psi} & \Gamma' & \bigzero \cr
\tilde{\bm{\psi}} & \bigzero & \rho^2_{\mathrm{net}} \Gamma^{(2\PN)}
},
\end{equation}
where $\Gamma'$ is a $(3N-2)$-dimensional square matrix for $\bm{\Psi} = (\delta \ln \rho_1, \delta \phi_1, \delta t_1, \dots, \delta \ln\rho_{N-1}, \delta \phi_{N-1}, \delta t_{N-1}, \overline{\ln\rho})$, and the components of $\tilde{\bm{\psi}}$ are $\tilde{\psi}^\alpha = \psi^\alpha~(\alpha=0,2,3,4)$, $\tilde{\psi}^5 = -\bar{\phi}$, $\tilde{\psi}^8=2 \pi \fref \bar{t}$.
$\tilde{J}$ is the transformation combining $J$, reordering of the parameters and rescaling of $\bar{\phi}$ and $\bar{t}$.

The marginalization of the posterior over $\thetaex$ can be written as
\begin{align}
&\int \posterior d \thetaex \nonumber \\
&=p(\thetain) \int p(\thetaex) \mathrm{exp}\left[ -\frac{1}{2} \Gamma'_{PQ} (\Psi^P - \hat{\Psi}^P) (\Psi^Q - \hat{\Psi}^Q) \right] \nonumber \\
& ~~\times \mathrm{exp}\left[ -\frac{1}{2} \rho^2_{\mathrm{net}} \Gamma^{(2\PN)}_{pq} (\tilde{\psi}^p - \hat{\tilde{\psi}}^p) (\tilde{\psi}^q - \hat{\tilde{\psi}}^q) \right] d \thetaex,
\end{align}
where $P$ and $Q$ are subscripts for $\bm{\Psi}$.
Here we assume that the prior can be decomposed into the prior on $\thetain$ and $\thetaex$, $\prior=p(\thetain) p(\thetaex)$.
$\Psi^P$ depends on the differences of constant phases or arrival times at different detectors, ratios of amplitudes at different detectors, or a geometrical mean of the amplitudes.
Therefore, it depends only on $\alpha,\delta,\iota,\psi,r, \Mc$.
In addition, typically the prior on $\phic$ and $\tc$ is uniform and does not have any correlations with the other parameters.
Therefore, the integration over $\phic$ and $\tc$ can be easily performed,
\begin{align}
&\int \mathrm{exp}\left[ -\frac{1}{2} \rho^2_{\mathrm{net}} \Gamma^{(2\PN)}_{pq} (\tilde{\psi}^p - \hat{\tilde{\psi}}^p) (\tilde{\psi}^q - \hat{\tilde{\psi}}^q) \right] d \phic d \tc \nonumber \\
&= - \frac{1}{2\pi\fref} \int \mathrm{exp}\left[ -\frac{1}{2} \rho^2_{\mathrm{net}} \Gamma^{(2\PN)}_{pq} (\tilde{\psi}^p - \hat{\tilde{\psi}}^p) (\tilde{\psi}^q - \hat{\tilde{\psi}}^q) \right] d \tilde{\psi}^5 d \tilde{\psi}^8 \nonumber \\
&\propto \mathrm{exp}\left[ -\frac{1}{2} \rho^2_{\mathrm{net}} \tilde{\Gamma}^{(2\PN)}_{\alpha \beta} (\psi^\alpha - \hat{\psi}^\alpha) (\psi^\beta - \hat{\psi}^\beta) \right].
\end{align}

On the other hand,  $\psi^\alpha$ does not depend on $\alpha,\delta,\iota,\psi,r$, and the integration over them can be done separately.
Finally, we get
\begin{align}
&\int \posterior d \thetaex \nonumber \\
&\propto C(\Mc) p(\thetain) \mathrm{exp}\left[-\frac{1}{2} \rho^2_{\mathrm{net}} \tilde{\Gamma}^{(2\PN)}_{\alpha \beta} (\psi^\alpha - \hat{\psi}^\alpha) (\psi^\beta - \hat{\psi}^\beta) \right], \label{eq:_marginalized_posterior}
\end{align}
where
\begin{align}
&C(\Mc) \equiv \nonumber \\
& \int p(\thetaex) \mathrm{exp}\left[ -\frac{1}{2} \Gamma'_{PQ} (\Psi^P - \hat{\Psi}^P) (\Psi^Q - \hat{\Psi}^Q) \right] d\alpha d \delta d \iota d\psi dr.
\end{align}
$C$ depends on $\Mc$ since $\overline{\ln\rho}$ depends on $\Mc$.
Reflecting the fact that $\Mc$ is measured mostly from the phase rather than the amplitude, $C(\Mc)$ should be a much smoother function than the Gaussian function of $\psi^\alpha$ in \eqref{eq:_marginalized_posterior}.
Therefore, we ignore its dependence on $\Mc$ and obtain \eqref{eq:marginalized_posterior}.

\bibliographystyle{apsrev4-1}
\bibliography{Paper_PRD}

\begin{thebibliography}{103}%
\makeatletter
\providecommand \@ifxundefined [1]{%
 \@ifx{#1\undefined}
}%
\providecommand \@ifnum [1]{%
 \ifnum #1\expandafter \@firstoftwo
 \else \expandafter \@secondoftwo
 \fi
}%
\providecommand \@ifx [1]{%
 \ifx #1\expandafter \@firstoftwo
 \else \expandafter \@secondoftwo
 \fi
}%
\providecommand \natexlab [1]{#1}%
\providecommand \enquote  [1]{``#1''}%
\providecommand \bibnamefont  [1]{#1}%
\providecommand \bibfnamefont [1]{#1}%
\providecommand \citenamefont [1]{#1}%
\providecommand \href@noop [0]{\@secondoftwo}%
\providecommand \href [0]{\begingroup \@sanitize@url \@href}%
\providecommand \@href[1]{\@@startlink{#1}\@@href}%
\providecommand \@@href[1]{\endgroup#1\@@endlink}%
\providecommand \@sanitize@url [0]{\catcode `\\12\catcode `\$12\catcode
  `\&12\catcode `\#12\catcode `\^12\catcode `\_12\catcode `\%12\relax}%
\providecommand \@@startlink[1]{}%
\providecommand \@@endlink[0]{}%
\providecommand \url  [0]{\begingroup\@sanitize@url \@url }%
\providecommand \@url [1]{\endgroup\@href {#1}{\urlprefix }}%
\providecommand \urlprefix  [0]{URL }%
\providecommand \Eprint [0]{\href }%
\providecommand \doibase [0]{http://dx.doi.org/}%
\providecommand \selectlanguage [0]{\@gobble}%
\providecommand \bibinfo  [0]{\@secondoftwo}%
\providecommand \bibfield  [0]{\@secondoftwo}%
\providecommand \translation [1]{[#1]}%
\providecommand \BibitemOpen [0]{}%
\providecommand \bibitemStop [0]{}%
\providecommand \bibitemNoStop [0]{.\EOS\space}%
\providecommand \EOS [0]{\spacefactor3000\relax}%
\providecommand \BibitemShut  [1]{\csname bibitem#1\endcsname}%
\let\auto@bib@innerbib\@empty
\bibitem [{\citenamefont {Harry}(2010)}]{Harry:2010zz}%
  \BibitemOpen
  \bibfield  {author} {\bibinfo {author} {\bibfnamefont {G.~M.}\ \bibnamefont
  {Harry}} (\bibinfo {collaboration} {LIGO Scientific}),\ }\href {\doibase
  10.1088/0264-9381/27/8/084006} {\bibfield  {journal} {\bibinfo  {journal}
  {Class. Quant. Grav.}\ }\textbf {\bibinfo {volume} {27}},\ \bibinfo {pages}
  {084006} (\bibinfo {year} {2010})}\BibitemShut {NoStop}%
\bibitem [{\citenamefont {Acernese}\ \emph {et~al.}(2015)\citenamefont
  {Acernese} \emph {et~al.}}]{TheVirgo:2014hva}%
  \BibitemOpen
  \bibfield  {author} {\bibinfo {author} {\bibfnamefont {F.}~\bibnamefont
  {Acernese}} \emph {et~al.} (\bibinfo {collaboration} {VIRGO}),\ }\href
  {\doibase 10.1088/0264-9381/32/2/024001} {\bibfield  {journal} {\bibinfo
  {journal} {Class. Quant. Grav.}\ }\textbf {\bibinfo {volume} {32}},\ \bibinfo
  {pages} {024001} (\bibinfo {year} {2015})},\ \Eprint
  {http://arxiv.org/abs/1408.3978} {arXiv:1408.3978 [gr-qc]} \BibitemShut
  {NoStop}%
\bibitem [{\citenamefont {Abbott}\ \emph
  {et~al.}(2017{\natexlab{a}})\citenamefont {Abbott} \emph
  {et~al.}}]{TheLIGOScientific:2017qsa}%
  \BibitemOpen
  \bibfield  {author} {\bibinfo {author} {\bibfnamefont {B.~P.}\ \bibnamefont
  {Abbott}} \emph {et~al.} (\bibinfo {collaboration} {LIGO Scientific,
  Virgo}),\ }\href {\doibase 10.1103/PhysRevLett.119.161101} {\bibfield
  {journal} {\bibinfo  {journal} {Phys. Rev. Lett.}\ }\textbf {\bibinfo
  {volume} {119}},\ \bibinfo {pages} {161101} (\bibinfo {year}
  {2017}{\natexlab{a}})},\ \Eprint {http://arxiv.org/abs/1710.05832}
  {arXiv:1710.05832 [gr-qc]} \BibitemShut {NoStop}%
\bibitem [{\citenamefont {Goldstein}\ \emph {et~al.}(2017)\citenamefont
  {Goldstein} \emph {et~al.}}]{Goldstein:2017mmi}%
  \BibitemOpen
  \bibfield  {author} {\bibinfo {author} {\bibfnamefont {A.}~\bibnamefont
  {Goldstein}} \emph {et~al.},\ }\href {\doibase 10.3847/2041-8213/aa8f41}
  {\bibfield  {journal} {\bibinfo  {journal} {Astrophys. J. Lett.}\ }\textbf
  {\bibinfo {volume} {848}},\ \bibinfo {pages} {L14} (\bibinfo {year}
  {2017})},\ \Eprint {http://arxiv.org/abs/1710.05446} {arXiv:1710.05446
  [astro-ph.HE]} \BibitemShut {NoStop}%
\bibitem [{\citenamefont {Savchenko}\ \emph {et~al.}(2017)\citenamefont
  {Savchenko} \emph {et~al.}}]{Savchenko:2017ffs}%
  \BibitemOpen
  \bibfield  {author} {\bibinfo {author} {\bibfnamefont {V.}~\bibnamefont
  {Savchenko}} \emph {et~al.},\ }\href {\doibase 10.3847/2041-8213/aa8f94}
  {\bibfield  {journal} {\bibinfo  {journal} {Astrophys. J. Lett.}\ }\textbf
  {\bibinfo {volume} {848}},\ \bibinfo {pages} {L15} (\bibinfo {year}
  {2017})},\ \Eprint {http://arxiv.org/abs/1710.05449} {arXiv:1710.05449
  [astro-ph.HE]} \BibitemShut {NoStop}%
\bibitem [{\citenamefont {Abbott}\ \emph
  {et~al.}(2017{\natexlab{b}})\citenamefont {Abbott} \emph
  {et~al.}}]{Monitor:2017mdv}%
  \BibitemOpen
  \bibfield  {author} {\bibinfo {author} {\bibfnamefont {B.~P.}\ \bibnamefont
  {Abbott}} \emph {et~al.} (\bibinfo {collaboration} {LIGO Scientific, Virgo,
  Fermi-GBM, INTEGRAL}),\ }\href {\doibase 10.3847/2041-8213/aa920c} {\bibfield
   {journal} {\bibinfo  {journal} {Astrophys. J. Lett.}\ }\textbf {\bibinfo
  {volume} {848}},\ \bibinfo {pages} {L13} (\bibinfo {year}
  {2017}{\natexlab{b}})},\ \Eprint {http://arxiv.org/abs/1710.05834}
  {arXiv:1710.05834 [astro-ph.HE]} \BibitemShut {NoStop}%
\bibitem [{\citenamefont {Abbott}\ \emph
  {et~al.}(2017{\natexlab{c}})\citenamefont {Abbott} \emph
  {et~al.}}]{GBM:2017lvd}%
  \BibitemOpen
  \bibfield  {author} {\bibinfo {author} {\bibfnamefont {B.}~\bibnamefont
  {Abbott}} \emph {et~al.} (\bibinfo {collaboration} {LIGO Scientific, Virgo,
  Fermi GBM, INTEGRAL, IceCube, AstroSat Cadmium Zinc Telluride Imager Team,
  IPN, Insight-Hxmt, ANTARES, Swift, AGILE Team, 1M2H Team, Dark Energy Camera
  GW-EM, DES, DLT40, GRAWITA, Fermi-LAT, ATCA, ASKAP, Las Cumbres Observatory
  Group, OzGrav, DWF (Deeper Wider Faster Program), AST3, CAASTRO, VINROUGE,
  MASTER, J-GEM, GROWTH, JAGWAR, CaltechNRAO, TTU-NRAO, NuSTAR, Pan-STARRS,
  MAXI Team, TZAC Consortium, KU, Nordic Optical Telescope, ePESSTO, GROND,
  Texas Tech University, SALT Group, TOROS, BOOTES, MWA, CALET, IKI-GW
  Follow-up, H.E.S.S., LOFAR, LWA, HAWC, Pierre Auger, ALMA, Euro VLBI Team, Pi
  of Sky, Chandra Team at McGill University, DFN, ATLAS Telescopes, High Time
  Resolution Universe Survey, RIMAS, RATIR, SKA South Africa/MeerKAT}),\ }\href
  {\doibase 10.3847/2041-8213/aa91c9} {\bibfield  {journal} {\bibinfo
  {journal} {Astrophys. J. Lett.}\ }\textbf {\bibinfo {volume} {848}},\
  \bibinfo {pages} {L12} (\bibinfo {year} {2017}{\natexlab{c}})},\ \Eprint
  {http://arxiv.org/abs/1710.05833} {arXiv:1710.05833 [astro-ph.HE]}
  \BibitemShut {NoStop}%
\bibitem [{\citenamefont {Coulter}\ \emph {et~al.}(2017)\citenamefont {Coulter}
  \emph {et~al.}}]{Coulter:2017wya}%
  \BibitemOpen
  \bibfield  {author} {\bibinfo {author} {\bibfnamefont {D.}~\bibnamefont
  {Coulter}} \emph {et~al.},\ }\href {\doibase 10.1126/science.aap9811}
  {\bibfield  {journal} {\bibinfo  {journal} {Science}\ }\textbf {\bibinfo
  {volume} {358}},\ \bibinfo {pages} {1556} (\bibinfo {year} {2017})},\ \Eprint
  {http://arxiv.org/abs/1710.05452} {arXiv:1710.05452 [astro-ph.HE]}
  \BibitemShut {NoStop}%
\bibitem [{\citenamefont {Drout}\ \emph {et~al.}(2017)\citenamefont {Drout}
  \emph {et~al.}}]{Drout:2017ijr}%
  \BibitemOpen
  \bibfield  {author} {\bibinfo {author} {\bibfnamefont {M.}~\bibnamefont
  {Drout}} \emph {et~al.},\ }\href {\doibase 10.1126/science.aaq0049}
  {\bibfield  {journal} {\bibinfo  {journal} {Science}\ }\textbf {\bibinfo
  {volume} {358}},\ \bibinfo {pages} {1570} (\bibinfo {year} {2017})},\ \Eprint
  {http://arxiv.org/abs/1710.05443} {arXiv:1710.05443 [astro-ph.HE]}
  \BibitemShut {NoStop}%
\bibitem [{\citenamefont {Kasliwal}\ \emph {et~al.}(2017)\citenamefont
  {Kasliwal} \emph {et~al.}}]{Kasliwal:2017ngb}%
  \BibitemOpen
  \bibfield  {author} {\bibinfo {author} {\bibfnamefont {M.}~\bibnamefont
  {Kasliwal}} \emph {et~al.},\ }\href {\doibase 10.1126/science.aap9455}
  {\bibfield  {journal} {\bibinfo  {journal} {Science}\ }\textbf {\bibinfo
  {volume} {358}},\ \bibinfo {pages} {1559} (\bibinfo {year} {2017})},\ \Eprint
  {http://arxiv.org/abs/1710.05436} {arXiv:1710.05436 [astro-ph.HE]}
  \BibitemShut {NoStop}%
\bibitem [{\citenamefont {Cowperthwaite}\ \emph {et~al.}(2017)\citenamefont
  {Cowperthwaite} \emph {et~al.}}]{Cowperthwaite:2017dyu}%
  \BibitemOpen
  \bibfield  {author} {\bibinfo {author} {\bibfnamefont {P.}~\bibnamefont
  {Cowperthwaite}} \emph {et~al.},\ }\href {\doibase 10.3847/2041-8213/aa8fc7}
  {\bibfield  {journal} {\bibinfo  {journal} {Astrophys. J. Lett.}\ }\textbf
  {\bibinfo {volume} {848}},\ \bibinfo {pages} {L17} (\bibinfo {year}
  {2017})},\ \Eprint {http://arxiv.org/abs/1710.05840} {arXiv:1710.05840
  [astro-ph.HE]} \BibitemShut {NoStop}%
\bibitem [{\citenamefont {Tanvir}\ \emph {et~al.}(2017)\citenamefont {Tanvir}
  \emph {et~al.}}]{Tanvir:2017pws}%
  \BibitemOpen
  \bibfield  {author} {\bibinfo {author} {\bibfnamefont {N.}~\bibnamefont
  {Tanvir}} \emph {et~al.},\ }\href {\doibase 10.3847/2041-8213/aa90b6}
  {\bibfield  {journal} {\bibinfo  {journal} {Astrophys. J. Lett.}\ }\textbf
  {\bibinfo {volume} {848}},\ \bibinfo {pages} {L27} (\bibinfo {year}
  {2017})},\ \Eprint {http://arxiv.org/abs/1710.05455} {arXiv:1710.05455
  [astro-ph.HE]} \BibitemShut {NoStop}%
\bibitem [{\citenamefont {Evans}\ \emph {et~al.}(2017)\citenamefont {Evans}
  \emph {et~al.}}]{Evans:2017mmy}%
  \BibitemOpen
  \bibfield  {author} {\bibinfo {author} {\bibfnamefont {P.}~\bibnamefont
  {Evans}} \emph {et~al.},\ }\href {\doibase 10.1126/science.aap9580}
  {\bibfield  {journal} {\bibinfo  {journal} {Science}\ }\textbf {\bibinfo
  {volume} {358}},\ \bibinfo {pages} {1565} (\bibinfo {year} {2017})},\ \Eprint
  {http://arxiv.org/abs/1710.05437} {arXiv:1710.05437 [astro-ph.HE]}
  \BibitemShut {NoStop}%
\bibitem [{\citenamefont {Arcavi}\ \emph {et~al.}(2017)\citenamefont {Arcavi}
  \emph {et~al.}}]{Arcavi:2017xiz}%
  \BibitemOpen
  \bibfield  {author} {\bibinfo {author} {\bibfnamefont {I.}~\bibnamefont
  {Arcavi}} \emph {et~al.},\ }\href {\doibase 10.1038/nature24291} {\bibfield
  {journal} {\bibinfo  {journal} {Nature}\ }\textbf {\bibinfo {volume} {551}},\
  \bibinfo {pages} {64} (\bibinfo {year} {2017})},\ \Eprint
  {http://arxiv.org/abs/1710.05843} {arXiv:1710.05843 [astro-ph.HE]}
  \BibitemShut {NoStop}%
\bibitem [{\citenamefont {Utsumi}\ \emph {et~al.}(2017)\citenamefont {Utsumi}
  \emph {et~al.}}]{Utsumi:2017cti}%
  \BibitemOpen
  \bibfield  {author} {\bibinfo {author} {\bibfnamefont {Y.}~\bibnamefont
  {Utsumi}} \emph {et~al.} (\bibinfo {collaboration} {J-GEM}),\ }\href
  {\doibase 10.1093/pasj/psx118} {\bibfield  {journal} {\bibinfo  {journal}
  {Publ. Astron. Soc. Jap.}\ }\textbf {\bibinfo {volume} {69}},\ \bibinfo
  {pages} {101} (\bibinfo {year} {2017})},\ \Eprint
  {http://arxiv.org/abs/1710.05848} {arXiv:1710.05848 [astro-ph.HE]}
  \BibitemShut {NoStop}%
\bibitem [{\citenamefont {Troja}\ \emph {et~al.}(2017)\citenamefont {Troja}
  \emph {et~al.}}]{Troja:2017nqp}%
  \BibitemOpen
  \bibfield  {author} {\bibinfo {author} {\bibfnamefont {E.}~\bibnamefont
  {Troja}} \emph {et~al.},\ }\href {\doibase 10.1038/nature24290} {\bibfield
  {journal} {\bibinfo  {journal} {Nature}\ }\textbf {\bibinfo {volume} {551}},\
  \bibinfo {pages} {71} (\bibinfo {year} {2017})},\ \Eprint
  {http://arxiv.org/abs/1710.05433} {arXiv:1710.05433 [astro-ph.HE]}
  \BibitemShut {NoStop}%
\bibitem [{\citenamefont {Hallinan}\ \emph {et~al.}(2017)\citenamefont
  {Hallinan} \emph {et~al.}}]{Hallinan:2017woc}%
  \BibitemOpen
  \bibfield  {author} {\bibinfo {author} {\bibfnamefont {G.}~\bibnamefont
  {Hallinan}} \emph {et~al.},\ }\href {\doibase 10.1126/science.aap9855}
  {\bibfield  {journal} {\bibinfo  {journal} {Science}\ }\textbf {\bibinfo
  {volume} {358}},\ \bibinfo {pages} {1579} (\bibinfo {year} {2017})},\ \Eprint
  {http://arxiv.org/abs/1710.05435} {arXiv:1710.05435 [astro-ph.HE]}
  \BibitemShut {NoStop}%
\bibitem [{\citenamefont {Langlois}\ \emph {et~al.}(2018)\citenamefont
  {Langlois}, \citenamefont {Saito}, \citenamefont {Yamauchi},\ and\
  \citenamefont {Noui}}]{Langlois:2017dyl}%
  \BibitemOpen
  \bibfield  {author} {\bibinfo {author} {\bibfnamefont {D.}~\bibnamefont
  {Langlois}}, \bibinfo {author} {\bibfnamefont {R.}~\bibnamefont {Saito}},
  \bibinfo {author} {\bibfnamefont {D.}~\bibnamefont {Yamauchi}}, \ and\
  \bibinfo {author} {\bibfnamefont {K.}~\bibnamefont {Noui}},\ }\href {\doibase
  10.1103/PhysRevD.97.061501} {\bibfield  {journal} {\bibinfo  {journal} {Phys.
  Rev. D}\ }\textbf {\bibinfo {volume} {97}},\ \bibinfo {pages} {061501}
  (\bibinfo {year} {2018})},\ \Eprint {http://arxiv.org/abs/1711.07403}
  {arXiv:1711.07403 [gr-qc]} \BibitemShut {NoStop}%
\bibitem [{\citenamefont {Creminelli}\ and\ \citenamefont
  {Vernizzi}(2017)}]{Creminelli:2017sry}%
  \BibitemOpen
  \bibfield  {author} {\bibinfo {author} {\bibfnamefont {P.}~\bibnamefont
  {Creminelli}}\ and\ \bibinfo {author} {\bibfnamefont {F.}~\bibnamefont
  {Vernizzi}},\ }\href {\doibase 10.1103/PhysRevLett.119.251302} {\bibfield
  {journal} {\bibinfo  {journal} {Phys. Rev. Lett.}\ }\textbf {\bibinfo
  {volume} {119}},\ \bibinfo {pages} {251302} (\bibinfo {year} {2017})},\
  \Eprint {http://arxiv.org/abs/1710.05877} {arXiv:1710.05877 [astro-ph.CO]}
  \BibitemShut {NoStop}%
\bibitem [{\citenamefont {Ezquiaga}\ and\ \citenamefont
  {Zumalacárregui}(2017)}]{Ezquiaga:2017ekz}%
  \BibitemOpen
  \bibfield  {author} {\bibinfo {author} {\bibfnamefont {J.~M.}\ \bibnamefont
  {Ezquiaga}}\ and\ \bibinfo {author} {\bibfnamefont {M.}~\bibnamefont
  {Zumalacárregui}},\ }\href {\doibase 10.1103/PhysRevLett.119.251304}
  {\bibfield  {journal} {\bibinfo  {journal} {Phys. Rev. Lett.}\ }\textbf
  {\bibinfo {volume} {119}},\ \bibinfo {pages} {251304} (\bibinfo {year}
  {2017})},\ \Eprint {http://arxiv.org/abs/1710.05901} {arXiv:1710.05901
  [astro-ph.CO]} \BibitemShut {NoStop}%
\bibitem [{\citenamefont {Baker}\ \emph {et~al.}(2017)\citenamefont {Baker},
  \citenamefont {Bellini}, \citenamefont {Ferreira}, \citenamefont {Lagos},
  \citenamefont {Noller},\ and\ \citenamefont {Sawicki}}]{Baker:2017hug}%
  \BibitemOpen
  \bibfield  {author} {\bibinfo {author} {\bibfnamefont {T.}~\bibnamefont
  {Baker}}, \bibinfo {author} {\bibfnamefont {E.}~\bibnamefont {Bellini}},
  \bibinfo {author} {\bibfnamefont {P.}~\bibnamefont {Ferreira}}, \bibinfo
  {author} {\bibfnamefont {M.}~\bibnamefont {Lagos}}, \bibinfo {author}
  {\bibfnamefont {J.}~\bibnamefont {Noller}}, \ and\ \bibinfo {author}
  {\bibfnamefont {I.}~\bibnamefont {Sawicki}},\ }\href {\doibase
  10.1103/PhysRevLett.119.251301} {\bibfield  {journal} {\bibinfo  {journal}
  {Phys. Rev. Lett.}\ }\textbf {\bibinfo {volume} {119}},\ \bibinfo {pages}
  {251301} (\bibinfo {year} {2017})},\ \Eprint
  {http://arxiv.org/abs/1710.06394} {arXiv:1710.06394 [astro-ph.CO]}
  \BibitemShut {NoStop}%
\bibitem [{\citenamefont {Abbott}\ \emph
  {et~al.}(2017{\natexlab{d}})\citenamefont {Abbott} \emph
  {et~al.}}]{Abbott:2017xzu}%
  \BibitemOpen
  \bibfield  {author} {\bibinfo {author} {\bibfnamefont {B.}~\bibnamefont
  {Abbott}} \emph {et~al.} (\bibinfo {collaboration} {LIGO Scientific, Virgo,
  1M2H, Dark Energy Camera GW-E, DES, DLT40, Las Cumbres Observatory, VINROUGE,
  MASTER}),\ }\href {\doibase 10.1038/nature24471} {\bibfield  {journal}
  {\bibinfo  {journal} {Nature}\ }\textbf {\bibinfo {volume} {551}},\ \bibinfo
  {pages} {85} (\bibinfo {year} {2017}{\natexlab{d}})},\ \Eprint
  {http://arxiv.org/abs/1710.05835} {arXiv:1710.05835 [astro-ph.CO]}
  \BibitemShut {NoStop}%
\bibitem [{\citenamefont {Tanaka}\ \emph {et~al.}(2017)\citenamefont {Tanaka}
  \emph {et~al.}}]{Tanaka:2017qxj}%
  \BibitemOpen
  \bibfield  {author} {\bibinfo {author} {\bibfnamefont {M.}~\bibnamefont
  {Tanaka}} \emph {et~al.},\ }\href {\doibase 10.1093/pasj/psx121} {\bibfield
  {journal} {\bibinfo  {journal} {Publ. Astron. Soc. Jap.}\ }\textbf {\bibinfo
  {volume} {69}},\ \bibinfo {pages} {psx12} (\bibinfo {year} {2017})},\ \Eprint
  {http://arxiv.org/abs/1710.05850} {arXiv:1710.05850 [astro-ph.HE]}
  \BibitemShut {NoStop}%
\bibitem [{\citenamefont {Alexander}\ \emph {et~al.}(2017)\citenamefont
  {Alexander} \emph {et~al.}}]{Alexander:2017aly}%
  \BibitemOpen
  \bibfield  {author} {\bibinfo {author} {\bibfnamefont {K.}~\bibnamefont
  {Alexander}} \emph {et~al.},\ }\href {\doibase 10.3847/2041-8213/aa905d}
  {\bibfield  {journal} {\bibinfo  {journal} {Astrophys. J. Lett.}\ }\textbf
  {\bibinfo {volume} {848}},\ \bibinfo {pages} {L21} (\bibinfo {year}
  {2017})},\ \Eprint {http://arxiv.org/abs/1710.05457} {arXiv:1710.05457
  [astro-ph.HE]} \BibitemShut {NoStop}%
\bibitem [{\citenamefont {Margutti}\ \emph {et~al.}(2017)\citenamefont
  {Margutti} \emph {et~al.}}]{Margutti:2017cjl}%
  \BibitemOpen
  \bibfield  {author} {\bibinfo {author} {\bibfnamefont {R.}~\bibnamefont
  {Margutti}} \emph {et~al.},\ }\href {\doibase 10.3847/2041-8213/aa9057}
  {\bibfield  {journal} {\bibinfo  {journal} {Astrophys. J. Lett.}\ }\textbf
  {\bibinfo {volume} {848}},\ \bibinfo {pages} {L20} (\bibinfo {year}
  {2017})},\ \Eprint {http://arxiv.org/abs/1710.05431} {arXiv:1710.05431
  [astro-ph.HE]} \BibitemShut {NoStop}%
\bibitem [{\citenamefont {Mooley}\ \emph {et~al.}(2018)\citenamefont {Mooley}
  \emph {et~al.}}]{Mooley:2017enz}%
  \BibitemOpen
  \bibfield  {author} {\bibinfo {author} {\bibfnamefont {K.}~\bibnamefont
  {Mooley}} \emph {et~al.},\ }\href {\doibase 10.1038/nature25452} {\bibfield
  {journal} {\bibinfo  {journal} {Nature}\ }\textbf {\bibinfo {volume} {554}},\
  \bibinfo {pages} {207} (\bibinfo {year} {2018})},\ \Eprint
  {http://arxiv.org/abs/1711.11573} {arXiv:1711.11573 [astro-ph.HE]}
  \BibitemShut {NoStop}%
\bibitem [{\citenamefont {Fairhurst}(2009)}]{Fairhurst:2009tc}%
  \BibitemOpen
  \bibfield  {author} {\bibinfo {author} {\bibfnamefont {S.}~\bibnamefont
  {Fairhurst}},\ }\href {\doibase 10.1088/1367-2630/11/12/123006} {\bibfield
  {journal} {\bibinfo  {journal} {New J. Phys.}\ }\textbf {\bibinfo {volume}
  {11}},\ \bibinfo {pages} {123006} (\bibinfo {year} {2009})},\ \bibinfo {note}
  {[Erratum: New J.Phys. 13, 069602 (2011)]},\ \Eprint
  {http://arxiv.org/abs/0908.2356} {arXiv:0908.2356 [gr-qc]} \BibitemShut
  {NoStop}%
\bibitem [{\citenamefont {Singer}\ and\ \citenamefont
  {Price}(2016)}]{Singer:2015ema}%
  \BibitemOpen
  \bibfield  {author} {\bibinfo {author} {\bibfnamefont {L.~P.}\ \bibnamefont
  {Singer}}\ and\ \bibinfo {author} {\bibfnamefont {L.~R.}\ \bibnamefont
  {Price}},\ }\href {\doibase 10.1103/PhysRevD.93.024013} {\bibfield  {journal}
  {\bibinfo  {journal} {Phys. Rev.}\ }\textbf {\bibinfo {volume} {D93}},\
  \bibinfo {pages} {024013} (\bibinfo {year} {2016})},\ \Eprint
  {http://arxiv.org/abs/1508.03634} {arXiv:1508.03634 [gr-qc]} \BibitemShut
  {NoStop}%
\bibitem [{\citenamefont {Singer}\ \emph {et~al.}(2016)\citenamefont {Singer}
  \emph {et~al.}}]{Singer:2016eax}%
  \BibitemOpen
  \bibfield  {author} {\bibinfo {author} {\bibfnamefont {L.~P.}\ \bibnamefont
  {Singer}} \emph {et~al.},\ }\href {\doibase 10.3847/2041-8205/829/1/L15}
  {\bibfield  {journal} {\bibinfo  {journal} {Astrophys. J. Lett.}\ }\textbf
  {\bibinfo {volume} {829}},\ \bibinfo {pages} {L15} (\bibinfo {year}
  {2016})},\ \Eprint {http://arxiv.org/abs/1603.07333} {arXiv:1603.07333
  [astro-ph.HE]} \BibitemShut {NoStop}%
\bibitem [{\citenamefont {Kapadia}\ \emph {et~al.}(2020)\citenamefont {Kapadia}
  \emph {et~al.}}]{Kapadia:2019uut}%
  \BibitemOpen
  \bibfield  {author} {\bibinfo {author} {\bibfnamefont {S.~J.}\ \bibnamefont
  {Kapadia}} \emph {et~al.},\ }\href {\doibase 10.1088/1361-6382/ab5f2d}
  {\bibfield  {journal} {\bibinfo  {journal} {Class. Quant. Grav.}\ }\textbf
  {\bibinfo {volume} {37}},\ \bibinfo {pages} {045007} (\bibinfo {year}
  {2020})},\ \Eprint {http://arxiv.org/abs/1903.06881} {arXiv:1903.06881
  [astro-ph.HE]} \BibitemShut {NoStop}%
\bibitem [{\citenamefont {Chatterjee}\ \emph {et~al.}(2019)\citenamefont
  {Chatterjee}, \citenamefont {Ghosh}, \citenamefont {Brady}, \citenamefont
  {Kapadia}, \citenamefont {Miller}, \citenamefont {Nissanke},\ and\
  \citenamefont {Pannarale}}]{Chatterjee:2019avs}%
  \BibitemOpen
  \bibfield  {author} {\bibinfo {author} {\bibfnamefont {D.}~\bibnamefont
  {Chatterjee}}, \bibinfo {author} {\bibfnamefont {S.}~\bibnamefont {Ghosh}},
  \bibinfo {author} {\bibfnamefont {P.~R.}\ \bibnamefont {Brady}}, \bibinfo
  {author} {\bibfnamefont {S.~J.}\ \bibnamefont {Kapadia}}, \bibinfo {author}
  {\bibfnamefont {A.~L.}\ \bibnamefont {Miller}}, \bibinfo {author}
  {\bibfnamefont {S.}~\bibnamefont {Nissanke}}, \ and\ \bibinfo {author}
  {\bibfnamefont {F.}~\bibnamefont {Pannarale}},\ }\href@noop {} {\  (\bibinfo
  {year} {2019})},\ \Eprint {http://arxiv.org/abs/1911.00116} {arXiv:1911.00116
  [astro-ph.IM]} \BibitemShut {NoStop}%
\bibitem [{\citenamefont {Veitch}\ \emph {et~al.}(2015)\citenamefont {Veitch}
  \emph {et~al.}}]{Veitch:2014wba}%
  \BibitemOpen
  \bibfield  {author} {\bibinfo {author} {\bibfnamefont {J.}~\bibnamefont
  {Veitch}} \emph {et~al.},\ }\href {\doibase 10.1103/PhysRevD.91.042003}
  {\bibfield  {journal} {\bibinfo  {journal} {Phys. Rev.}\ }\textbf {\bibinfo
  {volume} {D91}},\ \bibinfo {pages} {042003} (\bibinfo {year} {2015})},\
  \Eprint {http://arxiv.org/abs/1409.7215} {arXiv:1409.7215 [gr-qc]}
  \BibitemShut {NoStop}%
\bibitem [{\citenamefont {Ashton}\ \emph {et~al.}(2019)\citenamefont {Ashton}
  \emph {et~al.}}]{Ashton:2018jfp}%
  \BibitemOpen
  \bibfield  {author} {\bibinfo {author} {\bibfnamefont {G.}~\bibnamefont
  {Ashton}} \emph {et~al.},\ }\href {\doibase 10.3847/1538-4365/ab06fc}
  {\bibfield  {journal} {\bibinfo  {journal} {Astrophys. J. Suppl.}\ }\textbf
  {\bibinfo {volume} {241}},\ \bibinfo {pages} {27} (\bibinfo {year} {2019})},\
  \Eprint {http://arxiv.org/abs/1811.02042} {arXiv:1811.02042 [astro-ph.IM]}
  \BibitemShut {NoStop}%
\bibitem [{\citenamefont {Metropolis}\ \emph {et~al.}(1953)\citenamefont
  {Metropolis}, \citenamefont {Rosenbluth}, \citenamefont {Rosenbluth},
  \citenamefont {Teller},\ and\ \citenamefont {Teller}}]{Metropolis:1953am}%
  \BibitemOpen
  \bibfield  {author} {\bibinfo {author} {\bibfnamefont {N.}~\bibnamefont
  {Metropolis}}, \bibinfo {author} {\bibfnamefont {A.}~\bibnamefont
  {Rosenbluth}}, \bibinfo {author} {\bibfnamefont {M.}~\bibnamefont
  {Rosenbluth}}, \bibinfo {author} {\bibfnamefont {A.}~\bibnamefont {Teller}},
  \ and\ \bibinfo {author} {\bibfnamefont {E.}~\bibnamefont {Teller}},\ }\href
  {\doibase 10.1063/1.1699114} {\bibfield  {journal} {\bibinfo  {journal} {J.
  Chem. Phys.}\ }\textbf {\bibinfo {volume} {21}},\ \bibinfo {pages} {1087}
  (\bibinfo {year} {1953})}\BibitemShut {NoStop}%
\bibitem [{\citenamefont {Hastings}(1970)}]{10.1093/biomet/57.1.97}%
  \BibitemOpen
  \bibfield  {author} {\bibinfo {author} {\bibfnamefont {W.~K.}\ \bibnamefont
  {Hastings}},\ }\href {\doibase 10.1093/biomet/57.1.97} {\bibfield  {journal}
  {\bibinfo  {journal} {Biometrika}\ }\textbf {\bibinfo {volume} {57}},\
  \bibinfo {pages} {97} (\bibinfo {year} {1970})},\ \Eprint
  {http://arxiv.org/abs/https://academic.oup.com/biomet/article-pdf/57/1/97/23940249/57-1-97.pdf}
  {https://academic.oup.com/biomet/article-pdf/57/1/97/23940249/57-1-97.pdf}
  \BibitemShut {NoStop}%
\bibitem [{\citenamefont {Skilling}(2006)}]{skilling2006}%
  \BibitemOpen
  \bibfield  {author} {\bibinfo {author} {\bibfnamefont {J.}~\bibnamefont
  {Skilling}},\ }\href {\doibase 10.1214/06-BA127} {\bibfield  {journal}
  {\bibinfo  {journal} {Bayesian Anal.}\ }\textbf {\bibinfo {volume} {1}},\
  \bibinfo {pages} {833} (\bibinfo {year} {2006})}\BibitemShut {NoStop}%
\bibitem [{\citenamefont {Canizares}\ \emph {et~al.}(2015)\citenamefont
  {Canizares}, \citenamefont {Field}, \citenamefont {Gair}, \citenamefont
  {Raymond}, \citenamefont {Smith},\ and\ \citenamefont
  {Tiglio}}]{Canizares:2014fya}%
  \BibitemOpen
  \bibfield  {author} {\bibinfo {author} {\bibfnamefont {P.}~\bibnamefont
  {Canizares}}, \bibinfo {author} {\bibfnamefont {S.~E.}\ \bibnamefont
  {Field}}, \bibinfo {author} {\bibfnamefont {J.}~\bibnamefont {Gair}},
  \bibinfo {author} {\bibfnamefont {V.}~\bibnamefont {Raymond}}, \bibinfo
  {author} {\bibfnamefont {R.}~\bibnamefont {Smith}}, \ and\ \bibinfo {author}
  {\bibfnamefont {M.}~\bibnamefont {Tiglio}},\ }\href {\doibase
  10.1103/PhysRevLett.114.071104} {\bibfield  {journal} {\bibinfo  {journal}
  {Phys. Rev. Lett.}\ }\textbf {\bibinfo {volume} {114}},\ \bibinfo {pages}
  {071104} (\bibinfo {year} {2015})},\ \Eprint {http://arxiv.org/abs/1404.6284}
  {arXiv:1404.6284 [gr-qc]} \BibitemShut {NoStop}%
\bibitem [{\citenamefont {Smith}\ \emph {et~al.}(2016)\citenamefont {Smith},
  \citenamefont {Field}, \citenamefont {Blackburn}, \citenamefont {Haster},
  \citenamefont {Pürrer}, \citenamefont {Raymond},\ and\ \citenamefont
  {Schmidt}}]{Smith:2016qas}%
  \BibitemOpen
  \bibfield  {author} {\bibinfo {author} {\bibfnamefont {R.}~\bibnamefont
  {Smith}}, \bibinfo {author} {\bibfnamefont {S.~E.}\ \bibnamefont {Field}},
  \bibinfo {author} {\bibfnamefont {K.}~\bibnamefont {Blackburn}}, \bibinfo
  {author} {\bibfnamefont {C.-J.}\ \bibnamefont {Haster}}, \bibinfo {author}
  {\bibfnamefont {M.}~\bibnamefont {Pürrer}}, \bibinfo {author} {\bibfnamefont
  {V.}~\bibnamefont {Raymond}}, \ and\ \bibinfo {author} {\bibfnamefont
  {P.}~\bibnamefont {Schmidt}},\ }\href {\doibase 10.1103/PhysRevD.94.044031}
  {\bibfield  {journal} {\bibinfo  {journal} {Phys. Rev.}\ }\textbf {\bibinfo
  {volume} {D94}},\ \bibinfo {pages} {044031} (\bibinfo {year} {2016})},\
  \Eprint {http://arxiv.org/abs/1604.08253} {arXiv:1604.08253 [gr-qc]}
  \BibitemShut {NoStop}%
\bibitem [{\citenamefont {Pankow}\ \emph {et~al.}(2015)\citenamefont {Pankow},
  \citenamefont {Brady}, \citenamefont {Ochsner},\ and\ \citenamefont
  {O'Shaughnessy}}]{Pankow:2015cra}%
  \BibitemOpen
  \bibfield  {author} {\bibinfo {author} {\bibfnamefont {C.}~\bibnamefont
  {Pankow}}, \bibinfo {author} {\bibfnamefont {P.}~\bibnamefont {Brady}},
  \bibinfo {author} {\bibfnamefont {E.}~\bibnamefont {Ochsner}}, \ and\
  \bibinfo {author} {\bibfnamefont {R.}~\bibnamefont {O'Shaughnessy}},\ }\href
  {\doibase 10.1103/PhysRevD.92.023002} {\bibfield  {journal} {\bibinfo
  {journal} {Phys. Rev. D}\ }\textbf {\bibinfo {volume} {92}},\ \bibinfo
  {pages} {023002} (\bibinfo {year} {2015})},\ \Eprint
  {http://arxiv.org/abs/1502.04370} {arXiv:1502.04370 [gr-qc]} \BibitemShut
  {NoStop}%
\bibitem [{\citenamefont {Lange}\ \emph {et~al.}(2018)\citenamefont {Lange},
  \citenamefont {O'Shaughnessy},\ and\ \citenamefont {Rizzo}}]{Lange:2018pyp}%
  \BibitemOpen
  \bibfield  {author} {\bibinfo {author} {\bibfnamefont {J.}~\bibnamefont
  {Lange}}, \bibinfo {author} {\bibfnamefont {R.}~\bibnamefont
  {O'Shaughnessy}}, \ and\ \bibinfo {author} {\bibfnamefont {M.}~\bibnamefont
  {Rizzo}},\ }\href@noop {} {\  (\bibinfo {year} {2018})},\ \Eprint
  {http://arxiv.org/abs/1805.10457} {arXiv:1805.10457 [gr-qc]} \BibitemShut
  {NoStop}%
\bibitem [{\citenamefont {Wysocki}\ \emph {et~al.}(2019)\citenamefont
  {Wysocki}, \citenamefont {O'Shaughnessy}, \citenamefont {Lange},\ and\
  \citenamefont {Fang}}]{Wysocki:2019grj}%
  \BibitemOpen
  \bibfield  {author} {\bibinfo {author} {\bibfnamefont {D.}~\bibnamefont
  {Wysocki}}, \bibinfo {author} {\bibfnamefont {R.}~\bibnamefont
  {O'Shaughnessy}}, \bibinfo {author} {\bibfnamefont {J.}~\bibnamefont
  {Lange}}, \ and\ \bibinfo {author} {\bibfnamefont {Y.-L.~L.}\ \bibnamefont
  {Fang}},\ }\href {\doibase 10.1103/PhysRevD.99.084026} {\bibfield  {journal}
  {\bibinfo  {journal} {Phys. Rev. D}\ }\textbf {\bibinfo {volume} {99}},\
  \bibinfo {pages} {084026} (\bibinfo {year} {2019})},\ \Eprint
  {http://arxiv.org/abs/1902.04934} {arXiv:1902.04934 [astro-ph.IM]}
  \BibitemShut {NoStop}%
\bibitem [{\citenamefont {Vinciguerra}\ \emph {et~al.}(2017)\citenamefont
  {Vinciguerra}, \citenamefont {Veitch},\ and\ \citenamefont
  {Mandel}}]{Vinciguerra:2017ngf}%
  \BibitemOpen
  \bibfield  {author} {\bibinfo {author} {\bibfnamefont {S.}~\bibnamefont
  {Vinciguerra}}, \bibinfo {author} {\bibfnamefont {J.}~\bibnamefont {Veitch}},
  \ and\ \bibinfo {author} {\bibfnamefont {I.}~\bibnamefont {Mandel}},\ }\href
  {\doibase 10.1088/1361-6382/aa6d44} {\bibfield  {journal} {\bibinfo
  {journal} {Class. Quant. Grav.}\ }\textbf {\bibinfo {volume} {34}},\ \bibinfo
  {pages} {115006} (\bibinfo {year} {2017})},\ \Eprint
  {http://arxiv.org/abs/1703.02062} {arXiv:1703.02062 [gr-qc]} \BibitemShut
  {NoStop}%
\bibitem [{\citenamefont {Zackay}\ \emph {et~al.}(2018)\citenamefont {Zackay},
  \citenamefont {Dai},\ and\ \citenamefont {Venumadhav}}]{Zackay:2018qdy}%
  \BibitemOpen
  \bibfield  {author} {\bibinfo {author} {\bibfnamefont {B.}~\bibnamefont
  {Zackay}}, \bibinfo {author} {\bibfnamefont {L.}~\bibnamefont {Dai}}, \ and\
  \bibinfo {author} {\bibfnamefont {T.}~\bibnamefont {Venumadhav}},\
  }\href@noop {} {\  (\bibinfo {year} {2018})},\ \Eprint
  {http://arxiv.org/abs/1806.08792} {arXiv:1806.08792 [astro-ph.IM]}
  \BibitemShut {NoStop}%
\bibitem [{\citenamefont {Talbot}\ \emph {et~al.}(2019)\citenamefont {Talbot},
  \citenamefont {Smith}, \citenamefont {Thrane},\ and\ \citenamefont
  {Poole}}]{Talbot:2019okv}%
  \BibitemOpen
  \bibfield  {author} {\bibinfo {author} {\bibfnamefont {C.}~\bibnamefont
  {Talbot}}, \bibinfo {author} {\bibfnamefont {R.}~\bibnamefont {Smith}},
  \bibinfo {author} {\bibfnamefont {E.}~\bibnamefont {Thrane}}, \ and\ \bibinfo
  {author} {\bibfnamefont {G.~B.}\ \bibnamefont {Poole}},\ }\href {\doibase
  10.1103/PhysRevD.100.043030} {\bibfield  {journal} {\bibinfo  {journal}
  {Phys. Rev. D}\ }\textbf {\bibinfo {volume} {100}},\ \bibinfo {pages}
  {043030} (\bibinfo {year} {2019})},\ \Eprint
  {http://arxiv.org/abs/1904.02863} {arXiv:1904.02863 [astro-ph.IM]}
  \BibitemShut {NoStop}%
\bibitem [{\citenamefont {Smith}\ \emph {et~al.}(2019)\citenamefont {Smith},
  \citenamefont {Ashton}, \citenamefont {Vajpeyi},\ and\ \citenamefont
  {Talbot}}]{Smith:2019ucc}%
  \BibitemOpen
  \bibfield  {author} {\bibinfo {author} {\bibfnamefont {R.}~\bibnamefont
  {Smith}}, \bibinfo {author} {\bibfnamefont {G.}~\bibnamefont {Ashton}},
  \bibinfo {author} {\bibfnamefont {A.}~\bibnamefont {Vajpeyi}}, \ and\
  \bibinfo {author} {\bibfnamefont {C.}~\bibnamefont {Talbot}},\ }\href@noop {}
  {\  (\bibinfo {year} {2019})},\ \Eprint {http://arxiv.org/abs/1909.11873}
  {arXiv:1909.11873 [gr-qc]} \BibitemShut {NoStop}%
\bibitem [{\citenamefont {Abbott}\ \emph
  {et~al.}(2019{\natexlab{a}})\citenamefont {Abbott} \emph
  {et~al.}}]{LIGOScientific:2018mvr}%
  \BibitemOpen
  \bibfield  {author} {\bibinfo {author} {\bibfnamefont {B.}~\bibnamefont
  {Abbott}} \emph {et~al.} (\bibinfo {collaboration} {LIGO Scientific,
  Virgo}),\ }\href {\doibase 10.1103/PhysRevX.9.031040} {\bibfield  {journal}
  {\bibinfo  {journal} {Phys. Rev. X}\ }\textbf {\bibinfo {volume} {9}},\
  \bibinfo {pages} {031040} (\bibinfo {year} {2019}{\natexlab{a}})},\ \Eprint
  {http://arxiv.org/abs/1811.12907} {arXiv:1811.12907 [astro-ph.HE]}
  \BibitemShut {NoStop}%
\bibitem [{\citenamefont {Abbott}\ \emph {et~al.}(2020)\citenamefont {Abbott}
  \emph {et~al.}}]{Abbott:2020uma}%
  \BibitemOpen
  \bibfield  {author} {\bibinfo {author} {\bibfnamefont {B.~P.}\ \bibnamefont
  {Abbott}} \emph {et~al.} (\bibinfo {collaboration} {LIGO Scientific,
  Virgo}),\ }\href {\doibase 10.3847/2041-8213/ab75f5} {\bibfield  {journal}
  {\bibinfo  {journal} {Astrophys. J. Lett.}\ }\textbf {\bibinfo {volume}
  {892}},\ \bibinfo {pages} {L3} (\bibinfo {year} {2020})},\ \Eprint
  {http://arxiv.org/abs/2001.01761} {arXiv:2001.01761 [astro-ph.HE]}
  \BibitemShut {NoStop}%
\bibitem [{\citenamefont {Buonanno}\ \emph {et~al.}(2009)\citenamefont
  {Buonanno}, \citenamefont {Iyer}, \citenamefont {Ochsner}, \citenamefont
  {Pan},\ and\ \citenamefont {Sathyaprakash}}]{Buonanno:2009zt}%
  \BibitemOpen
  \bibfield  {author} {\bibinfo {author} {\bibfnamefont {A.}~\bibnamefont
  {Buonanno}}, \bibinfo {author} {\bibfnamefont {B.}~\bibnamefont {Iyer}},
  \bibinfo {author} {\bibfnamefont {E.}~\bibnamefont {Ochsner}}, \bibinfo
  {author} {\bibfnamefont {Y.}~\bibnamefont {Pan}}, \ and\ \bibinfo {author}
  {\bibfnamefont {B.}~\bibnamefont {Sathyaprakash}},\ }\href {\doibase
  10.1103/PhysRevD.80.084043} {\bibfield  {journal} {\bibinfo  {journal} {Phys.
  Rev. D}\ }\textbf {\bibinfo {volume} {80}},\ \bibinfo {pages} {084043}
  (\bibinfo {year} {2009})},\ \Eprint {http://arxiv.org/abs/0907.0700}
  {arXiv:0907.0700 [gr-qc]} \BibitemShut {NoStop}%
\bibitem [{\citenamefont {Hannam}\ \emph {et~al.}(2014)\citenamefont {Hannam},
  \citenamefont {Schmidt}, \citenamefont {Bohé}, \citenamefont {Haegel},
  \citenamefont {Husa}, \citenamefont {Ohme}, \citenamefont {Pratten},\ and\
  \citenamefont {Pürrer}}]{Hannam:2013oca}%
  \BibitemOpen
  \bibfield  {author} {\bibinfo {author} {\bibfnamefont {M.}~\bibnamefont
  {Hannam}}, \bibinfo {author} {\bibfnamefont {P.}~\bibnamefont {Schmidt}},
  \bibinfo {author} {\bibfnamefont {A.}~\bibnamefont {Bohé}}, \bibinfo
  {author} {\bibfnamefont {L.}~\bibnamefont {Haegel}}, \bibinfo {author}
  {\bibfnamefont {S.}~\bibnamefont {Husa}}, \bibinfo {author} {\bibfnamefont
  {F.}~\bibnamefont {Ohme}}, \bibinfo {author} {\bibfnamefont {G.}~\bibnamefont
  {Pratten}}, \ and\ \bibinfo {author} {\bibfnamefont {M.}~\bibnamefont
  {Pürrer}},\ }\href {\doibase 10.1103/PhysRevLett.113.151101} {\bibfield
  {journal} {\bibinfo  {journal} {Phys. Rev. Lett.}\ }\textbf {\bibinfo
  {volume} {113}},\ \bibinfo {pages} {151101} (\bibinfo {year} {2014})},\
  \Eprint {http://arxiv.org/abs/1308.3271} {arXiv:1308.3271 [gr-qc]}
  \BibitemShut {NoStop}%
\bibitem [{\citenamefont {Thorne}(1980)}]{Thorne:1980ru}%
  \BibitemOpen
  \bibfield  {author} {\bibinfo {author} {\bibfnamefont {K.}~\bibnamefont
  {Thorne}},\ }\href {\doibase 10.1103/RevModPhys.52.299} {\bibfield  {journal}
  {\bibinfo  {journal} {Rev. Mod. Phys.}\ }\textbf {\bibinfo {volume} {52}},\
  \bibinfo {pages} {299} (\bibinfo {year} {1980})}\BibitemShut {NoStop}%
\bibitem [{\citenamefont {Dietrich}\ \emph {et~al.}(2019)\citenamefont
  {Dietrich} \emph {et~al.}}]{Dietrich:2018uni}%
  \BibitemOpen
  \bibfield  {author} {\bibinfo {author} {\bibfnamefont {T.}~\bibnamefont
  {Dietrich}} \emph {et~al.},\ }\href {\doibase 10.1103/PhysRevD.99.024029}
  {\bibfield  {journal} {\bibinfo  {journal} {Phys. Rev. D}\ }\textbf {\bibinfo
  {volume} {99}},\ \bibinfo {pages} {024029} (\bibinfo {year} {2019})},\
  \Eprint {http://arxiv.org/abs/1804.02235} {arXiv:1804.02235 [gr-qc]}
  \BibitemShut {NoStop}%
\bibitem [{\citenamefont {Khan}\ \emph {et~al.}(2016)\citenamefont {Khan},
  \citenamefont {Husa}, \citenamefont {Hannam}, \citenamefont {Ohme},
  \citenamefont {Pürrer}, \citenamefont {Jiménez~Forteza},\ and\
  \citenamefont {Bohé}}]{Khan:2015jqa}%
  \BibitemOpen
  \bibfield  {author} {\bibinfo {author} {\bibfnamefont {S.}~\bibnamefont
  {Khan}}, \bibinfo {author} {\bibfnamefont {S.}~\bibnamefont {Husa}}, \bibinfo
  {author} {\bibfnamefont {M.}~\bibnamefont {Hannam}}, \bibinfo {author}
  {\bibfnamefont {F.}~\bibnamefont {Ohme}}, \bibinfo {author} {\bibfnamefont
  {M.}~\bibnamefont {Pürrer}}, \bibinfo {author} {\bibfnamefont
  {X.}~\bibnamefont {Jiménez~Forteza}}, \ and\ \bibinfo {author}
  {\bibfnamefont {A.}~\bibnamefont {Bohé}},\ }\href {\doibase
  10.1103/PhysRevD.93.044007} {\bibfield  {journal} {\bibinfo  {journal} {Phys.
  Rev. D}\ }\textbf {\bibinfo {volume} {93}},\ \bibinfo {pages} {044007}
  (\bibinfo {year} {2016})},\ \Eprint {http://arxiv.org/abs/1508.07253}
  {arXiv:1508.07253 [gr-qc]} \BibitemShut {NoStop}%
\bibitem [{\citenamefont {Dietrich}\ \emph {et~al.}(2017)\citenamefont
  {Dietrich}, \citenamefont {Bernuzzi},\ and\ \citenamefont
  {Tichy}}]{Dietrich:2017aum}%
  \BibitemOpen
  \bibfield  {author} {\bibinfo {author} {\bibfnamefont {T.}~\bibnamefont
  {Dietrich}}, \bibinfo {author} {\bibfnamefont {S.}~\bibnamefont {Bernuzzi}},
  \ and\ \bibinfo {author} {\bibfnamefont {W.}~\bibnamefont {Tichy}},\ }\href
  {\doibase 10.1103/PhysRevD.96.121501} {\bibfield  {journal} {\bibinfo
  {journal} {Phys. Rev. D}\ }\textbf {\bibinfo {volume} {96}},\ \bibinfo
  {pages} {121501} (\bibinfo {year} {2017})},\ \Eprint
  {http://arxiv.org/abs/1706.02969} {arXiv:1706.02969 [gr-qc]} \BibitemShut
  {NoStop}%
\bibitem [{\citenamefont {Smith}(2019)}]{pDNRT_roqdata}%
  \BibitemOpen
  \bibfield  {author} {\bibinfo {author} {\bibfnamefont {R.}~\bibnamefont
  {Smith}},\ }\href {\doibase 10.5281/zenodo.3255081} {\enquote {\bibinfo
  {title} {Imrphenomd-nrtidal-128s},}\ } (\bibinfo {year} {2019})\BibitemShut
  {NoStop}%
\bibitem [{\citenamefont {{Baylor}}\ \emph {et~al.}(2019)\citenamefont
  {{Baylor}}, \citenamefont {{Chase}},\ and\ \citenamefont
  {{Smith}}}]{pv2NRT_roqdata}%
  \BibitemOpen
  \bibfield  {author} {\bibinfo {author} {\bibfnamefont {A.}~\bibnamefont
  {{Baylor}}}, \bibinfo {author} {\bibfnamefont {E.}~\bibnamefont {{Chase}}}, \
  and\ \bibinfo {author} {\bibfnamefont {R.}~\bibnamefont {{Smith}}},\ }\href
  {\doibase 10.5281/zenodo.3478659} {\enquote {\bibinfo {title}
  {{IMRPhenomPv2NRT ROQ data set for GW190425}},}\ } (\bibinfo {year}
  {2019})\BibitemShut {NoStop}%
\bibitem [{\citenamefont {Ishii}\ \emph {et~al.}(2018)\citenamefont {Ishii},
  \citenamefont {Shigeyama},\ and\ \citenamefont {Tanaka}}]{Ishii:2018yjg}%
  \BibitemOpen
  \bibfield  {author} {\bibinfo {author} {\bibfnamefont {A.}~\bibnamefont
  {Ishii}}, \bibinfo {author} {\bibfnamefont {T.}~\bibnamefont {Shigeyama}}, \
  and\ \bibinfo {author} {\bibfnamefont {M.}~\bibnamefont {Tanaka}},\ }\href
  {\doibase 10.3847/1538-4357/aac385} {\bibfield  {journal} {\bibinfo
  {journal} {Astrophys. J.}\ }\textbf {\bibinfo {volume} {861}},\ \bibinfo
  {pages} {25} (\bibinfo {year} {2018})},\ \Eprint
  {http://arxiv.org/abs/1805.04909} {arXiv:1805.04909 [astro-ph.HE]}
  \BibitemShut {NoStop}%
\bibitem [{\citenamefont {Thorne}(1987)}]{Thorne:300yrs}%
  \BibitemOpen
  \bibfield  {author} {\bibinfo {author} {\bibfnamefont {K.}~\bibnamefont
  {Thorne}},\ }\href@noop {} {\emph {\bibinfo {title} {Three Hundred Years of
  Gravitation}}},\ edited by\ \bibinfo {editor} {\bibfnamefont {S.~W.}\
  \bibnamefont {Hawking}}\ and\ \bibinfo {editor} {\bibfnamefont
  {W.}~\bibnamefont {Israel}}\ (\bibinfo  {publisher} {Cambridge University
  Press},\ \bibinfo {year} {1987})\ pp.\ \bibinfo {pages}
  {330--458}\BibitemShut {NoStop}%
\bibitem [{\citenamefont {Allen}\ \emph {et~al.}(2012)\citenamefont {Allen},
  \citenamefont {Anderson}, \citenamefont {Brady}, \citenamefont {Brown},\ and\
  \citenamefont {Creighton}}]{Allen:2005fk}%
  \BibitemOpen
  \bibfield  {author} {\bibinfo {author} {\bibfnamefont {B.}~\bibnamefont
  {Allen}}, \bibinfo {author} {\bibfnamefont {W.~G.}\ \bibnamefont {Anderson}},
  \bibinfo {author} {\bibfnamefont {P.~R.}\ \bibnamefont {Brady}}, \bibinfo
  {author} {\bibfnamefont {D.~A.}\ \bibnamefont {Brown}}, \ and\ \bibinfo
  {author} {\bibfnamefont {J.~D.}\ \bibnamefont {Creighton}},\ }\href {\doibase
  10.1103/PhysRevD.85.122006} {\bibfield  {journal} {\bibinfo  {journal} {Phys.
  Rev. D}\ }\textbf {\bibinfo {volume} {85}},\ \bibinfo {pages} {122006}
  (\bibinfo {year} {2012})},\ \Eprint {http://arxiv.org/abs/gr-qc/0509116}
  {arXiv:gr-qc/0509116} \BibitemShut {NoStop}%
\bibitem [{\citenamefont {Owen}\ and\ \citenamefont
  {Sathyaprakash}(1999)}]{Owen:1998dk}%
  \BibitemOpen
  \bibfield  {author} {\bibinfo {author} {\bibfnamefont {B.~J.}\ \bibnamefont
  {Owen}}\ and\ \bibinfo {author} {\bibfnamefont {B.}~\bibnamefont
  {Sathyaprakash}},\ }\href {\doibase 10.1103/PhysRevD.60.022002} {\bibfield
  {journal} {\bibinfo  {journal} {Phys. Rev. D}\ }\textbf {\bibinfo {volume}
  {60}},\ \bibinfo {pages} {022002} (\bibinfo {year} {1999})},\ \Eprint
  {http://arxiv.org/abs/gr-qc/9808076} {arXiv:gr-qc/9808076} \BibitemShut
  {NoStop}%
\bibitem [{\citenamefont {Brown}\ \emph {et~al.}(2012)\citenamefont {Brown},
  \citenamefont {Harry}, \citenamefont {Lundgren},\ and\ \citenamefont
  {Nitz}}]{Brown:2012qf}%
  \BibitemOpen
  \bibfield  {author} {\bibinfo {author} {\bibfnamefont {D.~A.}\ \bibnamefont
  {Brown}}, \bibinfo {author} {\bibfnamefont {I.}~\bibnamefont {Harry}},
  \bibinfo {author} {\bibfnamefont {A.}~\bibnamefont {Lundgren}}, \ and\
  \bibinfo {author} {\bibfnamefont {A.~H.}\ \bibnamefont {Nitz}},\ }\href
  {\doibase 10.1103/PhysRevD.86.084017} {\bibfield  {journal} {\bibinfo
  {journal} {Phys. Rev. D}\ }\textbf {\bibinfo {volume} {86}},\ \bibinfo
  {pages} {084017} (\bibinfo {year} {2012})},\ \Eprint
  {http://arxiv.org/abs/1207.6406} {arXiv:1207.6406 [gr-qc]} \BibitemShut
  {NoStop}%
\bibitem [{\citenamefont {Harry}\ \emph {et~al.}(2009)\citenamefont {Harry},
  \citenamefont {Allen},\ and\ \citenamefont {Sathyaprakash}}]{Harry:2009ea}%
  \BibitemOpen
  \bibfield  {author} {\bibinfo {author} {\bibfnamefont {I.~W.}\ \bibnamefont
  {Harry}}, \bibinfo {author} {\bibfnamefont {B.}~\bibnamefont {Allen}}, \ and\
  \bibinfo {author} {\bibfnamefont {B.}~\bibnamefont {Sathyaprakash}},\ }\href
  {\doibase 10.1103/PhysRevD.80.104014} {\bibfield  {journal} {\bibinfo
  {journal} {Phys. Rev. D}\ }\textbf {\bibinfo {volume} {80}},\ \bibinfo
  {pages} {104014} (\bibinfo {year} {2009})},\ \Eprint
  {http://arxiv.org/abs/0908.2090} {arXiv:0908.2090 [gr-qc]} \BibitemShut
  {NoStop}%
\bibitem [{\citenamefont {Roy}\ \emph {et~al.}(2019)\citenamefont {Roy},
  \citenamefont {Sengupta},\ and\ \citenamefont {Ajith}}]{Roy:2017oul}%
  \BibitemOpen
  \bibfield  {author} {\bibinfo {author} {\bibfnamefont {S.}~\bibnamefont
  {Roy}}, \bibinfo {author} {\bibfnamefont {A.~S.}\ \bibnamefont {Sengupta}}, \
  and\ \bibinfo {author} {\bibfnamefont {P.}~\bibnamefont {Ajith}},\ }\href
  {\doibase 10.1103/PhysRevD.99.024048} {\bibfield  {journal} {\bibinfo
  {journal} {Phys. Rev.}\ }\textbf {\bibinfo {volume} {D99}},\ \bibinfo {pages}
  {024048} (\bibinfo {year} {2019})},\ \Eprint
  {http://arxiv.org/abs/1711.08743} {arXiv:1711.08743 [gr-qc]} \BibitemShut
  {NoStop}%
\bibitem [{\citenamefont {Allen}(2005)}]{Allen:2004gu}%
  \BibitemOpen
  \bibfield  {author} {\bibinfo {author} {\bibfnamefont {B.}~\bibnamefont
  {Allen}},\ }\href {\doibase 10.1103/PhysRevD.71.062001} {\bibfield  {journal}
  {\bibinfo  {journal} {Phys. Rev. D}\ }\textbf {\bibinfo {volume} {71}},\
  \bibinfo {pages} {062001} (\bibinfo {year} {2005})},\ \Eprint
  {http://arxiv.org/abs/gr-qc/0405045} {arXiv:gr-qc/0405045} \BibitemShut
  {NoStop}%
\bibitem [{\citenamefont {Messick}\ \emph {et~al.}(2017)\citenamefont {Messick}
  \emph {et~al.}}]{Messick:2016aqy}%
  \BibitemOpen
  \bibfield  {author} {\bibinfo {author} {\bibfnamefont {C.}~\bibnamefont
  {Messick}} \emph {et~al.},\ }\href {\doibase 10.1103/PhysRevD.95.042001}
  {\bibfield  {journal} {\bibinfo  {journal} {Phys. Rev. D}\ }\textbf {\bibinfo
  {volume} {95}},\ \bibinfo {pages} {042001} (\bibinfo {year} {2017})},\
  \Eprint {http://arxiv.org/abs/1604.04324} {arXiv:1604.04324 [astro-ph.IM]}
  \BibitemShut {NoStop}%
\bibitem [{\citenamefont {Usman}\ \emph {et~al.}(2016)\citenamefont {Usman}
  \emph {et~al.}}]{Usman:2015kfa}%
  \BibitemOpen
  \bibfield  {author} {\bibinfo {author} {\bibfnamefont {S.~A.}\ \bibnamefont
  {Usman}} \emph {et~al.},\ }\href {\doibase 10.1088/0264-9381/33/21/215004}
  {\bibfield  {journal} {\bibinfo  {journal} {Class. Quant. Grav.}\ }\textbf
  {\bibinfo {volume} {33}},\ \bibinfo {pages} {215004} (\bibinfo {year}
  {2016})},\ \Eprint {http://arxiv.org/abs/1508.02357} {arXiv:1508.02357
  [gr-qc]} \BibitemShut {NoStop}%
\bibitem [{\citenamefont {Biscoveanu}\ \emph {et~al.}(2019)\citenamefont
  {Biscoveanu}, \citenamefont {Vitale},\ and\ \citenamefont
  {Haster}}]{Biscoveanu:2019ugx}%
  \BibitemOpen
  \bibfield  {author} {\bibinfo {author} {\bibfnamefont {S.}~\bibnamefont
  {Biscoveanu}}, \bibinfo {author} {\bibfnamefont {S.}~\bibnamefont {Vitale}},
  \ and\ \bibinfo {author} {\bibfnamefont {C.-J.}\ \bibnamefont {Haster}},\
  }\href {\doibase 10.3847/2041-8213/ab479e} {\bibfield  {journal} {\bibinfo
  {journal} {Astrophys. J. Lett.}\ }\textbf {\bibinfo {volume} {884}},\
  \bibinfo {pages} {L32} (\bibinfo {year} {2019})},\ \Eprint
  {http://arxiv.org/abs/1908.03592} {arXiv:1908.03592 [astro-ph.HE]}
  \BibitemShut {NoStop}%
\bibitem [{\citenamefont {Creighton}\ and\ \citenamefont
  {Anderson}(2011)}]{Creighton:2011zz}%
  \BibitemOpen
  \bibfield  {author} {\bibinfo {author} {\bibfnamefont {J.~D.}\ \bibnamefont
  {Creighton}}\ and\ \bibinfo {author} {\bibfnamefont {W.~G.}\ \bibnamefont
  {Anderson}},\ }\href@noop {} {\emph {\bibinfo {title} {{Gravitational-wave
  physics and astronomy: An introduction to theory, experiment and data
  analysis}}}}\ (\bibinfo {year} {2011})\BibitemShut {NoStop}%
\bibitem [{\citenamefont {Apostolatos}\ \emph {et~al.}(1994)\citenamefont
  {Apostolatos}, \citenamefont {Cutler}, \citenamefont {Sussman},\ and\
  \citenamefont {Thorne}}]{Apostolatos:1994mx}%
  \BibitemOpen
  \bibfield  {author} {\bibinfo {author} {\bibfnamefont {T.~A.}\ \bibnamefont
  {Apostolatos}}, \bibinfo {author} {\bibfnamefont {C.}~\bibnamefont {Cutler}},
  \bibinfo {author} {\bibfnamefont {G.~J.}\ \bibnamefont {Sussman}}, \ and\
  \bibinfo {author} {\bibfnamefont {K.~S.}\ \bibnamefont {Thorne}},\ }\href
  {\doibase 10.1103/PhysRevD.49.6274} {\bibfield  {journal} {\bibinfo
  {journal} {Phys. Rev.}\ }\textbf {\bibinfo {volume} {D49}},\ \bibinfo {pages}
  {6274} (\bibinfo {year} {1994})}\BibitemShut {NoStop}%
\bibitem [{\citenamefont {Flanagan}\ and\ \citenamefont
  {Hinderer}(2008)}]{Flanagan:2007ix}%
  \BibitemOpen
  \bibfield  {author} {\bibinfo {author} {\bibfnamefont {E.~E.}\ \bibnamefont
  {Flanagan}}\ and\ \bibinfo {author} {\bibfnamefont {T.}~\bibnamefont
  {Hinderer}},\ }\href {\doibase 10.1103/PhysRevD.77.021502} {\bibfield
  {journal} {\bibinfo  {journal} {Phys. Rev.}\ }\textbf {\bibinfo {volume}
  {D77}},\ \bibinfo {pages} {021502} (\bibinfo {year} {2008})},\ \Eprint
  {http://arxiv.org/abs/0709.1915} {arXiv:0709.1915 [astro-ph]} \BibitemShut
  {NoStop}%
\bibitem [{\citenamefont {Field}\ \emph {et~al.}(2011)\citenamefont {Field},
  \citenamefont {Galley}, \citenamefont {Herrmann}, \citenamefont {Hesthaven},
  \citenamefont {Ochsner},\ and\ \citenamefont {Tiglio}}]{Field:2011mf}%
  \BibitemOpen
  \bibfield  {author} {\bibinfo {author} {\bibfnamefont {S.~E.}\ \bibnamefont
  {Field}}, \bibinfo {author} {\bibfnamefont {C.~R.}\ \bibnamefont {Galley}},
  \bibinfo {author} {\bibfnamefont {F.}~\bibnamefont {Herrmann}}, \bibinfo
  {author} {\bibfnamefont {J.~S.}\ \bibnamefont {Hesthaven}}, \bibinfo {author}
  {\bibfnamefont {E.}~\bibnamefont {Ochsner}}, \ and\ \bibinfo {author}
  {\bibfnamefont {M.}~\bibnamefont {Tiglio}},\ }\href {\doibase
  10.1103/PhysRevLett.106.221102} {\bibfield  {journal} {\bibinfo  {journal}
  {Phys. Rev. Lett.}\ }\textbf {\bibinfo {volume} {106}},\ \bibinfo {pages}
  {221102} (\bibinfo {year} {2011})},\ \Eprint {http://arxiv.org/abs/1101.3765}
  {arXiv:1101.3765 [gr-qc]} \BibitemShut {NoStop}%
\bibitem [{\citenamefont {Field}\ \emph {et~al.}(2014)\citenamefont {Field},
  \citenamefont {Galley}, \citenamefont {Hesthaven}, \citenamefont {Kaye},\
  and\ \citenamefont {Tiglio}}]{Field:2013cfa}%
  \BibitemOpen
  \bibfield  {author} {\bibinfo {author} {\bibfnamefont {S.~E.}\ \bibnamefont
  {Field}}, \bibinfo {author} {\bibfnamefont {C.~R.}\ \bibnamefont {Galley}},
  \bibinfo {author} {\bibfnamefont {J.~S.}\ \bibnamefont {Hesthaven}}, \bibinfo
  {author} {\bibfnamefont {J.}~\bibnamefont {Kaye}}, \ and\ \bibinfo {author}
  {\bibfnamefont {M.}~\bibnamefont {Tiglio}},\ }\href {\doibase
  10.1103/PhysRevX.4.031006} {\bibfield  {journal} {\bibinfo  {journal} {Phys.
  Rev.}\ }\textbf {\bibinfo {volume} {X4}},\ \bibinfo {pages} {031006}
  (\bibinfo {year} {2014})},\ \Eprint {http://arxiv.org/abs/1308.3565}
  {arXiv:1308.3565 [gr-qc]} \BibitemShut {NoStop}%
\bibitem [{\citenamefont {Blanchet}(2014)}]{Blanchet:2013haa}%
  \BibitemOpen
  \bibfield  {author} {\bibinfo {author} {\bibfnamefont {L.}~\bibnamefont
  {Blanchet}},\ }\href {\doibase 10.12942/lrr-2014-2} {\bibfield  {journal}
  {\bibinfo  {journal} {Living Rev. Rel.}\ }\textbf {\bibinfo {volume} {17}},\
  \bibinfo {pages} {2} (\bibinfo {year} {2014})},\ \Eprint
  {http://arxiv.org/abs/1310.1528} {arXiv:1310.1528 [gr-qc]} \BibitemShut
  {NoStop}%
\bibitem [{\citenamefont {Cutler}\ and\ \citenamefont
  {Flanagan}(1994)}]{Cutler:1994ys}%
  \BibitemOpen
  \bibfield  {author} {\bibinfo {author} {\bibfnamefont {C.}~\bibnamefont
  {Cutler}}\ and\ \bibinfo {author} {\bibfnamefont {E.~E.}\ \bibnamefont
  {Flanagan}},\ }\href {\doibase 10.1103/PhysRevD.49.2658} {\bibfield
  {journal} {\bibinfo  {journal} {Phys. Rev.}\ }\textbf {\bibinfo {volume}
  {D49}},\ \bibinfo {pages} {2658} (\bibinfo {year} {1994})},\ \Eprint
  {http://arxiv.org/abs/gr-qc/9402014} {arXiv:gr-qc/9402014 [gr-qc]}
  \BibitemShut {NoStop}%
\bibitem [{\citenamefont {Poisson}\ and\ \citenamefont
  {Will}(1995)}]{Poisson:1995ef}%
  \BibitemOpen
  \bibfield  {author} {\bibinfo {author} {\bibfnamefont {E.}~\bibnamefont
  {Poisson}}\ and\ \bibinfo {author} {\bibfnamefont {C.~M.}\ \bibnamefont
  {Will}},\ }\href {\doibase 10.1103/PhysRevD.52.848} {\bibfield  {journal}
  {\bibinfo  {journal} {Phys. Rev.}\ }\textbf {\bibinfo {volume} {D52}},\
  \bibinfo {pages} {848} (\bibinfo {year} {1995})},\ \Eprint
  {http://arxiv.org/abs/gr-qc/9502040} {arXiv:gr-qc/9502040 [gr-qc]}
  \BibitemShut {NoStop}%
\bibitem [{\citenamefont {Ohme}\ \emph {et~al.}(2013)\citenamefont {Ohme},
  \citenamefont {Nielsen}, \citenamefont {Keppel},\ and\ \citenamefont
  {Lundgren}}]{Ohme:2013nsa}%
  \BibitemOpen
  \bibfield  {author} {\bibinfo {author} {\bibfnamefont {F.}~\bibnamefont
  {Ohme}}, \bibinfo {author} {\bibfnamefont {A.~B.}\ \bibnamefont {Nielsen}},
  \bibinfo {author} {\bibfnamefont {D.}~\bibnamefont {Keppel}}, \ and\ \bibinfo
  {author} {\bibfnamefont {A.}~\bibnamefont {Lundgren}},\ }\href {\doibase
  10.1103/PhysRevD.88.042002} {\bibfield  {journal} {\bibinfo  {journal} {Phys.
  Rev.}\ }\textbf {\bibinfo {volume} {D88}},\ \bibinfo {pages} {042002}
  (\bibinfo {year} {2013})},\ \Eprint {http://arxiv.org/abs/1304.7017}
  {arXiv:1304.7017 [gr-qc]} \BibitemShut {NoStop}%
\bibitem [{\citenamefont {Owen}(1996)}]{Owen:1995tm}%
  \BibitemOpen
  \bibfield  {author} {\bibinfo {author} {\bibfnamefont {B.~J.}\ \bibnamefont
  {Owen}},\ }\href {\doibase 10.1103/PhysRevD.53.6749} {\bibfield  {journal}
  {\bibinfo  {journal} {Phys. Rev. D}\ }\textbf {\bibinfo {volume} {53}},\
  \bibinfo {pages} {6749} (\bibinfo {year} {1996})},\ \Eprint
  {http://arxiv.org/abs/gr-qc/9511032} {arXiv:gr-qc/9511032} \BibitemShut
  {NoStop}%
\bibitem [{\citenamefont {Dal~Canton}\ and\ \citenamefont
  {Harry}(2017)}]{DalCanton:2017ala}%
  \BibitemOpen
  \bibfield  {author} {\bibinfo {author} {\bibfnamefont {T.}~\bibnamefont
  {Dal~Canton}}\ and\ \bibinfo {author} {\bibfnamefont {I.~W.}\ \bibnamefont
  {Harry}},\ }\href@noop {} {\  (\bibinfo {year} {2017})},\ \Eprint
  {http://arxiv.org/abs/1705.01845} {arXiv:1705.01845 [gr-qc]} \BibitemShut
  {NoStop}%
\bibitem [{\citenamefont {Mukherjee}\ \emph {et~al.}(2018)\citenamefont
  {Mukherjee} \emph {et~al.}}]{Mukherjee:2018yra}%
  \BibitemOpen
  \bibfield  {author} {\bibinfo {author} {\bibfnamefont {D.}~\bibnamefont
  {Mukherjee}} \emph {et~al.},\ }\href@noop {} {\  (\bibinfo {year} {2018})},\
  \Eprint {http://arxiv.org/abs/1812.05121} {arXiv:1812.05121 [astro-ph.IM]}
  \BibitemShut {NoStop}%
\bibitem [{\citenamefont {Fairhurst}(2011)}]{Fairhurst:2010is}%
  \BibitemOpen
  \bibfield  {author} {\bibinfo {author} {\bibfnamefont {S.}~\bibnamefont
  {Fairhurst}},\ }\href {\doibase 10.1088/0264-9381/28/10/105021} {\bibfield
  {journal} {\bibinfo  {journal} {Class. Quant. Grav.}\ }\textbf {\bibinfo
  {volume} {28}},\ \bibinfo {pages} {105021} (\bibinfo {year} {2011})},\
  \Eprint {http://arxiv.org/abs/1010.6192} {arXiv:1010.6192 [gr-qc]}
  \BibitemShut {NoStop}%
\bibitem [{\citenamefont {Dudi}\ \emph {et~al.}(2018)\citenamefont {Dudi},
  \citenamefont {Pannarale}, \citenamefont {Dietrich}, \citenamefont {Hannam},
  \citenamefont {Bernuzzi}, \citenamefont {Ohme},\ and\ \citenamefont
  {Br\"ugmann}}]{Dudi:2018jzn}%
  \BibitemOpen
  \bibfield  {author} {\bibinfo {author} {\bibfnamefont {R.}~\bibnamefont
  {Dudi}}, \bibinfo {author} {\bibfnamefont {F.}~\bibnamefont {Pannarale}},
  \bibinfo {author} {\bibfnamefont {T.}~\bibnamefont {Dietrich}}, \bibinfo
  {author} {\bibfnamefont {M.}~\bibnamefont {Hannam}}, \bibinfo {author}
  {\bibfnamefont {S.}~\bibnamefont {Bernuzzi}}, \bibinfo {author}
  {\bibfnamefont {F.}~\bibnamefont {Ohme}}, \ and\ \bibinfo {author}
  {\bibfnamefont {B.}~\bibnamefont {Br\"ugmann}},\ }\href {\doibase
  10.1103/PhysRevD.98.084061} {\bibfield  {journal} {\bibinfo  {journal} {Phys.
  Rev. D}\ }\textbf {\bibinfo {volume} {98}},\ \bibinfo {pages} {084061}
  (\bibinfo {year} {2018})},\ \Eprint {http://arxiv.org/abs/1808.09749}
  {arXiv:1808.09749 [gr-qc]} \BibitemShut {NoStop}%
\bibitem [{\citenamefont {Rhoades}\ and\ \citenamefont
  {Ruffini}(1974)}]{Rhoades:1974fn}%
  \BibitemOpen
  \bibfield  {author} {\bibinfo {author} {\bibfnamefont {C.~E.}\ \bibnamefont
  {Rhoades}, \bibfnamefont {Jr.}}\ and\ \bibinfo {author} {\bibfnamefont
  {R.}~\bibnamefont {Ruffini}},\ }\href {\doibase 10.1103/PhysRevLett.32.324}
  {\bibfield  {journal} {\bibinfo  {journal} {Phys. Rev. Lett.}\ }\textbf
  {\bibinfo {volume} {32}},\ \bibinfo {pages} {324} (\bibinfo {year}
  {1974})}\BibitemShut {NoStop}%
\bibitem [{\citenamefont {Kalogera}\ and\ \citenamefont
  {Baym}(1996)}]{Kalogera:1996ci}%
  \BibitemOpen
  \bibfield  {author} {\bibinfo {author} {\bibfnamefont {V.}~\bibnamefont
  {Kalogera}}\ and\ \bibinfo {author} {\bibfnamefont {G.}~\bibnamefont
  {Baym}},\ }\href {\doibase 10.1086/310296} {\bibfield  {journal} {\bibinfo
  {journal} {Astrophys. J.}\ }\textbf {\bibinfo {volume} {470}},\ \bibinfo
  {pages} {L61} (\bibinfo {year} {1996})},\ \Eprint
  {http://arxiv.org/abs/astro-ph/9608059} {arXiv:astro-ph/9608059 [astro-ph]}
  \BibitemShut {NoStop}%
\bibitem [{\citenamefont {Abbott}\ \emph
  {et~al.}(2019{\natexlab{b}})\citenamefont {Abbott} \emph
  {et~al.}}]{Abbott:2018wiz}%
  \BibitemOpen
  \bibfield  {author} {\bibinfo {author} {\bibfnamefont {B.~P.}\ \bibnamefont
  {Abbott}} \emph {et~al.} (\bibinfo {collaboration} {LIGO Scientific,
  Virgo}),\ }\href {\doibase 10.1103/PhysRevX.9.011001} {\bibfield  {journal}
  {\bibinfo  {journal} {Phys. Rev.}\ }\textbf {\bibinfo {volume} {X9}},\
  \bibinfo {pages} {011001} (\bibinfo {year} {2019}{\natexlab{b}})},\ \Eprint
  {http://arxiv.org/abs/1805.11579} {arXiv:1805.11579 [gr-qc]} \BibitemShut
  {NoStop}%
\bibitem [{\citenamefont {Burgay}\ \emph {et~al.}(2003)\citenamefont {Burgay}
  \emph {et~al.}}]{Burgay:2003jj}%
  \BibitemOpen
  \bibfield  {author} {\bibinfo {author} {\bibfnamefont {M.}~\bibnamefont
  {Burgay}} \emph {et~al.},\ }\href {\doibase 10.1038/nature02124} {\bibfield
  {journal} {\bibinfo  {journal} {Nature}\ }\textbf {\bibinfo {volume} {426}},\
  \bibinfo {pages} {531} (\bibinfo {year} {2003})},\ \Eprint
  {http://arxiv.org/abs/astro-ph/0312071} {arXiv:astro-ph/0312071 [astro-ph]}
  \BibitemShut {NoStop}%
\bibitem [{\citenamefont {Lo}\ and\ \citenamefont {Lin}(2011)}]{Lo:2010bj}%
  \BibitemOpen
  \bibfield  {author} {\bibinfo {author} {\bibfnamefont {K.-W.}\ \bibnamefont
  {Lo}}\ and\ \bibinfo {author} {\bibfnamefont {L.-M.}\ \bibnamefont {Lin}},\
  }\href {\doibase 10.1088/0004-637X/728/1/12} {\bibfield  {journal} {\bibinfo
  {journal} {Astrophys. J.}\ }\textbf {\bibinfo {volume} {728}},\ \bibinfo
  {pages} {12} (\bibinfo {year} {2011})},\ \Eprint
  {http://arxiv.org/abs/1011.3563} {arXiv:1011.3563 [astro-ph.HE]} \BibitemShut
  {NoStop}%
\bibitem [{\citenamefont {Roulet}\ \emph {et~al.}(2019)\citenamefont {Roulet},
  \citenamefont {Dai}, \citenamefont {Venumadhav}, \citenamefont {Zackay},\
  and\ \citenamefont {Zaldarriaga}}]{Roulet:2019hzy}%
  \BibitemOpen
  \bibfield  {author} {\bibinfo {author} {\bibfnamefont {J.}~\bibnamefont
  {Roulet}}, \bibinfo {author} {\bibfnamefont {L.}~\bibnamefont {Dai}},
  \bibinfo {author} {\bibfnamefont {T.}~\bibnamefont {Venumadhav}}, \bibinfo
  {author} {\bibfnamefont {B.}~\bibnamefont {Zackay}}, \ and\ \bibinfo {author}
  {\bibfnamefont {M.}~\bibnamefont {Zaldarriaga}},\ }\href {\doibase
  10.1103/PhysRevD.99.123022} {\bibfield  {journal} {\bibinfo  {journal} {Phys.
  Rev. D}\ }\textbf {\bibinfo {volume} {99}},\ \bibinfo {pages} {123022}
  (\bibinfo {year} {2019})},\ \Eprint {http://arxiv.org/abs/1904.01683}
  {arXiv:1904.01683 [astro-ph.IM]} \BibitemShut {NoStop}%
\bibitem [{\citenamefont {{LIGO Scientific Collaboration}}(2018)}]{lalsuite}%
  \BibitemOpen
  \bibfield  {author} {\bibinfo {author} {\bibnamefont {{LIGO Scientific
  Collaboration}}},\ }\href {\doibase 10.7935/GT1W-FZ16} {\enquote {\bibinfo
  {title} {{LIGO} {A}lgorithm {L}ibrary - {LALS}uite},}\ }\bibinfo
  {howpublished} {free software (GPL)} (\bibinfo {year} {2018})\BibitemShut
  {NoStop}%
\bibitem [{\citenamefont {Abbott}\ \emph
  {et~al.}(2019{\natexlab{c}})\citenamefont {Abbott} \emph
  {et~al.}}]{Abbott:2019ebz}%
  \BibitemOpen
  \bibfield  {author} {\bibinfo {author} {\bibfnamefont {R.}~\bibnamefont
  {Abbott}} \emph {et~al.} (\bibinfo {collaboration} {LIGO Scientific,
  Virgo}),\ }\href@noop {} {\  (\bibinfo {year} {2019}{\natexlab{c}})},\
  \Eprint {http://arxiv.org/abs/1912.11716} {arXiv:1912.11716 [gr-qc]}
  \BibitemShut {NoStop}%
\bibitem [{\citenamefont {Sachdev}\ \emph {et~al.}(2019)\citenamefont {Sachdev}
  \emph {et~al.}}]{Sachdev:2019vvd}%
  \BibitemOpen
  \bibfield  {author} {\bibinfo {author} {\bibfnamefont {S.}~\bibnamefont
  {Sachdev}} \emph {et~al.},\ }\href@noop {} {\  (\bibinfo {year} {2019})},\
  \Eprint {http://arxiv.org/abs/1901.08580} {arXiv:1901.08580 [gr-qc]}
  \BibitemShut {NoStop}%
\bibitem [{\citenamefont {Chan}\ \emph {et~al.}(2018)\citenamefont {Chan},
  \citenamefont {Messenger}, \citenamefont {Heng},\ and\ \citenamefont
  {Hendry}}]{Chan:2018csa}%
  \BibitemOpen
  \bibfield  {author} {\bibinfo {author} {\bibfnamefont {M.~L.}\ \bibnamefont
  {Chan}}, \bibinfo {author} {\bibfnamefont {C.}~\bibnamefont {Messenger}},
  \bibinfo {author} {\bibfnamefont {I.~S.}\ \bibnamefont {Heng}}, \ and\
  \bibinfo {author} {\bibfnamefont {M.}~\bibnamefont {Hendry}},\ }\href
  {\doibase 10.1103/PhysRevD.97.123014} {\bibfield  {journal} {\bibinfo
  {journal} {Phys. Rev. D}\ }\textbf {\bibinfo {volume} {97}},\ \bibinfo
  {pages} {123014} (\bibinfo {year} {2018})},\ \Eprint
  {http://arxiv.org/abs/1803.09680} {arXiv:1803.09680 [astro-ph.HE]}
  \BibitemShut {NoStop}%
\bibitem [{\citenamefont {Mandel}\ \emph {et~al.}(2014)\citenamefont {Mandel},
  \citenamefont {Berry}, \citenamefont {Ohme}, \citenamefont {Fairhurst},\ and\
  \citenamefont {Farr}}]{Mandel:2014tca}%
  \BibitemOpen
  \bibfield  {author} {\bibinfo {author} {\bibfnamefont {I.}~\bibnamefont
  {Mandel}}, \bibinfo {author} {\bibfnamefont {C.~P.}\ \bibnamefont {Berry}},
  \bibinfo {author} {\bibfnamefont {F.}~\bibnamefont {Ohme}}, \bibinfo {author}
  {\bibfnamefont {S.}~\bibnamefont {Fairhurst}}, \ and\ \bibinfo {author}
  {\bibfnamefont {W.~M.}\ \bibnamefont {Farr}},\ }\href {\doibase
  10.1088/0264-9381/31/15/155005} {\bibfield  {journal} {\bibinfo  {journal}
  {Class. Quant. Grav.}\ }\textbf {\bibinfo {volume} {31}},\ \bibinfo {pages}
  {155005} (\bibinfo {year} {2014})},\ \Eprint {http://arxiv.org/abs/1404.2382}
  {arXiv:1404.2382 [gr-qc]} \BibitemShut {NoStop}%
\bibitem [{\citenamefont {Cook}\ \emph {et~al.}(2006)\citenamefont {Cook},
  \citenamefont {Gelman},\ and\ \citenamefont
  {Rubin}}]{doi:10.1198/106186006X136976}%
  \BibitemOpen
  \bibfield  {author} {\bibinfo {author} {\bibfnamefont {S.~R.}\ \bibnamefont
  {Cook}}, \bibinfo {author} {\bibfnamefont {A.}~\bibnamefont {Gelman}}, \ and\
  \bibinfo {author} {\bibfnamefont {D.~B.}\ \bibnamefont {Rubin}},\ }\href
  {\doibase 10.1198/106186006X136976} {\bibfield  {journal} {\bibinfo
  {journal} {Journal of Computational and Graphical Statistics}\ }\textbf
  {\bibinfo {volume} {15}},\ \bibinfo {pages} {675} (\bibinfo {year} {2006})},\
  \Eprint {http://arxiv.org/abs/https://doi.org/10.1198/106186006X136976}
  {https://doi.org/10.1198/106186006X136976} \BibitemShut {NoStop}%
\bibitem [{\citenamefont {{Talts}}\ \emph {et~al.}(2018)\citenamefont
  {{Talts}}, \citenamefont {{Betancourt}}, \citenamefont {{Simpson}},
  \citenamefont {{Vehtari}},\ and\ \citenamefont
  {{Gelman}}}]{2018arXiv180406788T}%
  \BibitemOpen
  \bibfield  {author} {\bibinfo {author} {\bibfnamefont {S.}~\bibnamefont
  {{Talts}}}, \bibinfo {author} {\bibfnamefont {M.}~\bibnamefont
  {{Betancourt}}}, \bibinfo {author} {\bibfnamefont {D.}~\bibnamefont
  {{Simpson}}}, \bibinfo {author} {\bibfnamefont {A.}~\bibnamefont
  {{Vehtari}}}, \ and\ \bibinfo {author} {\bibfnamefont {A.}~\bibnamefont
  {{Gelman}}},\ }\href@noop {} {\bibfield  {journal} {\bibinfo  {journal}
  {arXiv e-prints}\ ,\ \bibinfo {eid} {arXiv:1804.06788}} (\bibinfo {year}
  {2018})},\ \Eprint {http://arxiv.org/abs/1804.06788} {arXiv:1804.06788
  [stat.ME]} \BibitemShut {NoStop}%
\bibitem [{\citenamefont {Sidery}\ \emph {et~al.}(2014)\citenamefont {Sidery}
  \emph {et~al.}}]{Sidery:2013zua}%
  \BibitemOpen
  \bibfield  {author} {\bibinfo {author} {\bibfnamefont {T.}~\bibnamefont
  {Sidery}} \emph {et~al.},\ }\href {\doibase 10.1103/PhysRevD.89.084060}
  {\bibfield  {journal} {\bibinfo  {journal} {Phys. Rev. D}\ }\textbf {\bibinfo
  {volume} {89}},\ \bibinfo {pages} {084060} (\bibinfo {year} {2014})},\
  \Eprint {http://arxiv.org/abs/1312.6013} {arXiv:1312.6013 [astro-ph.IM]}
  \BibitemShut {NoStop}%
\bibitem [{\citenamefont {Berry}\ \emph {et~al.}(2015)\citenamefont {Berry}
  \emph {et~al.}}]{Berry:2014jja}%
  \BibitemOpen
  \bibfield  {author} {\bibinfo {author} {\bibfnamefont {C.~P.}\ \bibnamefont
  {Berry}} \emph {et~al.},\ }\href {\doibase 10.1088/0004-637X/804/2/114}
  {\bibfield  {journal} {\bibinfo  {journal} {Astrophys. J.}\ }\textbf
  {\bibinfo {volume} {804}},\ \bibinfo {pages} {114} (\bibinfo {year}
  {2015})},\ \Eprint {http://arxiv.org/abs/1411.6934} {arXiv:1411.6934
  [astro-ph.HE]} \BibitemShut {NoStop}%
\bibitem [{\citenamefont {Biwer}\ \emph {et~al.}(2019)\citenamefont {Biwer},
  \citenamefont {Capano}, \citenamefont {De}, \citenamefont {Cabero},
  \citenamefont {Brown}, \citenamefont {Nitz},\ and\ \citenamefont
  {Raymond}}]{Biwer:2018osg}%
  \BibitemOpen
  \bibfield  {author} {\bibinfo {author} {\bibfnamefont {C.}~\bibnamefont
  {Biwer}}, \bibinfo {author} {\bibfnamefont {C.~D.}\ \bibnamefont {Capano}},
  \bibinfo {author} {\bibfnamefont {S.}~\bibnamefont {De}}, \bibinfo {author}
  {\bibfnamefont {M.}~\bibnamefont {Cabero}}, \bibinfo {author} {\bibfnamefont
  {D.~A.}\ \bibnamefont {Brown}}, \bibinfo {author} {\bibfnamefont {A.~H.}\
  \bibnamefont {Nitz}}, \ and\ \bibinfo {author} {\bibfnamefont
  {V.}~\bibnamefont {Raymond}},\ }\href {\doibase 10.1088/1538-3873/aaef0b}
  {\bibfield  {journal} {\bibinfo  {journal} {Publ. Astron. Soc. Pac.}\
  }\textbf {\bibinfo {volume} {131}},\ \bibinfo {pages} {024503} (\bibinfo
  {year} {2019})},\ \Eprint {http://arxiv.org/abs/1807.10312} {arXiv:1807.10312
  [astro-ph.IM]} \BibitemShut {NoStop}%
\bibitem [{\citenamefont {Del~Pozzo}\ \emph {et~al.}(2018)\citenamefont
  {Del~Pozzo}, \citenamefont {Berry}, \citenamefont {Ghosh}, \citenamefont
  {Haines}, \citenamefont {Singer},\ and\ \citenamefont
  {Vecchio}}]{DelPozzo:2018dpu}%
  \BibitemOpen
  \bibfield  {author} {\bibinfo {author} {\bibfnamefont {W.}~\bibnamefont
  {Del~Pozzo}}, \bibinfo {author} {\bibfnamefont {C.}~\bibnamefont {Berry}},
  \bibinfo {author} {\bibfnamefont {A.}~\bibnamefont {Ghosh}}, \bibinfo
  {author} {\bibfnamefont {T.}~\bibnamefont {Haines}}, \bibinfo {author}
  {\bibfnamefont {L.}~\bibnamefont {Singer}}, \ and\ \bibinfo {author}
  {\bibfnamefont {A.}~\bibnamefont {Vecchio}},\ }\href {\doibase
  10.1093/mnras/sty1485} {\bibfield  {journal} {\bibinfo  {journal} {Mon. Not.
  Roy. Astron. Soc.}\ }\textbf {\bibinfo {volume} {479}},\ \bibinfo {pages}
  {601} (\bibinfo {year} {2018})},\ \Eprint {http://arxiv.org/abs/1801.08009}
  {arXiv:1801.08009 [astro-ph.IM]} \BibitemShut {NoStop}%
\bibitem [{\citenamefont {Romero-Shaw}\ \emph {et~al.}(2020)\citenamefont
  {Romero-Shaw} \emph {et~al.}}]{Romero-Shaw:2020owr}%
  \BibitemOpen
  \bibfield  {author} {\bibinfo {author} {\bibfnamefont {I.}~\bibnamefont
  {Romero-Shaw}} \emph {et~al.},\ }\href@noop {} {\  (\bibinfo {year}
  {2020})},\ \Eprint {http://arxiv.org/abs/2006.00714} {arXiv:2006.00714
  [astro-ph.IM]} \BibitemShut {NoStop}%
\bibitem [{\citenamefont {Ajith}\ \emph {et~al.}(2011)\citenamefont {Ajith}
  \emph {et~al.}}]{Ajith:2009bn}%
  \BibitemOpen
  \bibfield  {author} {\bibinfo {author} {\bibfnamefont {P.}~\bibnamefont
  {Ajith}} \emph {et~al.},\ }\href {\doibase 10.1103/PhysRevLett.106.241101}
  {\bibfield  {journal} {\bibinfo  {journal} {Phys. Rev. Lett.}\ }\textbf
  {\bibinfo {volume} {106}},\ \bibinfo {pages} {241101} (\bibinfo {year}
  {2011})},\ \Eprint {http://arxiv.org/abs/0909.2867} {arXiv:0909.2867 [gr-qc]}
  \BibitemShut {NoStop}%
\bibitem [{\citenamefont {Santamaria}\ \emph {et~al.}(2010)\citenamefont
  {Santamaria} \emph {et~al.}}]{Santamaria:2010yb}%
  \BibitemOpen
  \bibfield  {author} {\bibinfo {author} {\bibfnamefont {L.}~\bibnamefont
  {Santamaria}} \emph {et~al.},\ }\href {\doibase 10.1103/PhysRevD.82.064016}
  {\bibfield  {journal} {\bibinfo  {journal} {Phys. Rev. D}\ }\textbf {\bibinfo
  {volume} {82}},\ \bibinfo {pages} {064016} (\bibinfo {year} {2010})},\
  \Eprint {http://arxiv.org/abs/1005.3306} {arXiv:1005.3306 [gr-qc]}
  \BibitemShut {NoStop}%
\bibitem [{\citenamefont {Singer}\ \emph {et~al.}(2014)\citenamefont {Singer}
  \emph {et~al.}}]{Singer:2014qca}%
  \BibitemOpen
  \bibfield  {author} {\bibinfo {author} {\bibfnamefont {L.~P.}\ \bibnamefont
  {Singer}} \emph {et~al.},\ }\href {\doibase 10.1088/0004-637X/795/2/105}
  {\bibfield  {journal} {\bibinfo  {journal} {Astrophys. J.}\ }\textbf
  {\bibinfo {volume} {795}},\ \bibinfo {pages} {105} (\bibinfo {year}
  {2014})},\ \Eprint {http://arxiv.org/abs/1404.5623} {arXiv:1404.5623
  [astro-ph.HE]} \BibitemShut {NoStop}%
\bibitem [{\citenamefont {Vitale}\ and\ \citenamefont
  {Chen}(2018)}]{Vitale:2018wlg}%
  \BibitemOpen
  \bibfield  {author} {\bibinfo {author} {\bibfnamefont {S.}~\bibnamefont
  {Vitale}}\ and\ \bibinfo {author} {\bibfnamefont {H.-Y.}\ \bibnamefont
  {Chen}},\ }\href {\doibase 10.1103/PhysRevLett.121.021303} {\bibfield
  {journal} {\bibinfo  {journal} {Phys. Rev. Lett.}\ }\textbf {\bibinfo
  {volume} {121}},\ \bibinfo {pages} {021303} (\bibinfo {year} {2018})},\
  \Eprint {http://arxiv.org/abs/1804.07337} {arXiv:1804.07337 [astro-ph.CO]}
  \BibitemShut {NoStop}%
\bibitem [{\citenamefont {Tagoshi}\ \emph {et~al.}(2014)\citenamefont
  {Tagoshi}, \citenamefont {Mishra}, \citenamefont {Pai},\ and\ \citenamefont
  {Arun}}]{Tagoshi:2014xsa}%
  \BibitemOpen
  \bibfield  {author} {\bibinfo {author} {\bibfnamefont {H.}~\bibnamefont
  {Tagoshi}}, \bibinfo {author} {\bibfnamefont {C.~K.}\ \bibnamefont {Mishra}},
  \bibinfo {author} {\bibfnamefont {A.}~\bibnamefont {Pai}}, \ and\ \bibinfo
  {author} {\bibfnamefont {K.}~\bibnamefont {Arun}},\ }\href {\doibase
  10.1103/PhysRevD.90.024053} {\bibfield  {journal} {\bibinfo  {journal} {Phys.
  Rev. D}\ }\textbf {\bibinfo {volume} {90}},\ \bibinfo {pages} {024053}
  (\bibinfo {year} {2014})},\ \Eprint {http://arxiv.org/abs/1403.6915}
  {arXiv:1403.6915 [gr-qc]} \BibitemShut {NoStop}%
\end{thebibliography}%

\end{document}